# Roadmap on Perovskite Light-Emitting Diodes

*Ziming Chen*[1,*], *Robert L. Z. Hoye*[2,3,*], *Hin-Lap Yip*[4,5*], *Nadesh Fiuza-Maneiro*[6], *Iago López-Fernández*[6], *Clara Otero-Martínez*[6], *Lakshminarayana Polavarapu*[6], *Navendu Mondal*[1], *Alessandro Mirabelli*[7], *Miguel Anaya*[7], *Samuel D. Stranks*[7], *Hui Liu*[8], *Guangyi Shi*[8], *Zhengguo Xiao*[8], *Nakyung Kim*[9], *Yunna Kim*[9], *Byungha Shin*[9], *Jinquan Shi*[10,11], *Mengxia Liu*[10,11], *Qianpeng Zhang*[12], *Zhiyong Fan*[12], *James C. Loy*[13], *Lianfeng Zhao*[14], *Barry P. Rand*[14,15], *Habibul Arfin*[16], *Sajid Saikia*[16], *Angshuman Nag*[16], *Chen Zou*[17], *Lih Y. Lin*[18], *Hengyang Xiang*[19], *Haibo Zeng*[19], *Denghui Liu*[20], *Shi-Jian Su*[20], *Chenhui Wang*[21], *Haizheng Zhong*[21], *Tong-Tong Xuan*[22], *Rong-Jun Xie*[22], *Chunxiong Bao*[23], *Feng Gao*[24], *Xiang Gao*[25], *Chuanjiang Qin*[25], *Young-Hoon Kim*[26,27] and *Matthew C. Beard*[26]

[1] Department of Chemistry and Centre for Processible Electronics, Imperial College London, London W12 0BZ, United Kingdom.

[2] Inorganic Chemistry Laboratory, Department of Chemistry, University of Oxford, South Parks Road, Oxford OX1 3QR, United Kingdom.

[3] Department of Materials, Imperial College London, Exhibition Road, London SW7 2AZ, United Kingdom.

[4] Department of Materials Science and Engineering, City University of Hong Kong, Hong Kong, China.

[5] School of Energy and Environment, City University of Hong Kong, Hong Kong, China.

[6] CINBIO, Universidade de Vigo, Materials Chemistry and Physics Group, Department of Physical Chemistry Campus Universitario As Lagoas, Marcosende 36310, Vigo, Spain.

[7] Department of Chemical Engineering and Biotechnology, University of Cambridge, Cambridge CB3 0AS, United Kingdom.

[8] Department of Physics, CAS Key Laboratory of Strongly Coupled Quantum Matter Physics, University of Science and Technology of China, Hefei, Anhui 230026, China.

[9] Department of Materials Science and Engineering, Korea Advanced Institute of Science and Technology (KAIST), Daejeon 34141, Republic of Korea.

[10] Department of Electrical Engineering, Yale University, New Haven, Connecticut 06511, USA.

[11] Energy Sciences Institute, Yale University, West Haven, Connecticut 06516, USA.

[12] Department of Electronic & Computer Engineering, The Hong Kong University of Science and Technology, Clear Water Bay, Kowloon, Hong Kong SAR, China.

[13] Department of Physics, Princeton University, Princeton, NJ, 08544 USA.

[14] Department of Electrical and Computer Engineering, Princeton University, Princeton, NJ, 08544 USA.

[15] Andlinger Center for Energy and the Environment, Princeton University, Princeton, NJ, 08544 USA.

[16] Indian Institute of Science Education and Research (IISER) Pune 411008, India.

[17] Zhejiang University, Hangzhou, Zhejiang, 310058, China.

[18] University of Washington, Seattle, WA 98195, USA.

[19] MIIT Key Laboratory of Advanced Display Materials and Devices, Institute of Optoelectronics & Nanomaterials, College of Materials Science and Engineering, Nanjing University of Science and Technology, Nanjing, 210094, China.

[20] State Key Laboratory of Luminescent Materials and Devices and Institute of Polymer Optoelectronic Materials and Devices, South China University of Technology, 381 Wushan Road, Guangzhou, 510640, P. R. China.

[21] MIIT Key Laboratory for Low-Dimensional Quantum Structure and Devices, School of Materials Science & Engineering, Beijing Institute of Technology, Beijing 100081, People's Republic of China.

[22] Fujian Key Laboratory of Surface and Interface Engineering for High Performance Materials, College of Materials, Xiamen University, Xiamen 361005, China.

[23] National Laboratory of Solid State Microstructures, School of Physics, Nanjing University, Nanjing, 210093, China.

[24] Department of Physics, Chemistry and Biology (IFM), Linköping University, Linköping, Sweden.

[25] State Key Laboratory of Polymer Physics and Chemistry, Changchun Institute of Applied Chemistry, Chinese Academy of Sciences, Changchun 130022, P. R. China

[26] Materials, Chemical & Computational Sciences, National Renewable Energy Laboratory, USA

[27] Department of Energy Engineering, Hanyang University, Republic of Korea.

* Authors to whom any correspondence should be addressed.

**E-mail: z.chen@imperial.ac.uk, robert.hoye@chem.ox.ac.uk, and a.yip@cityu.edu.hk.**

**Abstract**

In recent years, the field of metal-halide perovskite emitters has rapidly emerged as a new community in solid-state lighting. Their exceptional optoelectronic properties have contributed to the rapid rise in external quantum efficiencies (EQEs) in perovskite light-emitting diodes (PeLEDs) from <1% (in 2014) to approaching 30% (in 2023) across a wide range of wavelengths. However, several challenges still hinder their commercialization, including the relatively low EQEs of blue/white devices, limited EQEs in large-area devices, poor device stability, as well as the toxicity of the easily accessible lead components and the solvents used in the synthesis and processing of PeLEDs. This roadmap addresses the current and future challenges in PeLEDs across fundamental and applied research areas, by sharing the community's perspectives. This work will provide the field with practical guidelines to advance PeLED development and facilitate more rapid commercialization.

# 0. Introduction

*Ziming Chen*[1], *Robert L. Z. Hoye*[1,2] and *Hin-Lap Yip*[3]

[1] Imperial College London, United Kingdom

[2] University of Oxford, United Kingdom

[3] City University of Hong Kong, Hong Kong, China

Over the past 9 years, metal-halide perovskites, with their excellent optoelectronic properties and compatibility with low-capital fabrication methods, have successfully joined the solid-state emitter family.[1] Perovskite light-emitting diodes (PeLEDs) share many similarities with organic LEDs (OLEDs), including processability with solution- and vapor-based techniques, as well as compatibility with flexible polymer/paper substrates and printing techniques for large-scale fabrication or micro-patterning.[2,3,4,5,6,7] But a critical advantage of PeLEDs over OLEDs is their much narrower emission peaks [both in photoluminescence (PL) and electroluminescence (EL)], and high color purity across the entire visible and near-infrared wavelength ranges.[8] Therefore, PeLEDs are considered to have great potential to incorporate into numerous lighting-related applications.

Compared with the mature OLED and quantum dot (QD) LED (QLED) fields with over 30 and 20 years of development, respectively, the history of PeLEDs with room-temperature emission is relatively short (less than 10 years), as shown in Figure 1a. Nonetheless, the PeLED field has grown rapidly, and the number of publications already exceeds that of the QLED field, forming a new solid-state-lighting community. There are two main reasons for PeLEDs to gain such considerable research interest. Firstly, PeLEDs are the fastest-developing technology in the history of LEDs in terms of performance (Figure 1b). Thanks to the knowledge and engineering skills developed in the OLED and QLED communities, PeLEDs have improved their devices' external quantum efficiencies (EQEs) from less than 1% to about 30% in just 8 years.[9,10] Secondly, PeLEDs have a low material cost and simple synthetic process. The estimated precursor material cost of representative emitters in PeLEDs is about \$2/g (Figure 1b, inset), while the cost of hole transport layers (HTLs)/electron transport layers (ETLs) is less than \$3/g. Furthermore, PeLEDs have potentially low manufacturing and overhead costs, making them an ideal choice for lighting devices with excellent cost performance in the future, particularly when compared to commercialized OLEDs that have emitter material costs of over \$45/g and HTLs/ETLs transport layer costs of over \$200/g.[11,12]

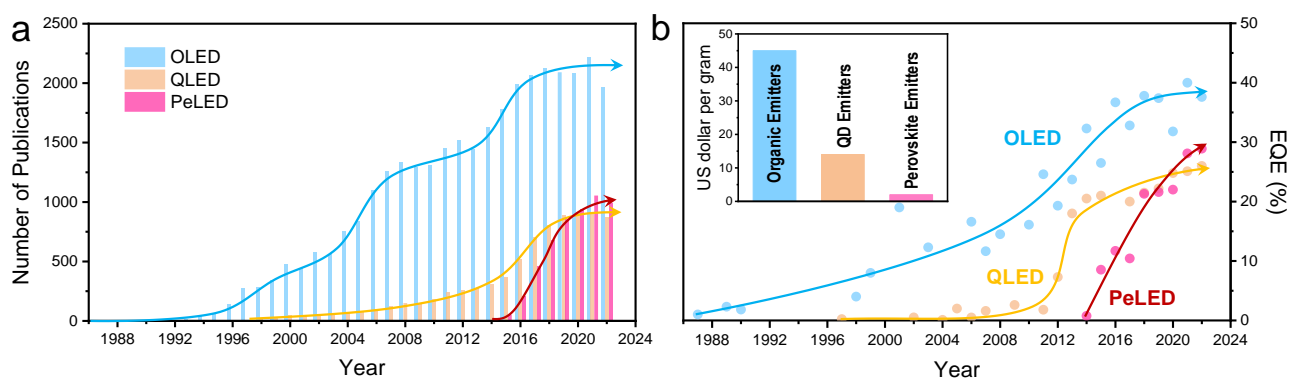

**Figure 1.** The development of OLEDs, QLEDs, and PeLEDs. (a) The number of publications in each year. Data was acquired from Web of Science (https://www.webofscience.com/) by searching with keywords of 'organic light-emitting diodes', 'quantum dot light-emitting diodes', and 'perovskite light-emitting diodes'. (b) Development of device performance. Data collected from Refs. [1,10,13,14,15,16,17,18,19,20,21,22,23]. The inset indicates the precursor material cost of representative emitters in each field.[11]

Although substantial progress in PeLEDs with various colors (i.e., near-infrared, red, green, blue, and white) and forms [i.e., thin-film perovskites and synthesized perovskite QDs/nanocrystals (NCs)] has been made in last decade (Figures 2a and 2b), there is still a long way to go to bring PeLEDs to the commercial level. We have evaluated the status of the current PeLED field from materials to devices to applications and summarized it in Figure 2c. To date, plenty of efforts have been made to develop perovskite materials by compositional and dimensional engineering, which contributes to a wide range of choices of perovskite emitters with various emission colors (from violet to near-infrared) and dimensionality [e.g., three-dimensional (3D) bulk perovskites, two-dimensional (2D)/quasi-2D perovskites, and zero-dimensional (0D) perovskite QDs/NCs].[1,24] With the assistance of organic ligands and charge/exciton confinement effect, the 2D/quasi-2D perovskites and perovskite QDs/NCs can achieve photoluminescence quantum yield (PLQY) of >90% and possess decent stability, while in general the PQLY and stability of 3D perovskites are poorer.[25,26] In terms of devices, the best green PeLEDs have achieved EQEs of >28%, meeting the demand for commercial products (~30%);[10,19] The EQEs of red (>25%) and near-infrared PeLEDs (>22%) are also approaching this criterion, but the development of blue PeLEDs and white PeLEDs (WPeLEDs) lags significantly, with EQEs of only >17% (for blue) and >12% (for white) achieved. More importantly, the overall stability of PeLEDs is still inferior at this stage and difficult to fulfil the commercial requirement (>11000 h for blue; >250000 h for red; >400000 h for green), although the best green and near-infrared PeLEDs show estimated half lifetime ($T_{50}$) of >30000 h.[10,27,28]

Therefore, the current challenge of PeLEDs is mainly in the device and application parts. The low efficiencies in blue and white devices, as well as the poor stability of PeLEDs, are the main bottlenecks to achieving practical applications such as displays, illumination, optical communication, electrically pumped lasers, and polarized-emission LEDs. Moreover, the use of toxic lead in PeLEDs commercially remains controversial.

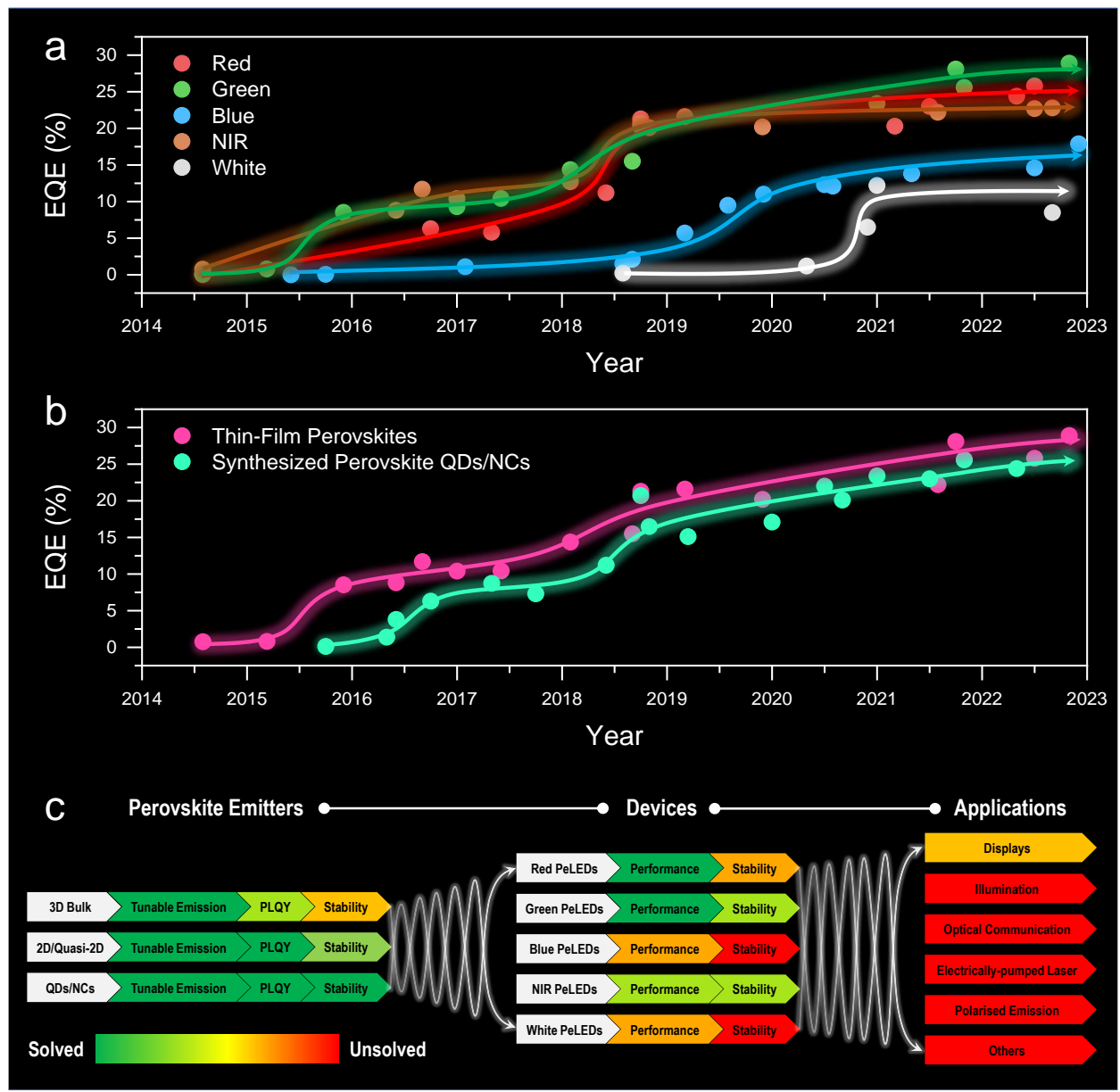

**Figure 2.** Efficiency chart of PeLEDs with (a) various colors and (b) various perovskite forms. Data was collected from Refs. [1,10,19,27,29,30,31,32,33,34,35,36,37,38,39, 40]. (c) Evaluation of perovskite lighting from emitters to devices to applications. The color gradient from green to red represents the status of each issue, ranging from solved to unsolved currently.

In this Roadmap, we have categorized the discussion and perspective into three main sections: i) perovskite emitter materials, ii) PeLED devices, and iii) potential applications.

In the perovskite emitter section (Chapters 1 and 2), we show a colorful world of the perovskite family established by various perovskite films and synthetic perovskite QDs/NCs. Then we delve into the radiative mechanism within different types of perovskites. The different dimensionality of perovskite crystals causes different charge-carrier dynamics that affect radiation.

In the PeLED device section (Chapters 3−12), we propose a standard measuring method for PeLEDs that will help institutionalize the reported EQEs in the field. Then large-scale and vacuum-deposited PeLEDs are demonstrated to exemplify the feature of solution processing and the compatibility with the industrial film deposition process, respectively. The subsequent discussion of two main optimization strategies at the device level nowadays (i.e., interfacial engineering and optical engineering) provides the field with hints to further improve the performance of PeLEDs, especially for the blue and white devices. Additionally, we explore other functional devices such as high-brightness devices, WPeLEDs, and perovskite-based hybrid LEDs to showcase the various research directions, possibilities, and concepts in the field. Finally, two main issues (i.e., poor stability and toxicity of lead) retarding the commercialization of PeLEDs are considered, showing a future route for the practical use of PeLEDs.

In the application section (Chapters 13−17), we show a wide range of possibilities for future applications of PeLEDs. With their high color purity, perovskite emitters are excellent for display applications, achieving the International Telecommunication Union Recommendation BT 2020 (Rec. 2020) standard. In fact, a liquid crystal display prototype based on perovskite light-emitting films was successfully demonstrated by Zhijing Nanotech and TCL in 2018.[41] Optical communication is another potential application for PeLEDs due to their high brightness and response speed. Finally, devices with other functional emissions like lasing and polarized emission are also discussed, which demonstrate the diverse possibilities of perovskite lighting.

In the near future, we envision a colorful PeLED world constructed from a wide range of perovskite emitters, devices, and potential applications.

**Acknowledgements**

Ziming Chen is a Marie Skłodowska-Curie Postdoctoral Fellow (Project No.: 101064229) funded by UK Research and Innovation (Grant Ref.: EP/X027465/1). R. L. Z. H. thanks the Royal Academy of Engineering through the Research Fellowships scheme (no.: RF \201718\1701). H.-L. Yip thanks for the support from the Hong Kong Research Grant Council for the GRF grant (No. 11314122)

# 1. 2D/3D Perovskite Thin Films and Colloidal Nanocrystal Perovskites

*Nadesh Fiuza-Maneiro*, *Iago López-Fernández*, *Clara Otero-Martínez* and *Lakshminarayana Polavarapu*

Universidade de Vigo, Spain

**Status**

*2D/3D halide perovskite thin films*: The preparation of an uniform light emissive layer with excellent charge carrier transport properties is critical in the fabrication process of LEDs. Numerous methods including spin coating, drop-casting, dual and single-source evaporation, pulsed laser deposition, and melt-processing have been developed to fabricate halide perovskite thin films. Among all, the solution-processed spin coating has been extensively used to obtain thin films of 3D and 2D perovskites and their mixtures (Figure 3). For more detailed information on this subject, we recommend the reader refer to a previous review.[42]

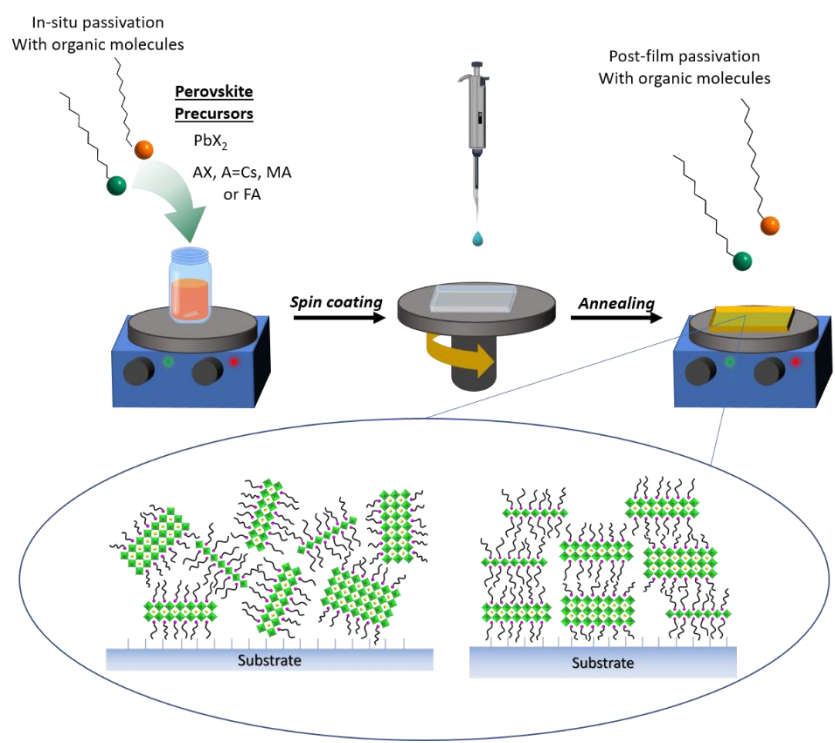

**Figure 3.** Schematic illustrations of the most commonly used spin coating approach for the preparation of halide perovskite thin films and the atomic view of some of the possible configurations of 2D and 3D perovskite layers separated by organic molecules. A dimethylformamide solution containing perovskite precursors (AX and $PbX_2$, X: Cl, Br or I, A: MA, FA or Cs) is drop cast on a desired substrate followed by thermal annealing, leading to homogeneous perovskite film. The PLQY of the films is often enhanced by introducing passivation molecules either in the precursor solution or through post-treatment of perovskite film.

One of the first reports of halide PeLEDs has demonstrated near-infrared and green EL with EQEs of 0.7% and 0.1%, respectively, using solution-processed 3D MAPbX$_3$ perovskites films (Figure 3, without using passivation molecules).[9] Subsequently, a large number of studies were carried out to improve the performance of PeLEDs by confining the injected carriers in nanograins or Ruddlesden-Popper 2D perovskite quantum wells with tunable bandgap, and the EQEs has already surpassed 20%.[43,44] The 2D perovskites exhibit high exciton binding energy and thus enhanced radiative recombination. As illustrated in Figure 3, the use of long-chain ammonium cations in the precursor solution leads to the formation of 2D or 2D/3D perovskite films because the molecules are too big to fit A-site and thus block the 3D crystal growth. The 2D perovskite quantum wells either form oriented or random stacks on the substrates depending on the experimental conditions. Researchers have tried to control the orientation of stacks by solvent engineering and study them with X-ray scattering techniques.

The organic spacer molecules not only confine the charge carriers in dielectric confinement but also passivates the surface of 2D perovskites to suppress nonradiative recombination as well as protect them from moisture-induced degradation. Thus, the organic molecules play a crucial role in obtaining 2D Ruddlesden-Popper perovskites with blue-shifted PL in comparison with that of 3D counterparts.[45] By controlling the ratio between the long-chain organic cations and A-cations [Cs, methylammonium (MA), or formamidinium (FA)], the thickness of the 2D perovskites is precisely controllable down to one monolayer ($n = 1$), enabling the bandgap tuning.[46] This approach enabled the fabrication of Br-based PeLEDs with emission tunable from pure green-to-blue wavelength regions with reasonable EQEs.[46] Various organic molecule additives that do not fit into the octahedral sites of perovskites, including long and short-chain alkyl (or conjugated organic molecular) amines, phosphine oxides, sulfonates, and polymers have been used for the passivation of 2D and 3D perovskite layers of the thin films (either by adding in precursor solution or by post-synthetic addition as shown in Figure 3).[45,47] Among all, phenylethylammonium (PEA) cation has been extensively exploited to obtain Br and I-based 2D perovskite quantum wells as well as mixed dimensional inorganic and hybrid perovskite thin films.[48]

The passivation methods have been effectively explored to reduce the trap density and suppression of thermal quenching in perovskite films. The ligands either bind at the A-sites or coordinate to the Pb$^{2+}$ defects of the surface of perovskites. The mixed dimensional perovskites refer to quantum wells with different thicknesses ranging from monolayer to bulk. These systems exhibit high PLQY under low excitation fluences and show enhancement in the electroluminescent efficiency (EQEs of >20%) due to an energy cascade process, i.e., energy transfer from higher energy bandgap quantum wells to lower energy bandgap 3D perovskites through energy funneling,[48] enabling efficient radiative recombination. The organic spaces not only passivate the perovskite surfaces and lead to 2D quantum wells but also enhances the stability against moisture and water. The spacer molecules with the capability to form intermolecular interactions improve the water stability of perovskites by acting as a barrier for water molecules to prevent their penetration into perovskites. Despite significant progress in the last decade, it is still extremely challenging to obtain perovskite thin films with high PLQY and long-term stability that is required for commercial applications.

*Colloidal halide perovskite NCs:* The first report on highly luminescent colloidal lead halide perovskites was first published in early 2014 by Pérez-Prieto and co-workers.[49] The synthesis is based on the acetone-induced reprecipitation of precursors into $MAPbX_3$ colloidal perovskite. In early 2015, inorganic lead halide perovskite ($CsPbX_3$) NCs with well-defined cubic morphology and high monodispersity were reported by Kovalenko and co-workers.[50] After these early reports, the research on metal halide perovskite NCs with different shapes and compositions (A, B, and X) has virtually exploded. Over the years, a wide range of synthesis methods has been reported for the controlled synthesis of metal halide perovskite NCs.[40] Among all, hot-injection synthesis and ligand-assisted reprecipitation approaches have received great attention owing to their versatility to obtain high-quality NCs of different morphologies and compositions (Figure 4). In these syntheses, an amine-acid ligand pair, especially oleylamine-oleic acid, has been used extensively to protect the metal halide perovskite NCs. The optical properties of lead halide perovskite NCs are easily tunable across visible to near-infrared by halide composition as well as by accessing the quantum-confinement effects. The halide and A-site composition is easily tunable by ion-exchange reactions at room temperature. The capping ligands play a crucial role in ion exchange reactions. The shape of the NCs is tunable from 3D nanocubes to quantum-confined 0D nanocubes and 2D nanoplatelets of different thicknesses, and nanowires to nanorods.[40] As illustrated in Figure 4, the shape of the NCs is tuned by controlling several reaction parameters such as reaction temperature, time, amine/acid ligand ratio, the ratio of A-cation to ligands, and long-chain to short-chain amines, etc.[40] On the other hand, dopants of different types can be introduced into metal halide perovskite NCs via in situ synthesis or post-synthetic doping.[40] Various dopants have been explored to improve phase stability as well as to induce new optical features and functions in metal halide perovskite NCs. In addition, as the stability of metal halide perovskite NCs is strongly dependent on the capping ligands, different types of ligands have been studied and found that multidentate ligands with several binding sites have great potential for enhanced stability.

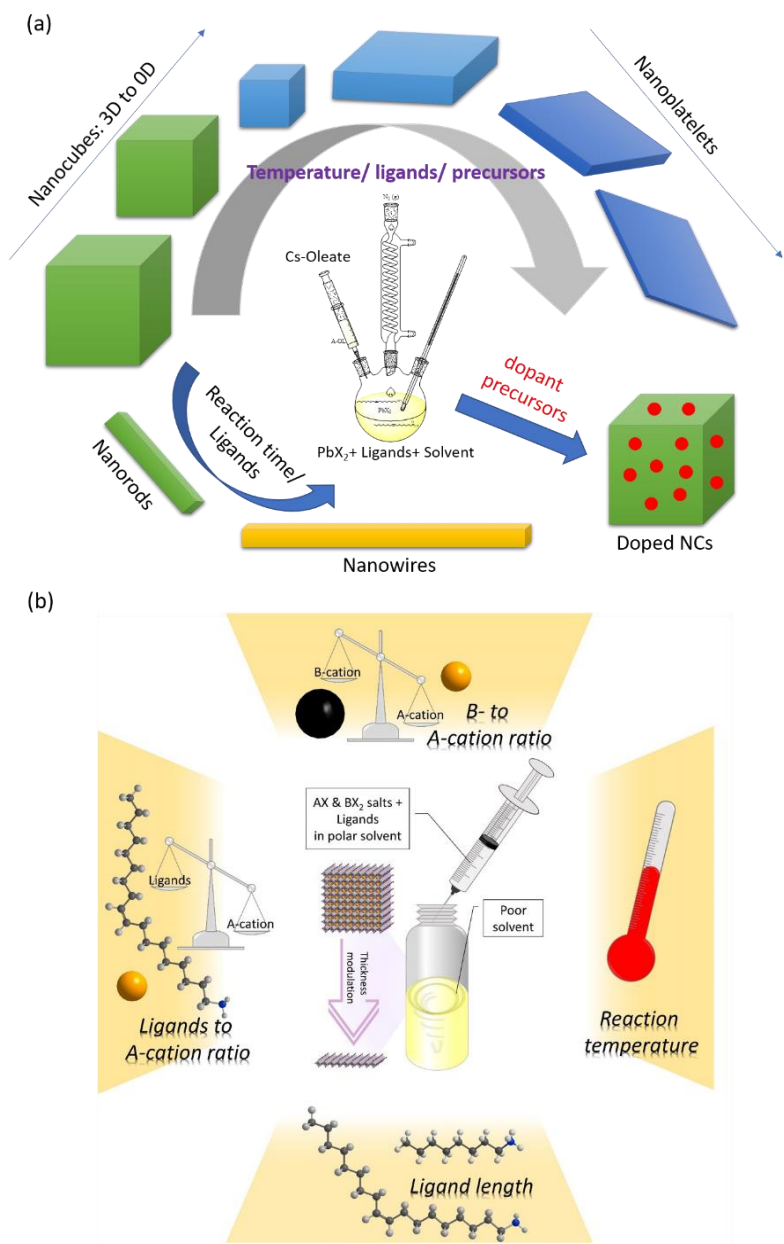

**Figure 4**. Schematic illustration of the shape-controlled synthesis of halide perovskite NCs by hot-injection (a) and reprecipitation (b) approaches. The shape of the NCs is controllable from 3D nanocubes to quantum confined nanocubes and nanoplatelets of different thicknesses as well as nanowires and nanorods, and beyond. The shape is controllable by various reaction parameters as illustrated in the figure. Doped NCs can be obtained by either in-situ synthesis or by post-synthetic doping.

**Current and Future Challenges**

*2D/3D perovskite thin films*: Whereas significant progress was made in achieving high-efficiency thin films for green and red emissive LEDs, blue emissive thin films are currently lagging with low PLQY.[46] Blue emitters can be prepared either with Cl-perovskites or using 2D quantum wells of Br-perovskites. However, both suffer from low PLQY. While Cl-perovskites are less defect tolerant unlike Br- and I-perovskites, Br-perovskite quantum wells exist high defect density because of the high surface-to-volume ratio.[51] Therefore, the fabrication of metal halide perovskite-based blue emitters with efficiency as high as green and red emitters is extremely challenging. Although the emissive color is tunable by varying the thickness of the quantum wells, precise control over the average thickness of quantum wells in thin films is difficult and often leads to mixed-dimensional films with multiple emissive peaks. Another challenge associated with 2D perovskite thin films is their thermal and environmental stability. Although hydrophobic organic spacer molecules act as barriers to water molecules that degrade perovskites, they can also block the charge carriers and thus affects the efficiency of LEDs. Moreover, they can't provide complete waterproofing. Besides, reducing toxicity using Pb-free perovskites is another challenge as the Pb-free perovskites are either inefficient emitters (e.g., double perovskites) or relatively less stable (e.g., Sn-perovskites).[40]

*Colloidal halide perovskite NCs:* The challenges associated with 2D/3D thin-film perovskites regarding blue emitters, long-term stability, and toxicity is also applicable to colloidal perovskite NCs. Besides, precise control over the emissive wavelength of quantum-confined nanocubes or nanoplatelets is challenging as the synthesis often results in NCs with mixed dimensions with multiple emissive peaks in the PL spectra. Although various shapes of NCs have been reported, nanocubes and nanoplatelets have been significantly studied regarding their use as emitters in LEDs.[40] While nanocubes are efficient and relatively easy to obtain in large quantities, nanoplatelet LEDs (blue or red) generally exhibit low EQE and are difficult to process their thin films, and show poor stability compared to nanocubes. Another important challenge associated with NCs is their processing into thin films. This step requires excess ligand-free colloidal solutions that can be obtained by antisolvent-induced washing or multiple centrifugation cycles.[51] However, these process often results in the removal of surface ligands and lead to a reduction in PLQY and even degradation of NCs.[40] Therefore, it is challenging to find strong-binding short-chain ligands that can stabilize perovskite NCs without affecting the charge carrier transport. Despite various reported morphologies, $CsPbX_3$ nanocubes and nanoplatelets have been significantly in the fabrication of perovskite NC-based LEDs.[40] It would be interesting to make use of thr anisotropic shape of perovskite NCs to achieve polarized emission and the corresponding LEDs. While an EQE of over 23% has been achieved for nanocubes, the maximum EQEs obtained for nanoplatelet LEDs are still in the range of 1-2%. Besides, organic-inorganic hybrid perovskite NCs have been relatively less explored as compared with all-inorganic systems due to the challenges associated with the stability of hybrid systems. However, a few studies have demonstrated the potential of hybrid perovskite NCs for making efficient LEDs,[31] which are encouraging for the exploration of hybrid systems in the future.

**Advances in Science and Technology to Meet Challenges**

Over the years, researchers have made great progress in understanding the science and technology of halide perovskite materials. The fabrication methods for thin films and colloidal NCs have been well developed, and the origin of luminescence from these materials and the degradation mechanism are being extensively investigated. To improve the blue luminescence of Cl-perovskites, the surface of perovskites should be strongly passivated with strong binding ligands. In this regard, an increase of PLQY from 1.1% to 19.8% of blue LEDs was reported through the incorporation of phenylethylammonium chloride (PEACl) into 3D $CsPbBr_3$ perovskites.[52] The addition PEACl results in 2D/3D mixed-dimensional perovskite. The efficiency could be improved to 49.7 % through the incorporation of yttrium (III) chloride into the $CsPbBr_3$ perovskite. The yttrium increases the bandgap and confines the charge carriers leading to efficient radiative recombination, reaching EQE of 11.0% and 4.8% for sky-blue and blue LEDs.[52] The $PEA^+$ is also used in the fabrication of 2D quantum wells of $PEA_2(Rb_xCs_{1-x})_{n-1}Pb_nBr_{3n+1}$ that emits pure blue PL with a PLQY as high as 82%. However, the EQE of the LEDs made out of these materials is still very low (1.35%) compared to 3D counterparts.[46] Therefore, further scientific advances require a better understanding of the origin of low EQE despite the high PLQY of thin films as well as 2D nanoplatelets. One reason could be due to the high density of spacer molecules that blocks the charge carrier transport in 2D quantum well layers. It would be interesting to study carrier transport with different organic spacers in 2D quantum well layers of thin films as well as quantum-confined colloidal nanoplatelets.

In regard to waterproofing, improving the intermolecular interactions between organic spacers could improve the stability against water. Also, it is important to further advance the film encapsulation strategies for long-term stability.[53] To reduce the toxicity, there is a need to develop strategies to use Sn-based as well as doped double perovskite systems (thin films and colloidal NCs) as alternatives to Pb-based emitters. Device encapsulation could help to stabilize the Sn-based perovskites against oxidation. In regard to NC stability against washing, multidentate ligands that have several binding sites need to be explored without compromising the charge carrier transport in corresponding NC films.[51] It would be interesting to develop conjugated organic multidentate ligands to enhance stability as well as charge transport.

**Concluding Remarks**

The preparation of highly luminescent thin films and colloidal NC perovskites is a primary step in the fabrication of PeLEDs. Various methods have been developed for both thin films and colloidal perovskites. 2D/3D mixed perovskites with proper organic spacers are efficient for the fabrication of LEDs with over 20% EQE. Despite great progress in green and red LEDs, blue LEDs are lagging due to the low PLQY of Cl-based perovskites as well as Br-based quantum wells. Recent studies on 2D quantum wells showed promising results with high PLQY, but the EQE is still low. Besides, Sn-based and double perovskites need to be explored as an alternative to Pb to address toxicity issues. In addition, molecular engineering of organic spacers and encapsulation strategies need to be explored for long-term stability. On the other hand, an EQE of over 20% is already achieved for perovskite NC-based LEDs through appropriate surface passivation. Therefore, strong binding ligands need to be explored for effective surface passivation of metal halide perovskite NCs even after the antisolvent-induced washing process. In addition, the research on Pb-free perovskite NCs needs to be accelerated to explore them with their full potential.

**Acknowledgements**

L.P. acknowledges the support from the Spanish Ministerio de Ciencia e Innovación through the Ramón y Cajal grant (RYC2018-026103-I) and the Spanish State Research Agency (Grant No. PID2020-117371RA-I00; TED2021-131628A-I00), as well as the grant from the Xunta de Galicia (ED431F2021/05).

## 2. Radiative mechanism of Perovskite Light-Emitting Diodes

*Navendu Mondal* and *Ziming Chen*

Imperial College London, United Kingdom

**Status**

EQE, which is one of the critical parameters to gauge the quality of a LED, can be described by the following equation:[1]

$$EQE = B_{e-h} \times L_{e-h} \times LEE \times ELQY \tag{1}$$

where $B_{e-h}$, $L_{e-h}$, and $LEE$ describe the balance of injected electrons and holes, loss of injected electrons and holes, and light extraction efficiency, respectively, which are mostly governed by the device architecture design. While the EL quantum yield (ELQY) is mainly determined by the intrinsic emissive properties of the emitter under electrical injection. It directly relates to the ratio between radiative and non-radiative recombination within the emitter. Therefore, understanding the radiative mechanism of LEDs is important to improve their overall performance.

However, in the perovskite light-emitting field, due to the difficulty of directly measuring ELQY, researchers largely rely on understanding the radiative mechanism from the viewpoint of PLQY. Therefore, initial research efforts in the field were directed towards identifying key factors that could enhance PLQYs of perovskites, e.g., improving the radiative emission rates and/or mitigating the competing non-radiative emission (loss) pathways. Various transient PL and absorption spectroscopies have been implemented on wide variety of perovskites in rationalizing the luminescence mechanism through understanding the dynamics of charge carriers (excitons) following photogeneration.

Free carriers, being the major photoexcited species in 3D perovskites, undergo various non-radiative processes (i.e., first-order trap-assisted and three-body Auger recombination) along with the second-order radiative recombination (Figure 5a).[1,54,55] While in low-dimensional perovskites where exciton being the main photoexcited species undergo first-order radiative excitonic recombination competing with non-radiative first-order trap-assisted recombination and second-order exciton-exciton annihilation related Auger type process (insets of Figure 5b).[56] The overall charge-carrier or exciton dynamics can be represented by the following Equations 2 and 3, respectively. Here, $n$ and $N$ represent the density of charge-carriers (electron/hole) and exciton, respectively; $k_1$, $k_2$, and $k_3$ represent monomolecular, bimolecular, and Auger recombination rate constants, respectively; $k_{trap}$, $k_{ex}$, and $k_{ex-ex}$ represent trap-assisted recombination, exciton recombination and exciton-exciton annihilation rate constants, respectively.[1,54,57,58,59] Impact of these processes in overall dynamics is largely influenced by the density of carriers/excitons generated in the system (as shown in Figure 5c) and therefore PLQY also becomes their density dependent as expressed below.

$$-\frac{dn}{dt} = k_1 n + k_2 n^2 + k_3 n^3, \qquad PLQY = \frac{k_2 n}{k_1 + k_2 n + k_3 n^2} \qquad (2)$$

$$-\frac{dN}{dt} = (k_{trap} + k_{ex})N + k_{ex-ex}N^2, \qquad PLQY = \frac{k_{ex}}{(k_{trap} + k_{ex}) + k_{ex-ex}N} \qquad (3)$$

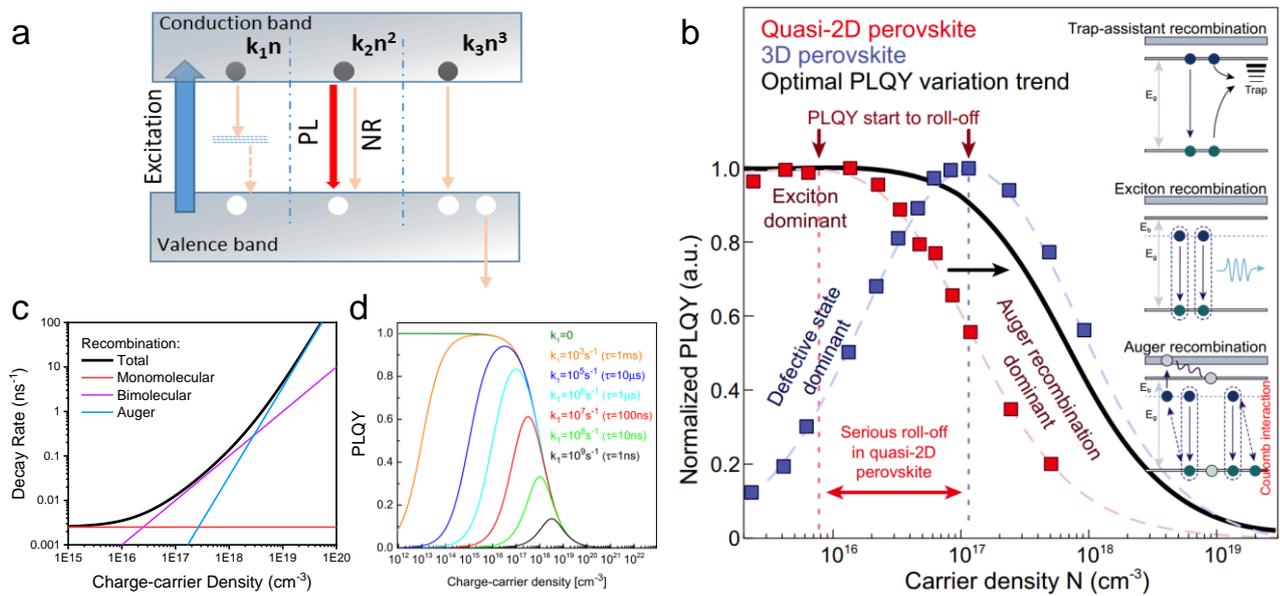

**Figure 5**. (a) Schematic description of typical relaxation pathways of carriers generated in 3D perovskites following photoexcitaion. (b) Carrier-density-dependent PLQY for 3D and quasi-2D perovskites. (c) Simulated curves for complete decay rate (comprises three different orders) of charge-carriers in 3D perovskites. (d) Variations of PLQY in 3D perovskites upon changing the monomolecular recombination rate constant. (b) Reprinted with permission from [56], Copyright 2021, Springer Nature. (c) Reprinted with permission from [58], Copyright 2018, WILEY-VCH Verlag GmbH & Co. KGaA, Weinheim. (d) Reprinted with permission from [59], Copyright 2015, American Chemical Society.

At a low-carrier density regime (<$10^{16}$ cm$^{-3}$), trap-assisted non-radiative recombination and radiative excitonic recombination generally dominate in 3D and 2D perovskites, respectively. Therefore, unlike 2D perovskites, maximization of PLQY values for 3D perovskites can be achieved under relatively higher carrier density as evident from Figures 5b and 5c.[56,60] However, such level of carrier density ($10^{16}$–$10^{18}$ cm$^{-3}$) is even higher than that required under normal working conditions of PeLEDs (<$10^{16}$ cm$^{-3}$). This points to the fact that unlike bulk 3D perovskites, perovskites NCs and 2D perovskites wherein excitonic recombination dominates, could be potential candidates for the PeLED applications. According to Equations 2 and 3, the following strategies can be considered in terms of PLQY improvement for various class of perovskites.

*Reducing $k_1$ or $k_{trap}$ value.* Metal halide perovskites known for their defect tolerance nature, generally possess shallow traps instead of deep traps. Although shallow traps are less harmful to the excited carriers, they still induce non-radiative recombination loss. For example, as evident from simulated results in Figure 5d, that significant reduction of monomolecular recombination rate constant (i.e., reducing the trap density) is a key strategy towards achieving maximized PLQY in 3D perovskites.

*Improving $k_2$ or $k_{ex}$ value.* Confinement of electrons and holes within a small volume generally enable better overlap of their wavefunction, which facilitates faster radiative recombination according to Fermi's golden rule. Therefore, reducing crystal size has been adopted as a key strategy to achieve spatial confinement for 3D perovskites. While the formation of multi-quantum well for 2D perovskites and type-I core-shell structure for perovskite QDs would be the methods to confine excitons spatially and energetically.

*Reducing the $k_3$ or $k_{ex-ex}$ value.* When Auger recombination is dominant (under the high density of carriers/excitons) in the overall carrier/exciton dynamics, reduction in PLQY is expected and results in efficiency roll-off. On the other hand, although confinement of carriers is highly desired to improve the $k_2$ or $k_{ex}$ values, the greater electron-hole wavefunction overlap also enables faster (efficient) Auger recombination. This effect becomes more obvious at relatively lower excitation density for those systems with higher exciton binding energies. Therefore, efficiency roll-off for 2D/quasi-2D perovskites occurs at a much lower excitation threshold compared to 3D perovskites as evident from Figure 5b. However, attempts towards suppression of Auger recombination to weaken the efficiency roll-off is still limited to a handful of literatures.[56,61]

**Current and Future Challenges**

Even though measurement of PLQY and associated identification of radiative and non-radiative routes are important characteristic markers to predict the ELQY of PeLEDs, significant discrepancies between actual ELQY values and measured PLQYs may occur due to several reasons:

(i) The discrepancy of EL and PL behavior due to the electron-hole-balance difference, for

instance in PL, identical electron and hole population is always generated by means of optical excitation, while in EL, their population ratio highly depends on the respective injection capability of electrons and holes in the device. For instance, recently Vacha et al. demonstrated that EL intermittency behavior is quite different from the PL intermittency in single perovskite nanoparticles due to imbalanced charge-injection/accumulation in the perovskite layers or their interfaces under electrical biasing condition.[62] This imbalance of charges resulting to additional non-radiative Auger recombination associated with various charged-exciton (also known as trions) complexes which often reduce the brightness of the emitter.

(ii) The charge carrier dynamics of the emitting layer under photoexcitation might be different from the operando carrier dynamics under electrical injection. For instance, interfacial trap filling process would be critical in EL as all injected carriers have to pass every interface in the device while such an effect would be less obvious in PL as carriers have to successfully diffuse to interfaces first. In addition, funneling of charges from large-bandgap to small-bandgap components is usually observed in quasi-2D perovskite under photoexcitation, while such processes would become much less significant as majority of the carriers should be directly injected into the low-bandgap components due to their low electric resistance.

(iii) The ratio of singlet- and triplet-excitons generated by photoexcitation and electrical injection might be different. In typical organic materials, the ratio of singlet and triplet under photoexcitation is generally based on the competition among singlet decay rate, intersystem crossing rate and reverse intersystem crossing rate; while such ratio originally becomes 1:3 under electrical injection according to the spin statistics, although processes like thermally-activated delayed fluorescence (TADF)[63] and triplet-triplet annihilation could further amend this ratio.[64] Therefore, such a discrepancy might become more obvious in the perovskite systems containing large portion of organic ligands (e.g., 2D and 0D perovskites), especially in the case where singlet/triplet transfer between perovskite and organic counterpart is expected.

The lack of proper characteristic techniques and the inferior stability of PeLEDs under applied voltage retard the field in studying the aforementioned areas. Therefore, investigation towards the radiative mechanism in PeLED working devices remains fairly unexplored to date.

**Advances in Science and Technology to Meet Challenges**

After well establishing the radiative mechanism under photoexcitation for excitonic and non-excitonic perovskite emitters, research interest is encouraged to shift towards the investigation of radiative mechanism under electrical injection for the PeLED devices. To meet this end, novel characterization technologies need to be developed.

Steady-state and time-resolved optical spectroscopies integrated with electrical pumping sources might be one of the potential options. For instance, a combination of electrical excitation and optical detection (e.g., time-resolved EL, electrical pump-optical probe, electrical pump-THz probe) could resolve a wide range of charge carrier dynamics under electrical injection, which also would help in minimizing the knowledge gap towards device physics.

Additionally, we consider following research aspects would become indispensable puzzles to develop the whole picture of the radiative mechanism of PeLEDs:

(i) Development of proper device structure to mitigate the charge imbalance which would allow to suppress/regulate recombination from various other undesired excitonic complexes like trions. In short, careful consideration towards band-edge energetic alignment and charge carrier mobility among functional layers in the device is essential to achieve adequate charge balance as indicated previously by Klimov & co-workers in their finding on QLEDs.[65]

(ii) More systematic efforts towards exploring the photophysics of quasi-2D in composition and confinement space are required to unveil the critical role of multiple layers on exciton funneling to tuning the exciton binding energy. This would allow finding prospective directions towards high ELQY.

(iii) Development of appropriate kinetic models (beyond the conventional ABC model) is required to isolate various other non-radiative events like evidence of Auger-assisted trapping reported previously[66] in 3D perovskites. Also, existence of both carriers and excitons (like in quasi-2D perovskites, weakly confined perovskite NCs) may also complicate the simplest form of kinetic equations (like Equations 2 and 3).

(iv) As discussed previously, presence of trap centers often limits both the PLQY and ELQY due to the trap-assisted non-radiative recombination processes, while in few instances evidence of trapping-detrapping events found to facilitate the radiative recombination process by means of delayed PL.[67] Therefore, careful control of energetic positioning of those trap-states and their origin could be beneficial for realization of PeLEDs analogous to typical organic TADF materials.

(v) Exploring the triplet-exciton of perovskites for TADF could be an interesting avenue. Evidence suggests that low-lying triplet excitons of perovskites are involved in the TADF process, either directly or by forming a hybrid composite with low-lying organic triplets.[68,69,70,71] Therefore, manipulating singlet-triplet exciton energy spacing in wide range of perovskites and hybrid perovskite composites could provide the design guidelines for hybrid TADF materials in potential light-emitting applications.

**Concluding Remarks**

Understanding and carefully controlling the radiative and non-radiative processes in various classes of perovskites is the prerequisite towards obtaining efficient performance of PeLEDs. There is already significant progress towards introducing confinement effect to improve radiative processes and eliminating various non-radiative trap centers to eventually enhance the PLQY and ELQY. While more attention required in managing the efficiency roll-off through manipulating Auger recombination processes. In general, inconsistency between PLQY and ELQY values urges intense effort towards exploring the operando photophysics and comparing them with the intrinsic photophysics of perovskite emitters is highly essential to bridge the existing knowledge gap. To meet this end, efforts on novel characterization techniques and stable device development are highly encouraged, and some open research directions proposed here could be key for comprehensive understanding of radiative mechanism of PeLEDs.

**Acknowledgements**

Navendu Mondal acknowledges funding support from the European Commission through the Marie Skłodowska-Curie Actions (Project PeroVIB, H2020-MSCA-IF-2020-101018002). Ziming Chen is a Marie Skłodowska-Curie Postdoctoral Fellow (Project No.: 101064229) funded by UK Research and Innovation (Grant Ref.: EP/X027465/1).

## 3. Measuring Methods of Perovskite Light-Emitting Diodes

*Alessandro Mirabelli*, *Miguel Anaya* and *Samuel D. Stranks*

University of Cambridge, United Kingdom

**Status**

As an emerging class of light sources, PeLEDs have aroused significant interest for their precise color control, brightness, and cheap fabrication processes.[72] However, measurement protocols in use for other LED technologies do not capture the various peculiar phenomena behind the working nature of halide perovskite devices, such as transient effects or photon recycling.[54] There has not been consistent use of measurement protocols for PeLEDs, resulting in differences between the figures of merit and setups employed to assess their performance. Although improving in recent times,[73] this lack of uniformity has complicated comparisons between literature reports. This section will provide an accessible guideline for the characterization of PeLEDs, which should be widely used across the field to ensure a rigorous and hastened development of the technology.

An important parameter for assessing the quality of a PeLED is the EQE. This dimensionless metric is defined as the ratio of photons escaping the device over the number of charge carriers flowing through the external circuit. This value should be reported in conjunction with the current density $J$ (mA cm$^{-2}$) over many orders of magnitudes of current, i.e., $10^0$–$10^3$ mA cm$^{-2}$. It is common to obtain the peak EQE value at relatively low voltages (low current densities, e.g., on the order of $10^0$ mA cm$^{-2}$) due to a combination of factors, including the fact that in that recombination regime there are minimal Auger recombination events occurring, good charge injection balance, and low Joule heating.[74,75] Therefore, while this metric gives valuable information on the upper limit of the PeLED performance, it tells little about how useful the device is in terms of brightness. In this regard, it is important to also assess the emissivity of a PeLED. This can be done by calculating the radiance (W sr$^{-1}$ m$^{-2}$), which is the photon power emitted per unit of area per unit solid angle. However, this quantity is usually reshaped in terms of luminance (cd m$^{-2}$) for devices emitting in the visible range, which accounts for the photopic response of the human eye. Following this reasoning, it is also convenient to recast the EQE in terms of current efficacy (cd A$^{-1}$) to evaluate the ability of a device to achieve a practical value of luminance – around 1000 cd m$^{-2}$ for modern outdoor displays and light applications. Finally, the emission color and its purity should be given in terms of the chromatic coordinates (x,y) determined by the Commission Internationale de l'Eclairage (CIE), an established system that takes into account the behavior of the cone cells in the human eye.

**Current and Future Challenges**

Halide perovskites are sensitive to different external agents such as air and moisture. Such factors have a negative impact on the emitter as they accelerate transient phenomena, such as ion migration, that lead to degradation and stability issues. For this reason, it is important to characterize PeLEDs in an inert atmosphere, e.g., a $N_2$-filled glovebox, or after being encapsulated by glass slides in such a glovebox; provided the encapsulation is robust, this approach nominally eliminates environmental effects that introduce artefacts in the metrics and otherwise limit access to the fundamental processes governing PeLED operation. The ionic nature of halide perovskites makes halide segregation and phase heterogeneity impactful processes that affect structural stability and device integrity of PeLEDs. Correctly assessing these compositional and structural properties can be difficult because such transient phenomena will appear as soon as voltage is applied, and given there is currently no standardized stress tests among different research groups for PeLEDs, this complicates drawing of conclusions. Joule heating is another pertinent obstacle towards the advance of perovskite emitters. The active material layer thickness is usually in the order of a few tens of nanometers and, without tackling heat dissipation, the aforementioned transient degradation processes are accelerated.[76] Therefore, while a *J-V* sweep can reveal the current/voltage characteristics of a device, important performance changes occur during operation. Measurements must consider this potential evolution of the optoelectronic qualities of PeLEDs during acquisition. Determining the optimal voltage range for which the PeLED does not change and conducting both a fast scan followed by a slower one, say at 0.5 V and 0.1 V step size, respectively, should highlight eventual material structural issues. A key note here is that scan rate should be carefully selected to avoid inducing measurement artefacts from the ionic movement of the halides.

Another pressing issue concerns the transient nature of the EL intensity and the color stability. The PeLED EL spectrum should be reported at the conditions corresponding to peak EQE and at the maximum luminance, which typically are not at the same operational bias. This is particularly important to evaluate color stability in blue PeLEDs that typically shift to green wavelengths under operation (Figure 6).[77] The same device could maintain true blue emission (<470 nm) at 4 V, but the spectrum already shifts to green emission at 5 V or higher, giving a false statement of its color stability. Therefore, when reporting particular PeLED stability and lifetime metrics one must also report the luminance as a function of time and highlight typical metrics such as $T_{50}$ and $T_{95}$ (time at which the LED has dropped to 50% and 95% of its initial luminescence, respectively) at different brightness values. Reporting these luminance metrics for appropriate outdoor (>1000 cd m$^{-2}$) and indoor (>100 cd m$^{-2}$) display applications returns a comprehensive overview of the device limitations.

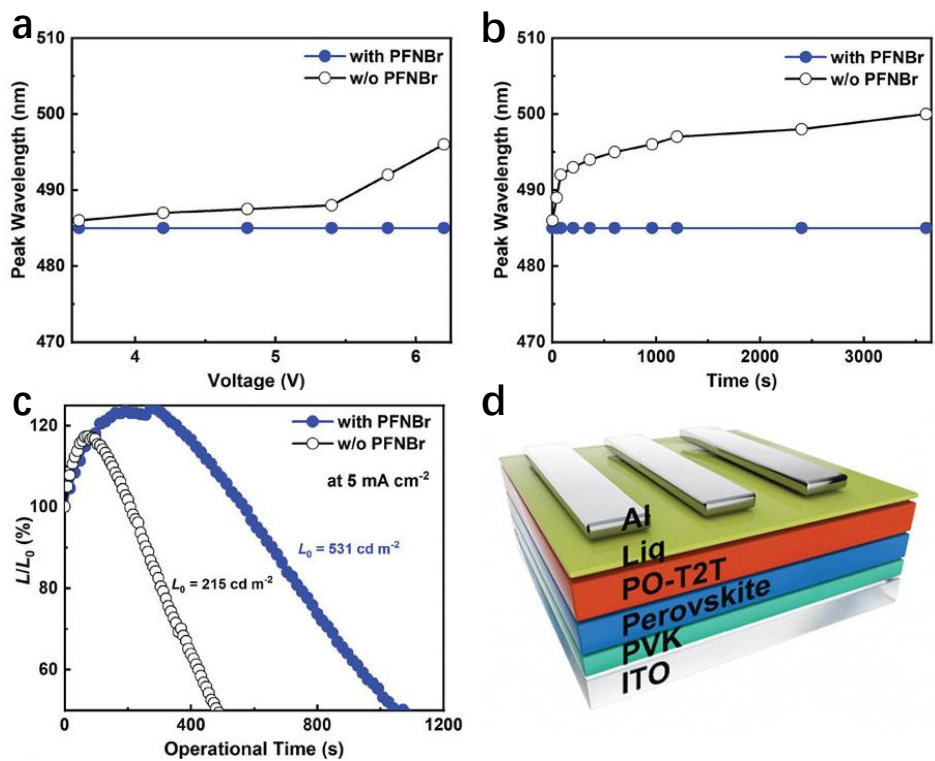

**Figure 6.** (a) Operating voltage-dependent changes of the EL peak wavelength in the sky-blue PeLEDs with (blue) and without (black) PFNBr modification. (b) Temporal evolution of the EL peak wavelength in the sky-blue PeLEDs with and without PFNBr modification operating under a constant voltage of 4.0 V. (c) Operational lifetime measurement for the sky-blue PeLEDs with and without PFNBr modification measured under a constant current density of 5 mA cm$^{-2}$. (d) Device architecture. (a–d) are adapted from Ref. [77].

**Advances in Science and Technology to Meet Challenges**

Various methods and equipment exist to properly assess the optoelectronic characteristics of PeLEDs. We suggest that the best route involves collection of emitted photons just in the forward direction by employing an integrating sphere in combination with a spectrometer and a sourcemeter (Figure 7). Using a 2π geometry, where the LED is placed tangentially to the surface of the sphere, makes handling of devices inside an inert glovebox straightforward compared to applying 'leg' contacts to an LED, then mounting it onto a holder, measuring the emitted photons with a photodetector and successively integrating over the forward solid angle. Using an integrating sphere along with carefully obtained calibration factors, one can convert counts read by the spectrometer into number of photons, as one can assume that the spherical geometry makes the light isotropic and the same fraction of light is collected. One must be careful to perform both measurements and calibrations under the same conditions, to avoid incorrect estimations of the number of photons emitted. These three components allow the user to precisely obtain the EQE value, *J-V* characteristics, and EL spectra of the measured PeLED.

It is also possible to carry out a more detailed characterization by analyzing the angular distribution of the emitted photons. Lambertian emission profiles are usually assumed and reported for most PeLEDs. However, with the increasing interest in 2D perovskites and new additives, coupled with processing techniques to control crystal orientation, a novel class of emitters with substantial distribution variations of its emission profile may become relevant. Therefore, it becomes necessary to assess the directionality of the emission as well. Such a requirement can be carried out by mounting the PeLED on a rail connected to the integrating sphere and have the device sit on a goniometer. By positioning the PeLED far away from the sphere and rotating the sample stage, one can record the emission profile. Alternatively, the LED can be mounted in a fixed position and the detector can rotate in the plane normal to the device.

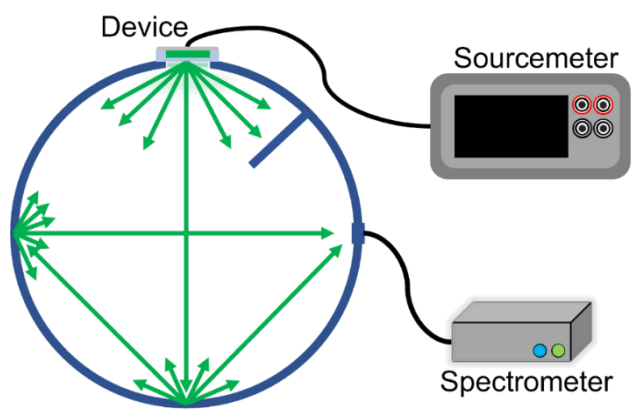

**Figure 7.** Measurement setup consisting of integrating sphere, calibrated spectrometer and sourcemeter.

**Concluding Remarks**

By adopting common and controlled measurement protocols for PeLEDs, the exchange of information and scientific knowledge between laboratories will benefit. It is important to also highlight eventual failures or considerable issues in specific PeLED architectures as this will inspire new studies in the community to investigate these important phenomena. Detailed device fabrication methods as well as material synthesis should also be reported as it contributes to the advance of the field and reproducibility. It is also good practice to measure and characterize one's own PeLEDs in different setups and even laboratories to better validate the results obtained.

**Acknowledgments**

The work has received funding from the European Research Council under the European Union's Horizon 2020 research and innovation programme (HYPERION, grant agreement no. 756962). M.A. acknowledges funding from the Marie Skłodowska-Curie Actions (grant agreement no. 841386) under the European Union's Horizon 2020 research and innovation programme, a Leverhulme Trust Early Career Fellowship, and support by the Royal Academy of Engineering under the Research Fellowship programme. S.D.S. acknowledges the Royal Society and Tata Group (grant no. UF150033).

## 4. Large-area Perovskite Light-Emitting Diodes Based on Solution Processes

*Hui Liu, Guangyi Shi, Zhengguo Xiao*

University of Science and Technology of China, China

**Status**

Large-area fabrication of PeLEDs using industry-scale techniques is critical for their commercialization. In the past few years, solution methods such as spin coating, blade coating, and inkjet printing have been used to make large-area PeLEDs (Figure 8). Spin-coating was first attempted to make large-area PeLEDs using formamidinium lead iodide (FAPbI$_3$) in 2020. A decent EQE of 12.1% was achieved with a device area of 9 cm$^2$,[78] and the devices have been tested for medical applications. Recently, Yuan et al. introduced an additive, L-norvaline, into the precursor to adjust the crystallization process of quasi-2D perovskite, PEA$_2$(FA$_{0.7}$Cs$_{0.3}$)$_2$Pb$_3$Br$_{10}$.[79] Very uniform films with a low roughness of 1.2 nm were obtained without the help of an antisolvent, and the EQE of the PeLEDs was boosted to 16.4% with a device area of 9 cm$^2$.

Blade coating enables quick and uniform thin films over large rigid or flexible substrates without a large amount of solution waste. To accelerate the perovskite crystallization process for improved morphology, various strategies, such as additive engineering,[80] using supersaturated precursors,[81] and vacuum quenching,[82] have been developed. Ultrauniform methylammonium lead iodide (MAPbI$_3$) films with a roughness of 1 nm were prepared using blade coating by regulating the sol-gel process using a diluted, organoammonium-excessed perovskite precursor. A high EQE of 16.1% was achieved in small-area PeLEDs (4 mm$^2$), and large-area PeLEDs (28 cm$^2$) also show very uniform emission.[80] Large-area sky-blue PeLEDs (emission peak 489 nm) based on CsPb(Br$_{0.84}$Cl$_{0.16}$)$_3$ were also reported using blade coating, and the EQE reached 10.3% (Figure 8e).[81] Recently, Lee et al. reported a very high EQE of 21.46% (device area 9 cm$^2$) using a modified bar coating of presynthesized perovskite NCs.[83]

The inkjet-printing technique provides a promising way to fabricate large-area PeLEDs with patterns. Zeng et al. proposed a universal ternary-solvent-ink strategy toward highly luminescent and stable ink using all-inorganic CsPbX$_3$ (X = I, Br, Br$_x$Cl$_{1-x}$) perovskite QDs,[84] and a high EQE of 8.54% was achieved. Very recently, Sun et al. improved the morphology of inkjet-printed CsPbBr$_3$ films by inserting a polyvinylpyrrolidone wetting layer on top of the HTL and controlling the printing temperature. A high EQE of 9.0% was obtained with a device area of 0.09 cm$^2$.[85]

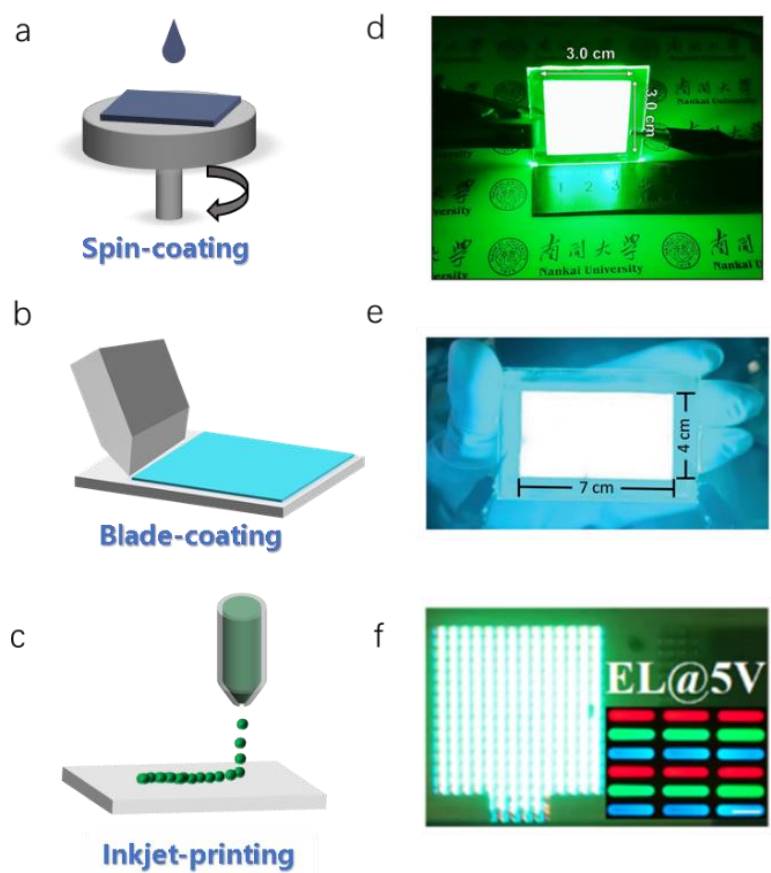

**Figure 8.** (a)–(c) Schematic diagrams of the fabrication methods for large-area PeLEDs based on solution processes and (d)–(f) the corresponding large-area EL images of the PeLEDs. EL large-area image based on spin-coating reprinted with permission from Ref. [79]. EL large-area image based on blade coating reprinted with permission from [81]. EL large-area image based on inkjet printing reprinted with permission from [86].

**Current and Future Challenges**

Despite the great progress that has been made in large-area PeLEDs, there are still many challenges in the above solution processes in making large-area PeLEDs. Antisolvent-assisted spin-coating has been proven to be the most effective approach to control the morphology of spin-coated films. However, the inhomogeneous diffusion of antisolvent in a narrow window-time deteriorates the performance of large-area PeLEDs. To date, the device area of spin-coated PeLEDs is still limited to 9 cm$^2$, and larger PeLEDs have not been reported. Another challenge is the precursor waste, making spin-coating likely not economical for the commercialization of PeLEDs.

For blade coating, the fine control of the crystallization process and film morphology of different perovskite compositions is still very challenging. All-inorganic perovskites have much lower solubility and faster crystallization speed than organic-inorganic hybrid perovskites, making the blade-coating process more difficult to control. The bromide- or chloride-based perovskites also have much faster crystallization speeds and tend to be less tolerant to defects. Therefore, film quality control becomes more challenging and more important. The different crystallization processes of different perovskite compositions cause the optimization process of blade coating to be very time consuming. For ink-jet printing, developing printable ink and stable droplets are required to avoid nozzle clogging and ensure pattern accuracy. However, droplet manipulation is difficult to control, and inkjet-printed PeLEDs show inferior device performance with a low EQE due to nonideal film quality and the coffee ring effect.[86]

In addition to the challenges in the solution techniques mentioned above, there are still some other challenges in the commercialization process of large-area PeLEDs, such as the low conductivity of transparent electrodes, higher leakage current, less efficient thermal dissipation, and nondestructive encapsulation. For example, there is a significant loss in performance and operation stability with increasing device area, most likely due to the abovementioned issues.

**Advances in Science and Technology to Meet Challenges**

Special advances in science and technology are needed to address the challenges on the road of commercialization of large-area PeLEDs. First, a more robust optimization strategy is needed for the solution process to deposit large-area pinhole-free perovskite films. The optoelectronic properties of large-area films also need to be optimized. To this end, searching or synthesizing functional additives that can regulate the perovskite crystallization process and passivate the traps may be a feasible approach. It would be better for the new strategies to be compatible with different perovskite compositions. In addition, the HTLs/ETLs also need to be deposited by scalable techniques. Their film thickness and uniformity and conductivity also need to be optimized.

The conductivity of transparent electrodes, such as ITO, needs to be improved for large-area PeLEDs. The voltage and luminance drop and heat builds up locally as the current flows laterally due to the low conductivity of the transparent electrode. This issue becomes more severe as the device area increases. Inspired by OLED panel design, this problem may be solved by applying metal grid electrodes in PeLEDs. In this case, conformal growth of the HTLs/ETLs and the perovskite layer are needed.

**Concluding remarks**

In conclusion, this section summarizes the recent advances and challenges of solution-processed large-area PeLEDs. From the analysis above, blade-coating or similar techniques, such as bar-coating and slot-die coating, should be the most promising techniques for large-area fabrication of PeLEDs, while ink-jet printing is required for patterning. Fine control of the morphology and optoelectronic property uniformity plays a critical role in ensuring the performance of large-area PeLEDs. In addition, the conductivity of the transparent electrode should be improved to ensure luminance uniformity over a large area. In addition, proper encapsulation with good thermal dissipation techniques should be developed to improve the operational stability of large-area PeLEDs. Finally, as one of the most important research directions of PeLEDs, WPeLEDs need to be developed whose structure could refer to the more sophisticated OLEDs and QLEDs.

**Acknowledgments**

We gratefully acknowledge the support from the National Natural Science Foundation of China (51872274, 62175226) and the Fundamental Research Funds for the Central Universities (WK2060190100).

## 5. Perovskite Light-Emitting Diodes Based on Vacuum Deposition

*Nakyung Kim, Yunna Kim, Byungha Shin*

Korea Advanced Institute of Science and Technology, Republic of Korea

**Status**

Vacuum deposition has been the most commonly used method in the OLED industry since the 1980s. It was in 2016 when the first report on vacuum-deposited PeLED was published.[87] Compared with the solution process, vacuum deposition has several advantages: 1) It allows perovskite synthesis without solvent, giving freedom in the choice of precursor chemicals. For example, inorganic precursors like CsBr and $PbCl_2$ have poor solubility in common solvents such as dimethyl sulfoxide and dimethylformamide. Low solubility affects perovskite morphology and composition ratio due to the fast crystallization. 2) It is free of the solvent orthogonality issue, allowing stacking of any materials. In the case of the solution process, a layer can be damaged during the preparation of the next layer, and therefore the selection of solvents is oftentimes limited. 3) Fine control of perovskite thickness down to nanometer is possible. As PeLEDs usually require a thinner perovskite layer (~tens of nm) than perovskite solar cells (~hundreds of nm), thickness needs to be precisely controlled. 4) Considering the need for large-area scalability, vacuum deposition has been proven as a successful industrial fabrication technique in the display industry. Moreover, the vacuum conditions provide well-controlled deposition environments and allow reproducible production of high-quality films, which is compulsory for commercialization.

The first report of vacuum-deposited PeLEDs demonstrated an EQE of 0.35% at the emission wavelength of 768 nm.[87] Since this first report, EQEs have improved to 10.9%.[88] The progress is summarized in Figure 9. In the early stage, MA-based perovskites were studied because of existing research efforts on solar cells based on MA-cation. However, the organic components lead to inferior thermal and chemical stability. Therefore, Cs-based all-inorganic perovskites have become the primary choice for a light emitter of PeLEDs. The main strategy for improving the performance of vacuum-deposited $CsPbBr_3$ PeLEDs has been fine control of the precursor ratio (CsBr : $PbBr_2$). For example, EQE of $CsPbBr_3$ PeLEDs of the same device structure (ITO/$NiO_x$/perovskite/1,3,5-tris(1-phenyl-1H-benzimidazol-2-yl)-benzene (TPBi)/LiF/Al) improved from 3.26%[89] to 8.0%[90] only by adjusting the composition of two precursors. Another important strategy to further increase efficiency includes the application of a passivation layer such as 9-(4-tert-butylphenyl)-3,6-bis(triphenylsilyl)-9H-carbazole and guanidinium (GA) bromide, respectively, which was used to achieve the record EQE of PeLEDs of 10.9%.[88]

Additionally, due to the safety concern of Pb, there have been reports on PeLEDs based on Pb-free perovskites using Cu,[91] and Eu[92] as a B-site cation, and their performance has achieved EQE of 7.4% and 6.5% at the emission wavelength of 568 nm and 448 nm, respectively.

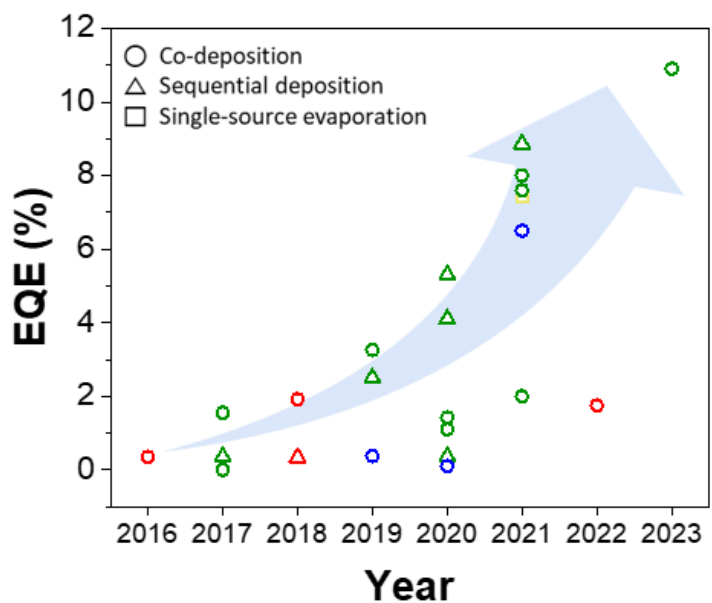

**Figure 9.** Development of vacuum-deposited PeLEDs; each color of the symbols represents the light emission color of PeLEDs.

**Current and Future Challenges**

Among various kinds of vacuum deposition methods, thermal evaporation is the most widely studied in the field of vacuum-deposited PeLEDs. There are three main approaches to depositing perovskite via thermal evaporation; co-evaporation, sequential deposition, and single-source evaporation. First, co-evaporation, or multiple-source deposition, is the most common method to thermally evaporate perovskite. Each precursor, such as CsBr and $PbBr_2$, is contained in a separate crucible. The relative deposition rate of each source determines the precursor ratio. Second, in sequential deposition, precursor layers are sequentially stacked, and the thickness of each layer controls the final precursor ratio of the film. Lastly, single-source evaporation is the method that directly evaporates a pre-synthesized perovskite powder with a chosen composition. The pros and cons of each method are illustrated in Figure 10.

The performance and stability of solution-processed PeLEDs have improved by changing the structure from 3D to low-dimensional, including NCs, and by combining with various passivation strategies. However, most of the research on vacuum-deposited PeLEDs has been focused on 3D perovskite with limited passivation strategies. Fu et al. demonstrated hybrid 2D/3D PeLEDs by sequentially depositing PbBr$_2$, BABr, and CsBr to fabricate (BA)$_2$Cs$_{n-1}$Pb$_n$Br$_{3n+1}$ perovskite.[93] This is the only report that produced vacuum-deposited PeLEDs with low-dimensional perovskite. Therefore, various attempts are needed to fabricate low-dimension perovskite.

The solution process is easy to incorporate minuscule amounts of passivation additives by simply mixing additives with the perovskite solution. On the other hand, considering that the additive ratio is usually less than 10% of the perovskite precursors amount, the flux rate of an additive source should be under 0.1 Å/s for the co-evaporation method, where fine control of this level of flux is difficult. Therefore, instead of the direct incorporation of passivation material, a separate passivation layer such as PEABr[94] or PEO[95] was introduced on top and/or underneath a perovskite emitting layer in vacuum-deposited PeLEDs. However, these passivation layers are prepared by spin-coating.

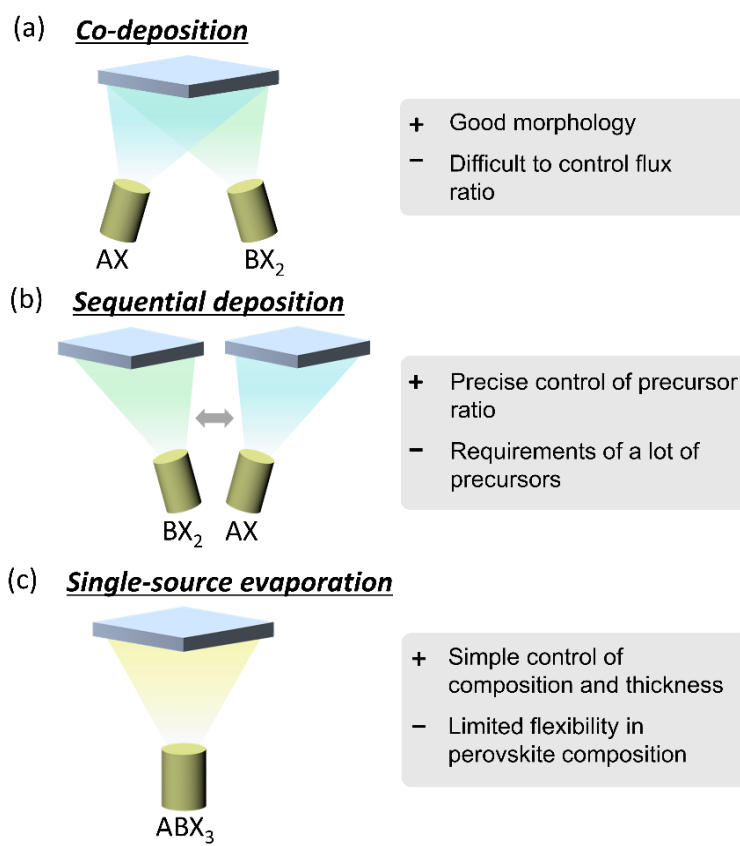

**Figure 10.** Schematic diagrams of three approaches to depositing perovskite via thermal evaporation. (a) co-deposition, (b) sequential deposition, and (c) single-source evaporation.

To maximize the biggest merits of vacuum deposition, i.e., scalability, it is recommended that all layers of LED devices be deposited via vacuum techniques. Considering that most of the PeLEDs adopted the p-i-n structure and electron-transport layers, such as TPBi, are mostly deposited through thermal evaporation, HTLs should be vacuum-processible. However, only a few HTLs have been reported to be processed with vacuum deposition. Li et al. demonstrated a LED device with an EQE of 1.45% by using thermally-evaporated LiF as a HTL.[89] With the replacement of an insulating LiF layer with sputter-deposited $NiO_x$, the EQE was improved to 3.26%. The choice of HTL significantly influences device performance, therefore, an exhaustive search for vacuum-processible HTL should be carried out.

**Advances in Science and Technology to Meet Challenges**

Considering that vacuum-deposited PeLEDs have been less studied compared to the solution process, we believe there is a large potential for technological advancements in both emitting layers and device structures. First of all, there is plenty of room to explore various kinds of vacuum deposition methods other than thermal evaporation, such as chemical vapor deposition and atomic layer deposition (ALD). ALD is a powerful tool to deposit high-quality thin films based on a unique self-limiting growth mechanism, resulting in a high conformality and uniformity in a large area. The ALD process has been used not only for the perovskite layer but also for the charge transport layer or passivation layer in perovskite solar cells. By combining various vacuum deposition techniques, it would be easier to realize all-vacuum deposited PeLEDs. Second, a more detailed investigation of unraveling the crystallization mechanism during the vacuum deposition process is required to fully understand the thin film growth. Shin et al. studied the relation between deposition rate and grain growth of $CsPbBr_3$ film.[96] Furthermore, an in-depth analysis on the crystallization of different A-site cations, i.e., MA and FA, or different halide compositions by changing the various condition such as evaporation rate, substrate temperature, and evaporation time should be conducted to reveal processes occurring inside the vacuum chamber during the deposition. Third, a subdivision of the deposition stage can more precisely control the precursor stoichiometry when the number of precursors increases. Co-evaporation or sequential evaporation has been used so far because most of the researchers used only 2 or 3 precursors. However, a more complicated deposition process is needed when the passivation materials are incorporated into perovskite due to the different sticking coefficients of different materials.

**Concluding Remarks**

Vacuum deposition is a mature technology widely adopted in the OLED display industry. Although most of the research on PeLEDs is concentrated on the solution process, vacuum-deposited PeLEDs have attracted considerable attention due to their solvent-free process, scalability, and being capable of fine control of film thickness. However, there are still many hurdles to achieving comparable performance to OLEDs. It is important to extend the study to low-dimensional perovskite with various passivation strategies. Also, finding vacuum-processible HTL is necessary in fabricating all-vacuum-deposited PeLEDs. We believe that there is a lot of potentials to further enhance the performance of vacuum-deposited PeLEDs and take a step forward to commercialization.

**Acknowledgements**

This work was supported by the National Research Foundation of Korea (NRF) grant funded by the Korea government (MSIT) (No. 2020R1A2C3008111). The work was supported by development of smart chemical materials for IoT devices (KRICT SI 1921-20). This work was supported by the National Research Foundation of Korea (NRF) grant funded by the Korea government (MSIT) (No. NRF-2022K1A4A8A02080275).

## 6. Interfacial Engineering of Perovskite Light-Emitting Diodes

*Jinquan Shi* and *Mengxia Liu*

Yale University, USA

**Status**

PeLEDs have achieved remarkable progress in recent years, reaching high color purity and impressive EQE of greater than 20% for green,[19] red,[30] and near-infrared emission wavelengths.[29] However, PeLEDs still suffer from poor operational stability, efficiency roll-off at high current density,[97] and low EQE at blue emission.[98,99] Interfacial engineering has been identified as a successful strategy for enhancing both efficiency and stability of PeLEDs.

A typical PeLED device consists of a perovskite light emitting layer (EML) sandwiched between an electron and a hole injection layer (HIL), a transparent conductive layer as anode, and metal layers as cathode. Each interface plays an important role in controlling charge recombination, injection and degradation behavior of PeLEDs (Figure 11). Notably, interfacial defects inevitably exist between the perovskite EML and carrier injection layers (CILs) and result in nonradiative recombination of carriers. These defect sites, in particular vacancies and interstitials, are responsible for ion migration across the interfaces,[100] initiating irreversible degradation upon exposure to moisture, oxygen or heat.[101]

In addition to defects, energy level alignment between different layers determines the carrier injection and transport dynamics at interfaces. Energy level misalignment induces injection barriers, which result in undesired charge accumulation and unbalanced charge transport. A good CIL with intrinsically high charge mobility and ideal electronic band structure results in efficient charge injection, while protecting perovskite EML from ion diffusion between adjacent layers.

Moreover, electrochemical reactions driven by external bias can happen at both the anode and cathode interfaces. Redox reactions at the cathode interface reduce $Pb^{2+}$ interstitials into $Pb^0$ and lead to the decomposition of perovskite.[102] Chemical reactions between perovskites and metal contacts or CILs cause the corrosion of electrodes and substantially deteriorate device performance.[103,104]

Appropriate selection of CILs and electrode materials is a key to control charge transport and support the high-quality growth of the perovskite layer. Poly(3,4-ethylenedioxythiophene) polystyrene sulfonate (PEDOT:PSS) is one of the most commonly used materials for HILs in perovskite optoelectronics. However, PEDOT:PSS induces strong carrier quenching and hole injection barriers due to the large energy level misalignment.[105,106] New materials, dopants, or modifications are expected to increase the stability and conductivity of HIL and improve carrier injection efficiency. Moreover, the nucleation and growth of perovskite EMLs are highly dependent on the selected CIL underneath. Hydrophilic CIL materials such as metal oxides[107,108] improve the wettability of perovskite precursor solution and enable the growth of high-quality perovskite films with reduced nonradiative recombination and current leakage.

Another strategy is to add interface modifiers between the perovskite layer and CILs. Lithium halides (LiX, X = Cl$^-$, Br$^-$, or I$^-$) have been widely used to passivate surface defects and suppress nonradiative recombination.[109] Inserting a buffer layer of perfluorinated polymeric acid such as tetrafluoroethylene-perfluoro-3,6-dioxane-4-methyl-7-octene-sulfonic acid copolymer (PFI) between perovskite and HIL can reduce charge transport barriers and facilitate hole injection.[110] Hydrophilic polymers for example polyvinylpyridine (PVP) have been used on top of the ZnO HIL to modify the interfacial wettability, leading to uniform growth of perovskite film and good reproducibility.[111]

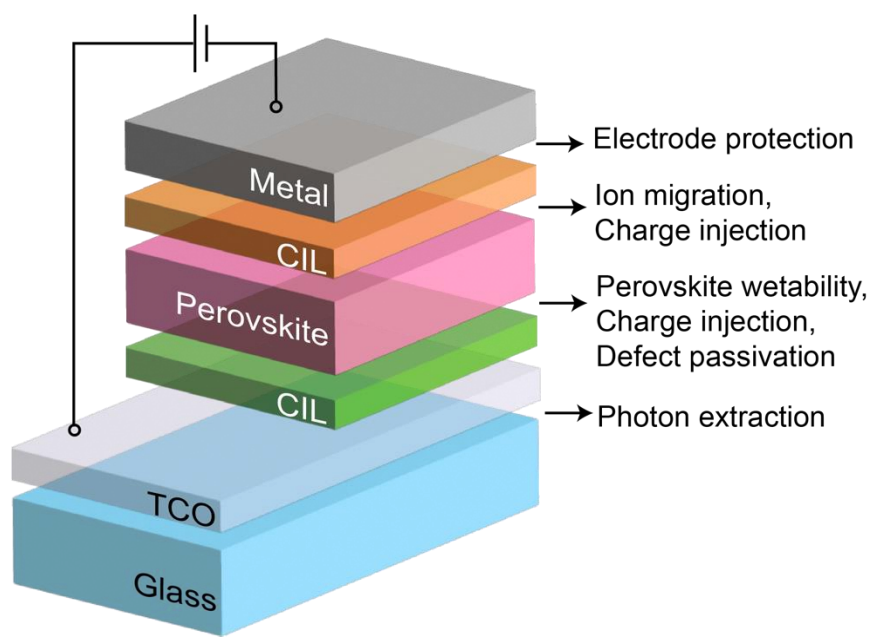

**Figure 11.** A sketch of a PeLED device. The role of each interface playing in a device is listed on the right. TCO, transparent conductive oxide. CIL, charge injection layer.

**Current and future challenges**

Understanding the physical and chemical interactions at the buried interfaces in a device is a vital but challenging task.[112] The correlation between interfacial structure and carrier dynamics remains unclear and requires in-depth research integrating both experimental and theoretical study. Novel imaging and spectroscopy techniques bring new insights to the structural and carrier dynamics of the materials.[29,113] The combination of ultrafast optical spectroscopy with microscopy approaches provides unconventional details of charge and exciton behavior with spatial, temporal, and spectral resolution.[114,115] To boost the device performance and stability, more emphasis should be placed on the investigation and analyses of interfacial transport and degradation mechanisms, especially under voltage bias, to guide materials innovation and device fabrication.

The versatile roles of each interface and the sophisticated interplay between adjacent layers impose another challenge to the design and selection of novel interface materials. It is crucial to find effective interfacial engineering strategies that allow for efficient injection, balanced transport, while maintaining good perovskite crystallization compatibility and device stability. This requires the interface materials to be conductive and thin enough to fulfill the required transport properties, but stable and thick enough to avoid pinholes, ion migration, and interfacial reactions. The commonly employed organic carrier injection materials suffer from poor operating stability. Joule heating and interfacial electrochemical reactions coupled with charge carrier injection lead to rapid degradation of organic CIL. Inorganic CILs with appropriate doping levels or surface modifications[116,117] are promising candidates to achieve significantly improved stability without compromising efficiency. In addition, the ionic nature and the soft lattice of perovskite render it more difficult for contact engineering, as the perovskite presents strong chemical interactions with many established interfacial materials.

Interfacial engineering in specific types of devices encountered more difficulties. Blue-light PeLEDs suffer from insufficient hole injection. Due to the large bandgap of blue-emission perovskites, their valance band maximum is usually deeper than the highest occupied molecular orbital (HOMO) of commonly used HIL materials, producing large hole injection barriers. To address this challenge, HILs with deeper HOMO level and higher carrier mobility are required to facilitate hole injection into blue perovskite EMLs and achieve more balanced charge injection. Interfacial engineering strategies for WPeLEDs and flexible PeLEDs are also desired. In single-emissive-layer WPeLEDs, the large bandgap of EML – such as cesium copper halides[118] – proposes great difficulties in carrier injection. Devices with multiple emissive layers are more demanding in interface design and light out-coupling strategies considering the complex electrical and optical properties of each layer. In flexible PeLEDs, finding CIL and electrode materials compatible with mechanical deformation is a major challenge. Interfacial slippage and delamination are prone to occur at the interfaces and are detrimental to device functionalities. Innovative strategies should be designed to reduce the interfacial strain and increase the adhesion between different layers.

**Advances in science and technology to meet challenges**

The exploration of new interface modifiers is a dominant approach towards effective interfacial engineering. Zhu et al. introduced organic molecules with high triplet energy levels as an EIL modifier to block undesirable exciton transfer in a sky-blue PeLED, improving the EQE of the device.[119] Enhanced EQE has also been achieved by passivating uncoordinated $Pb^{2+}$ with a series of self-assembled monolayers fabricated on an ultrathin layer of Au.[120] Zhao et al. modified the poly(9,9-dioctylfluorene-alt-N-(4-sec-butylphenyl)diphenylamine) (TFB) HIL with an ultrathin (~1 nm) layer of LiF, which significantly improved the wettability of HIL and thus resulted in perovskite films with uniform grain size and longer carrier lifetime (Figure 12a).[121]

New materials that enable optimized CILs and electrodes are also essential for improving the performance of PeLEDs. Conjugated materials having hydrophilic nature are promising CIL materials that support the crystallization of perovskite (Figure 12b).[122] Tunable energy levels and interface characteristics can be achieved by varying the functional group. Metal oxides with proper doping or surface modification) can lead to desired performance and show potential for mass production (Figure 12c).[111,123,124] For instance, Li-doped $NiO_x$ HIL has been reported in large-area all-vacuum-deposited PeLEDs.[90] As for flexible devices, stretchable electrodes compatible with mechanical deformation should be developed. Self-organized conducting polymers[125] and MXene-based materials[126,127] have been successfully used in flexible PeLEDs and shown competitive performance compared with rigid PeLEDs (Figures 12d and 12e).

Efficient photon extraction via interfacial engineering is also an important way to achieve extreme efficiency. Perovskite EMLs possess a higher refractive index compared to that of the organic CILs and transparent electrodes, trapping the emitted photons via the total internal reflection at interfaces inside PeLEDs. Efficient photon extraction can be achieved when optical constants of different layers are fully matched. Chen et al. introduced a light out-coupling strategy that utilized a layer of red-emitting perovskite NCs on top of a blue-light PeLED with a semi-transparent cathode.[128] Through photon tunneling and evanescent wave coupling, the top layer successfully extracts the blue photons trapped within the device and converts them into red emission, enabling a white light PeLEDs of EQE above 12% (Figure 12f). Shen et al. obtained a nanostructured front-electrode/perovskite interface by patterning the ZnO and PEDOT:PSS layer and thus achieved an enhanced outcoupling efficiency of the waveguided light.[129] An array of microlens were mounted on the glass substrate to further reduce the internal reflection at the substrate/air interface, enabling an EQE of 28.2% and 88.7 cd $A^{−1}$ in green PeLEDs.

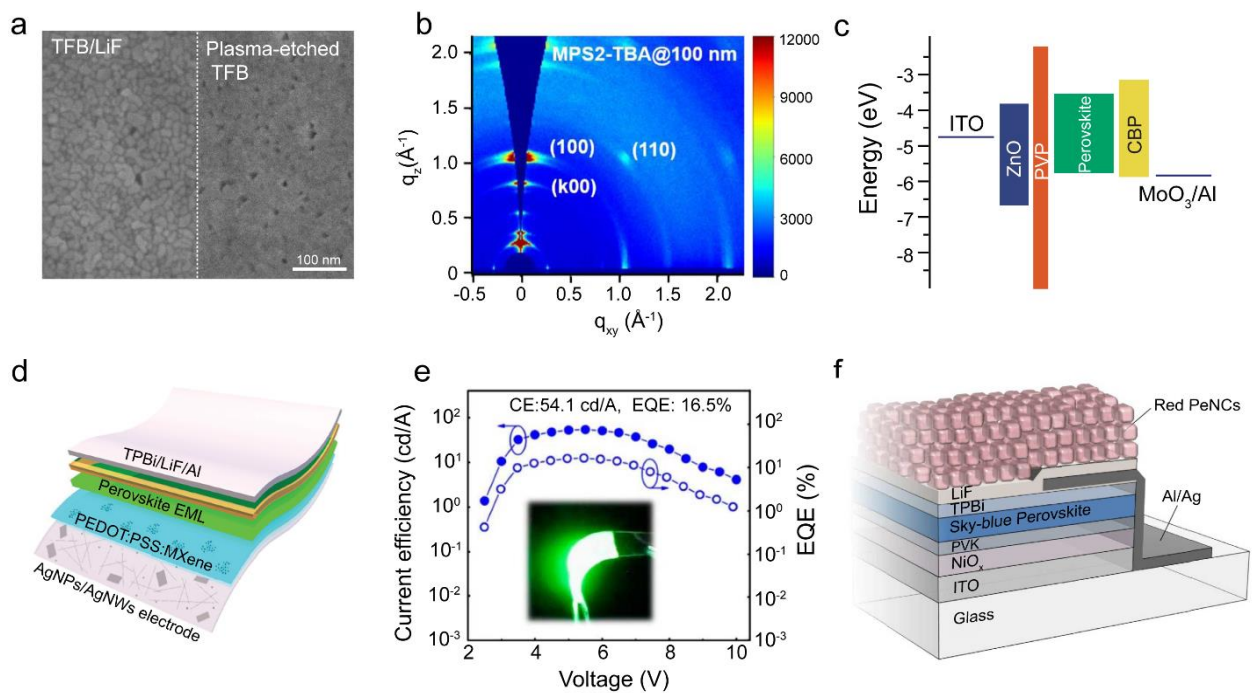

**Figure 12.** Interfacial engineering strategies for PeLEDs. (a) The morphology of perovskite films deposited on TFB/LiF (left) and pristine plasma-treated TFB surfaces (right). Lithium fluoride significantly improved the morphology and crystallinity of EML. Reproduced from Ref. [121]. (b) The grazing incident wide-angle X-ray scattering pattern of 100 nm-thick perovskite film deposited on a conjugated polyelectrolyte HIL. Enhanced perovskite growth was achieved compared to a PEDOT:PSS HIL. Reproduced from Ref. [122]. (c) The energy band diagram of an ultrabright PeLED based on PVP-modified metal oxide HIL. PVP acted as a passivator while also helped with balanced injection and perovskite morphology. Reproduced from Ref. [111]. (d) A sketch of the device architecture and the (e) current efficiency and EQE of a flexible PeLED using MXene-based composite electrodes. Reproduced from Ref. [127]. (f) The sketch of a high-performance WPeLED with enhanced light extraction efficiency by adding a layer of red perovskite NCs on top of the LED device. Reproduce from Ref. [128].

**Concluding remarks**

Despite the great potential of PeLEDs in lighting and display technologies, the commercial application of PeLEDs is still hindered by the low efficiency of blue PeLEDs and poor operational stability. Interfacial engineering has been proven as a promising path towards efficient and stable PeLEDs. Future improvement requires more comprehensive physical understandings of the concentration and distribution of defects across the interfaces, carrier dynamics and transport behaviors, and the degradation mechanism of PeLEDs. To alleviate the EQE roll-off and operational stability issue, the interfaces should be optimized to block interfacial electrochemical reactions and achieve barrier-free charge transport, balanced charge injection, and low device operating voltages. To further increase the EQE of PeLEDs, more efforts should be paid to the development of photon extraction strategies through electrode patterning and interface modification. In addition, the engineering of a single interface or one component of PeLEDs is not adequate to achieve the next milestone in efficiency and stability. Simultaneous modification of various interfaces will provide synergistic advantages which are more beneficial to the overall device performance.

## 7. Optical Engineering of Perovskite Light-Emitting Diodes

*Qianpeng Zhang* and *Zhiyong Fan*

The Hong Kong University of Science and Technology, Hong Kong SAR, China

**Status**

Since the EQE of PeLEDs has already reached a high level of over 20% and the IQE is close to unity, the hope for further improving the EQE lies in optical engineering that enhances light extraction. Due to the total internal reflection caused by the high-index nature inside the planar PeLED, most of the light generated cannot be extracted but remains in a waveguide mode.[130] Other non-extracted photons can be dissipated in surface plasmon polariton (SPP) mode, substrate mode, parasitic absorption, etc.[131]

Adding nanostructures outside or inside thin-film PeLEDs is a promising strategy for lighting purposes (Figure 13a).[132,133] Mao et al. performed direct nanopatterning on perovskite material to form periodic nanograting structures, which can couple out the guided mode with the help of Bloch mode.[134] By applying moth-eye shaped PEDOT:PSS layer, Shen et al. improved the EQE of $CsPbBr_3$ LEDs from 13.4% to 20.3%.[129] A typical example of nanostructures inside perovskite EML is the spontaneously formed submicron structures reported by Cao et al.[135] The structures increased light outcoupling efficiency from 21.8% to ~30%, compared to the flat control. More importantly, the EQE of 30% at low temperature (6 K) verified their outcoupling efficiency experimentally.

Even without additional nanostructures, the light outcoupling can be also optimized. For example, by using micro-cavity effect with a bottom Au reflector and a top semi-transparent Au electrode, Miao et al. successfully increased the EQE of the near-infrared multiple-quantum-well PeLEDs from 14.5% to 20.2%.[136] Moreover, according to Zhao et al, the thickness of perovskite layer also has an effect on the outcoupling efficiency.[137]

On the other hand, the intrinsic optical engineering of PeLEDs can be the control of the dipole directions, as has been proved to be quite effective in OLEDs.[138,139] For instance, Kumar et al. used the anisotropic NC superlattices to increase the ratio of horizontal dipole in PeLEDs to 75%, resulting in an enhanced light outcoupling efficiency of 30%.[140] Cui et al. obtained perovskite nanoplatelet film with horizontal dipole ratio of ~84%, corresponding to a light outcoupling efficiency of ~31% and an EQE of 23.6%.[141] The orientation of transition dipole moments can be measured by angular dependent PL and back focal plane imaging, as reported by Jurow et al.[142]

Intriguingly, Zou and Lin found that the dipole direction in CsPbBr$_3$ films is more vertically oriented by *p*-polarized PL measurements.[143] This finding makes the optical engineering more demanding in order to reduce the ratio of vertical (out-of-plane) dipoles.

**Current and future challenges**

As we know, one of the most appealing advantages of PeLED is its facile fabrication with solution methods. Adding extra nanostructures increases the fabrication difficulties and requires a delicate design and optimization process, making optical engineering not that popular in current PeLED research community. Moreover, additional structures will increase the volume of light emitting devices and therefore are not favored for the flat-panel displays.

Despite the above concerns, the optical limit on the light outcoupling efficiency restrains the upper bound of PeLEDs' EQEs, and optical engineering can overcome that limit. The future perovskite optoelectronic systems need more complicated optical integrations of different types of photonic devices,[144] which makes optical engineering more and more significant than in simple planar devices,[145] especially when the PeLED performance approaches the limits.

It's also worth mentioning that photon recycling is an important process when considering the light outcoupling processes. The non-extracted photons can experience the absorption and re-emission processes multiple times, therefore the light outcoupling efficiency is also affected by the radiative and non-radiative recombination rates.[66,146] According to Cho and Greenham, the re-absorption of dipole emission cannot be ignored in perovskite when it comes to the accurate optical modeling of PeLEDs.[147]

**Advances in science and technology to meet challenges**

A well-designed nanophotonic substrate can significantly enhance the light extraction, and we reported a nanophotonic structure with light outcoupling efficiency of up to 73% in 2019 (Figure 13b).[132] We also showed that only optimized structure is favored for light extraction improvement, making the optical design mandatory for devices with different materials and configurations. The nanodomes below thin film MAPbBr$_3$ LED can effectively couple light from perovskite EML to the bottom TiO$_2$ nanowire arrays in porous alumina membrane (PAM). Then the nanowire/PAM structure works as the optical antennas to convert the guided modes to leaky modes for light extraction to air. With the assistance of Finite-Difference Time-Domain simulations, we could also visualize the light propagation processes with different structure geometries, which provides us with an intuitive understanding of the light extraction mechanisms.

Another method of optical engineering is directly modifying the perovskite EML. As the light outcoupling efficiency is inversely proportional to the square of the emitting material's refractive index, reducing the index can be a straightforward and effective way. For example, by embedding perovskite quantum wire (diameter 6.4 nm) arrays in PAM,[148,149] the effective index of EML can be reduced, given that the refractive indexes of perovskite and aluminum oxide are about 2.2 and 1.7, respectively. Zhao et al reported the perovskite-polymer bulk heterojunction LEDs with an EQE of 20.1% in 2018.[47] By mixing perovskite ($n \sim 2.7$) with polymer ($n \sim 1.5$), the effective index of this system was reduced to 1.9 and the outcoupling efficiency was therefore increased. In terms of modifying the EML, there is another interesting report from Hou et al.,[150] in which they showed that the perovskite photonic crystals contributed to the light energy redistribution from 2D guided modes to vertical direction in perovskite photonic crystals thin films, leading to a 23.5-fold enhanced PL.

Because waveguide mode and SPP mode both can dissipate about 20−30% photons, Chen et al. designed a WPeLED device structure by depositing a thick layer of red perovskite NCs on the metal electrode of blue PeLED device, which can utilize the evanescent fields of waveguide mode and SPP mode.[128] The WPeLED's average EQE was more than 1.5-fold higher than the sky-blue counterpart, thanks to the effective extraction of trapped sky-blue photons by red perovskite NCs.

In terms of the dipole control, Walters et al. studied the directional light emission from layered metal halide perovskite crystals with the ratio of in-plane dipole up to 90% and the theoretical EQE can reach 45%.[151] Adding nanostructures to the EML can also affect the ratio of the horizontal dipole. For instance, by applying nanohole arrays inside EML, Jeon et al. found that the ratio of horizontal dipole was increased to 71% by angle dependent PL measurement, a bit higher than 67% for isotropic dipole direction.[152]

In addition to optical engineering of the EML, modifying the optical property (e.g., thickness and index) of injection layer can also affect the light outcoupling. The ultrathin PEDOT:PSS can enhance both the light outcoupling and balanced charge injection, as reported by Lu et al.[153]

Intriguingly, since the optical engineering leads to extraction of more photons, fewer phonons can be generated internally, in order to maintain high overall luminance. This makes optical engineering not only useful for outcoupling efficiency improvement but also for an easier heat management.[154] Considering this reason, researchers might pay more attention to optical engineering for the sake of achieving long device operational life-time.

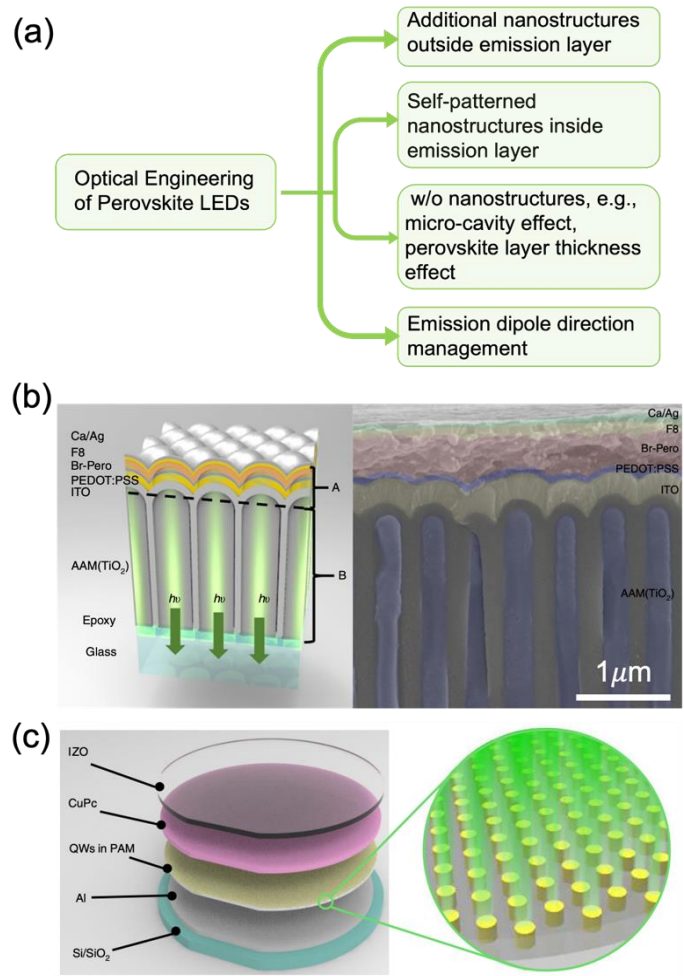

**Figure 13.** Optical engineering of PeLEDs. (a) General strategies of optical engineering of PeLEDs: (1) additional nanostructures outside EML; (2) self-patterned nanostructures inside EML; (3) w/o nanostructures, e.g., micro-cavity effect, perovskite layer thickness effect; (4) emission dipole direction management. (b) example of additional nanostructures outside EML: nanodomes as light couplers and nanowire arrays as optical antennas. Image reproduced with permission from Ref. [132]. Copywrite Springer Nature 2019. (c) example of self-patterned nanostructures inside EML, perovskite quantum wire arrays inside porous alumina membrane, and the EML's effective refractive index is reduced. Image reproduced with permission from Ref. [148]. Copywrite Springer Nature 2022.

**Concluding remarks**

Optical engineering of PeLEDs can address the limited light extraction problem. A universal strategy is tuning the EML layer's properties such as thickness, refractive index, the dipole ratio, and so on, which does not change the overall planar morphology and is more compatible with flat panel displays. As for integrated optical systems, delicate optical design will play an important role.

Both efficient light outcoupling from individual devices and light coupling between interconnected optical components will be required. The non-extracted photons can become phonons which can increase the working temperature of the devices. This makes optical engineering especially important in harsh circumstances.

When designing the nanophotonic structures, we should also notice that the light extraction is not an isolated process, it is also related to photon recycling, recombination rate, heat (phonon) generation, etc. We can expect that more investigations on this topic related to optical engineering of PeLED devices will be conducted in the future. In fact, for inorganic GaN based LEDs, optical engineering strategy has been adopted in patterned/nano-patterned sapphire substrates for better performances in practical devices.

**Acknowledgements**

The authors acknowledge the support from Hong Kong Research Grant Council (GRF 16214619, 16205321), Guangdong-Hong Kong-Macao Intelligent Micro-Nano Optoelectronic Technology Joint Laboratory (grant no. 2020B1212030010), and the support from The State Key Laboratory of Advanced Displays and Optoelectronics Technologies at HKUST.

## 8. Stability of Perovskite Light-Emitting Diodes

*James C. Loy*, *Lianfeng Zhao* and *Barry P. Rand*

Princeton University, USA.

**Status**

The performance of PeLEDs has dramatically improved within the last few years, with peak EQE above 20% and device lifetime of hundreds of hours being repeatedly reported. The latter is inadequate for practical application, and lags far behind reported lifetimes of OLEDs or QLEDs. Additionally, most reported peak EQEs of PeLEDs are achieved at relatively low current densities, and device lifetimes are usually tested under conditions with relatively low initial luminance specifically to avoid known stressors on the device. For the successful commercialization of PeLEDs, further improvement in their operational stability is required to make the best devices also those that are most stable and reproducible. Several mechanisms responsible for PeLED instability are discussed below, which are closely related to the unique material properties of metal halide perovskite semiconductors.

**Current and Future Challenges**

The stability of PeLEDs is very sensitive to temperature. It has been reported that by increasing the environmental temperature from 10 to 40 °C, the device lifetime is reduced by an order of magnitude.[137] This temperature sensitivity is sourced from the unique properties of metal halide perovskite semiconductors, chief among these is that the halide perovskite medium is a mixed electronic and ionic conductor. The ionic property, in particular, distinguishes halide perovskites from many other thin-film semiconductors such as organics and colloidal QDs. Concurrently, metal halide perovskites are chemically active, and chemical reactions are involved in numerous device degradation processes. Notably, all these ionic and chemical processes are thermally activated, resulting in the high temperature sensitivity of PeLEDs during device operation. Attempts to enter higher current regimes are therefore thwarted in performance by operational temperature increases stemming from joule heating. Exacerbating these effects is the fact that the science of doping to increase the free charge population and reduce resistance has not been successfully uncovered.

Given the current-injection operation of LEDs, metal halide perovskites are further subject to electrical stressors that promote degradation through various forms of ionic migration.[155] These issues can arise within the emitter layer itself through various forms of perovskite decomposition. In one example, halide segregation of an initially homogeneously mixed composition is a major issue for mixed halide materials that threatens the potential tunability and color purity of operating devices. These perovskites tend to separate into respective halide-rich regions, and leading to broadened and red-shifted emission. The cause of this behavior is still an active research area, but appears to be related to the trap-state density of the crystal.[156] One proposed model that manages to explain photo-assisted as well as voltage-induced segregation attributes halide segregation to a natural disparity between the ease of oxidation between the mixed halide constituents, and proposes various (thermodynamic as well as kinetic) mechanisms over different length scales that lead the various oxidized species to displace throughout the crystal and redistribute during operation.[157] Further, perovskites can be subject to photo-assisted degradation internally,[158] and reactive degradation at its interfaces with charge transport layers.[159]

Beyond the emitter layer, metal halide perovskites can also compromise adjacent layers. For example, ionic transport away from the perovskite occurs when volatile halide species are allowed to permeate the device stack. This has been demonstrated to dope hole transport materials and corrode device electrodes, which can alter or destroy devices during operation.[160]

In summary, for PeLEDs to be stable we must surmount the difficult task of protecting the adjacent charge transport layers, metal contacts, and the perovskite itself during continuous operation, and these solutions should ideally be applicable to the full spectrum of perovskite stoichiometries and deal with the plethora of different charge transport layers and electrodes. Harmonizing strategies to mitigate distinct sources of degradation is likewise not guaranteed.

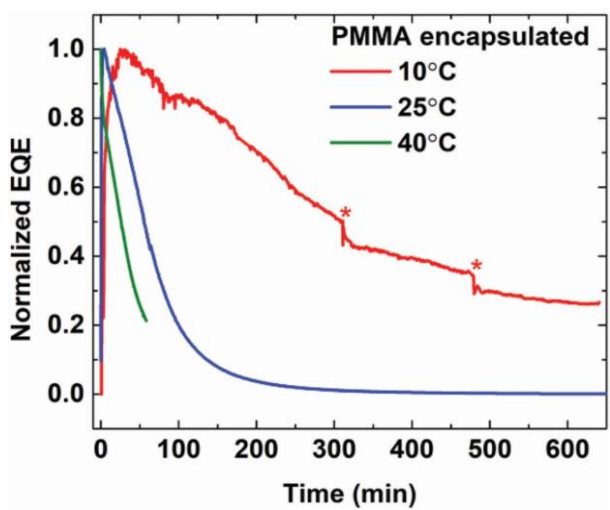

**Figure 14.** Operational stability of PeLEDs working at various environmental temperatures operating at a constant current density of 10 mA/cm$^2$. Reprinted with permission from [137].

**Advances in Science and Technology to Meet Challenges**

Given the temperature sensitivity of PeLEDs, an obvious strategy to improve the device lifetime is through efficient thermal management. Several thermal management strategies have been shown to be effective to keep PeLEDs cool and improve device lifetime, such as the use of more electrically conductive charge transport layers to reduce voltage drop and joule heating, the addition of heat sinks, and the use of more thermally conductive substrates for better heat dissipation.[76] Figure 14 shows the promise of improved EQE roll-off versus time stemming from cooler environmental temperatures, but these are difficult to expect in real-world applications. Addressing the pixel scale directly is more likely to be commercially viable and warrants further consideration.

Electrode corrosion and adjacent charge transport layer doping can be addressed through prudent selection of charge transport layers. A sufficiently deep HOMO level has been shown capable of reducing ionic transfer through hole transport materials, reducing halide uptake and reaction with Au electrodes.[160] This is further capable of prolonging perovskite layer integrity. Further suppression of ion migration has been directly demonstrated in 2D Ruddlesden-Popper perovskites when compared to 3D counterparts, indicating that some form of layer engineering may be employed to inhibit out-of-plane ionic transport.[161] Perovskite decomposition has been shown to be stymied via clever additions of dicarboxylic acids which form stable amide barriers to prevent interfacial reactions between the perovskite and adjacent charge transport layers.[159]

Halide segregation has received considerable research attention and has created a few promising avenues to pursue, primarily involving optimally tuned ratios of ions introduced in the A-cation site, larger crystalline grains, and reducing trap densities with additives.[156] The A-cation site, in particular, tunes the perovskite lattice's bond lengths, angles, and molecular orbital energy levels to a degree; it may then be possible to engineer the site to alter halide oxidation rates (with oxidized halide migration being a known avenue for segregation).[157] Cross-linked passivation of grain boundaries has recently been shown to suppress halide mobility within green PeLEDs granting a promising 200 h of stability in a device above 15% EQE (see Figure 15).[162]

The A-site cation holds further importance; perovskite degradation via reaction with deprotonated organo-ammonium species (i.e., amines) on this site has been shown to lead to formation of $Pb^0$ under stress through a decomposition of a formed lead-amide complex. This reaction has been studied and design parameters required to prevent this degradation pathway are known, but this further restriction requires more study to generate a known collection of safe ammonium cationic species.[158]

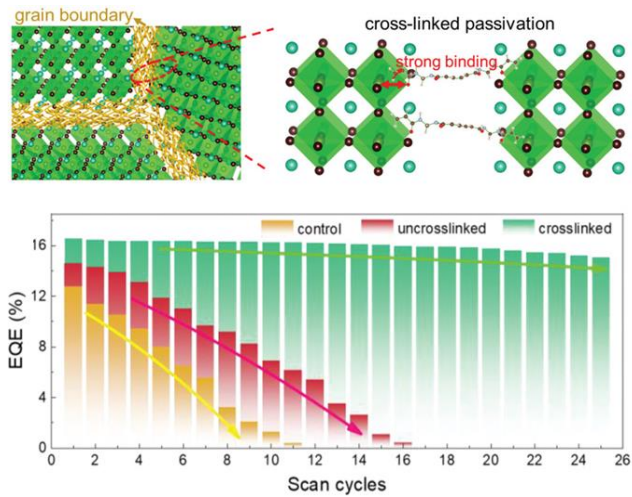

**Figure 15.** Cross-linking strategy in a green PeLED interferes directly with ionic transport at the grain boundaries, demonstrating improvement in repeated run stability and leads to a device with a $T_{50}$ of over 200 h. Reprinted with permission from [162].

**Concluding Remarks**

The promise of PeLEDs as a disruptive technology hinges on the ability to stabilize the material in a practical device setting. Making this challenge more intimidating is the apparent competition between currently known ways to address various instabilities. Finding cooperation between novel stabilizing methods is also not guaranteed, further obscuring the end-goal. However, we are fortunate that most ways to address stability issues lie in the direction of also creating higher performance devices, a direction the field is already pursuing and refining.[156] Further efficiency improvements are expected to naturally produce new techniques of stabilization in the lab setting, be that through novel charge transport layers to improve electron-hole charge balance, new electrical dopants to increase device conductivity. Already, we have shown examples of new additives that alter perovskite crystal grain sizes and produced candidate methods with fewer trap states, reduced ion transport, less redox active surfaces, and improved thermal management. With more options that become available, the higher the likelihood that combinations facilitate future success. It is clearly worthwhile that as these potential breakthroughs are found, we also conduct further study to better clarify the mechanisms of improvement and how it can be balanced with the myriad demands for a complete working device.

**Acknowledgements**

The authors acknowledge funding from DARPA under Award No. N66001-20-1-4052.

## 9. Lead-Free Halide Perovskite Light-Emitting Diodes

*Habibul Arfin*, *Sajid Saikia* and *Angshuman Nag*

Indian Institute of Science Education and Research Pune, India

**Status**

Lead halide perovskites are wonderful optoelectronic materials. But what is so special about $Pb^{2+}$? The existing wisdom suggests $6s^2$ (two $s$ electrons in the outermost orbital) electrons and strong spin-orbit coupling of $Pb^{2+}$ ions do miracles in electronic band structure, leading to interesting optoelectronic properties. These findings raise obvious fundamental curiosity about why can't one develop lead-free halide perovskites with electronic band structure similar to the lead-halide perovskites. Importantly, such lead-free materials might have advantages of being environmentally benign and more stability, compared lead-halide perovskites. A combination of this fundamental curiosity and technological relevance has led to the exploration of various lead-free halide perovskites for LEDs.

Lead-free halide perovskites have been explored for both EL LED, and PL based phosphor-converted LED (pc-LED).[163,164,165] Figure 16a shows that $CsSnI_3$ EL LED that emit near infrared light (peak at 932 nm). The radiance reduced to 50% of its maximum radiance after, $T_{50}$ = 23.6 h of operation inside $N_2$ filled glovebox.[163] By optimizing the film quality, device structure and controlled charge injection in the device, the EQE of 5.4% is achieved with maximum radiance of 162 W sr$^{-1}$ m$^{-2}$. To improve the stability, 2D layered hybrid Sn-halide perovskites are used increasing the $T_{50}$ > 150 h inside $N_2$ filled glovebox, but EQE reduced to 3.3% and the emission peak wavelength blue shifted to ~605 nm.[166]

The other strategy is to develop lead-free halide perovskite phosphors, that can be coated on a commercial ultraviolet or blue LED chip. Double perovskites like $Cs_2AgInCl_6$ and $Cs_2NaInCl_6$ are doped with suitable metal ions, yielding pc-LEDs of desired light emission (see Figure 16b). Doping s-electrons ($Bi^{3+}$, $Sb^{3+}$), d-electron ($Mn^{2+}$), f-electrons ($Sm^{3+}$, $Yb^{3+}$, $Er^{3+}$, $Ce^{3+}$) yields emission of a wide variety of visible lights with desired color and spectral width.[167,168,169] Interestingly, $Yb^{3+}$, $Nd^{3+}$, $Er^{3+}$ and $Cr^{3+}$ doped in double perovskites yields short-wave infrared radiation of peak wavelengths in the range of 880 to 1550 nm and spectral width in the range of 0.25 to 0.01 eV.[164,170,171] These results provide huge scope for developing pc-LED, particularly in the short-wave infrared region.

Note that there are perovskite derivative structures like 2D $FA_3Bi_2Br_9$, $Cs_3Sb_2Br_9$ NCs, impurity ion doped 0D $Cs_2SnCl_6$ and $Cs_2ZrCl_6$, that show intense emissions in the visible region. Non perovskite $Cs_3Cu_2I_5$ have PLQY close to 90%.[165] Such metal halides are also good candidates for pc-LED applications.

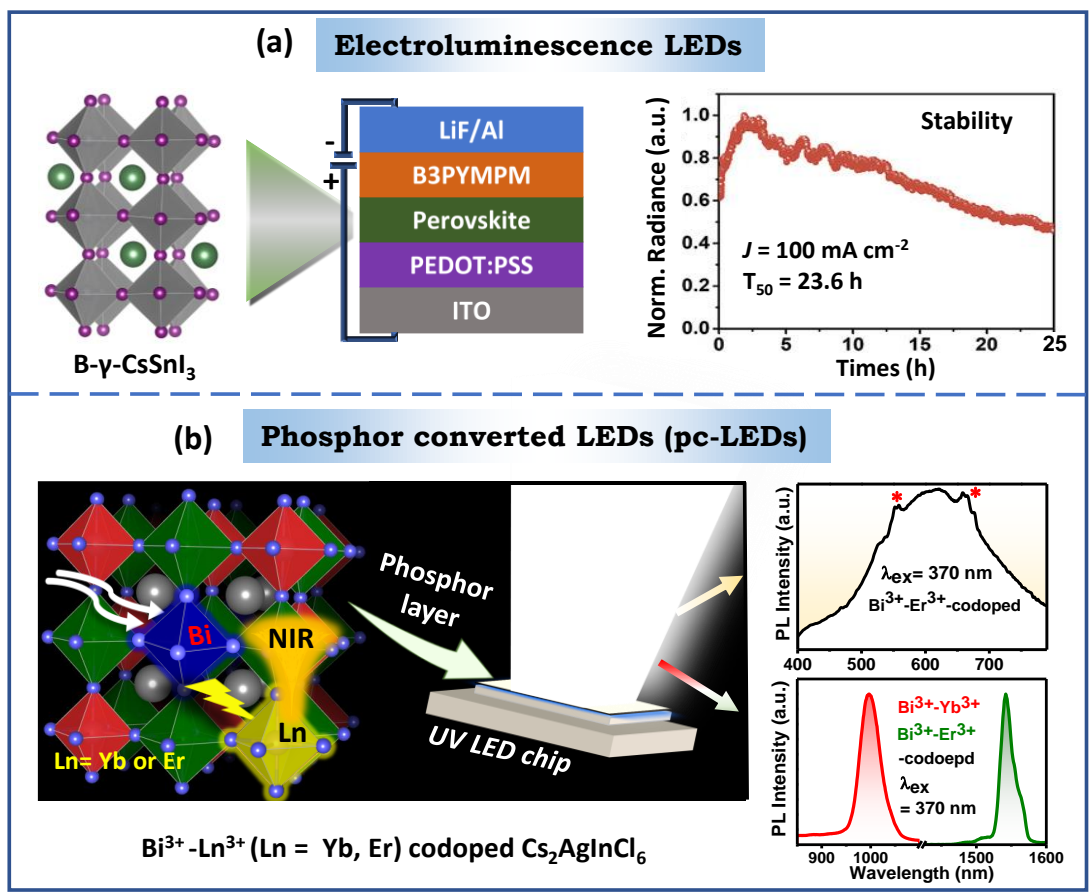

**Figure 16.** (a) Schematics of crystal structure of orthorhombic black B-γ-$CsSnI_3$, which has been used as active layer in fabricating an EL LED device. The right panel shows the radiance stability of the device operating at high current density of 100 mA cm$^{-2}$. (b) Schematics of $Bi^{3+}$-$Ln^{3+}$ (Ln = Yb, Er) codoped $Cs_2AgInCl_6$ double perovskite phosphor, that can be coated on an ultraviolet chip, fabricating pc-LED that show both broad white light emission and narrow short-wave infrared radiation. Stability plot in (a) is reprinted with permission from [163], and (b) is reprinted with permission from [164].

**Current and Future Challenges**

EL LED: Poor stability and/or poor charge transport is the major problem. EL operates at large diffusion current, where $Sn^{2+}$ of $CsSnX_3$ (X = $Cl^-$, $Br^-$, $I^-$) rapidly oxidizes to $Sn^{4+}$. Initially the $Sn^{4+}$ defects reduce both PL and EL efficiency, and subsequently, sufficient oxidation leads to structural phase transition from 3D $CsSnX_3$ to 0D $Cs_2SnX_6$ perovskite derivative structure. Such 0D structures suffers from poor charge transport. Other perovskite derivatives or metal halides like $Cs_3Cu_2I_5$, $Cs_2ZrCl_6$, and $Cs_3Bi_2X_9$ (X = $Cl^-$, $Br^-$, $I^-$) also possess the low- dimensional structure, hindering the charge transport. Double perovskites like $Cs_2AgInCl_6$ and $Cs_2NaInCl_6$ possess the 3D perovskite structure but exhibit both wide band gap (>3.6 eV) and poor carrier mobility. All these reasons limit the materials space of lead-free halide perovskites for EL LED applications. The obtained EQE and $T_{50}$ values remain significantly inferior compared to lead-halide perovskites and other bench mark LED materials.

pc-LED: White and other visible light emitting pc-LEDs have been demonstrated for doped double perovskites, and metal halides with lower dimensional perovskite derivative structure. Importantly, there are huge numbers of oxide, nitride and phosphate-based phosphor materials along with a few commercial ones exists.[172] Internal PLQY of lead-free perovskite phosphors compares well with the previously studied phosphor materials. But the high quantum yield is not the sufficient criterion for real-life applications. The perovskite phosphors need to remain stable under prolonged operation at high temperatures around 100−200 ºC. For high power pc-LEDs, the PL spectral shape and quantum yield of phosphors need to remain stable at high excitation powers in the range of 1 to 10 W $mm^{-2}$. For example, the commercially successful phosphor $Y_3Al_5O_{12}$:$Ce^{3+}$ (YAG:$Ce^{3+}$) show 90% PLQY at room temperature, which decreases by only 12% at 200 ºC, and stable with excitation power up to around 7.5 W $mm^{-2}$.[172] Such a high stability is attributed to high structural rigidity of the YAG lattice as indicated by their high Debye temperatures. Unfortunately, such rigorous stability study has not yet been reported yet for halide perovskites. Achieving similar stability is a key requirement before the halide perovskites find applications in real pc-LEDs. Another issue is that the doped double perovskites that emit short-wave infrared lights are excited by ultraviolet LED chips. Consequently, the electrical to optical power conversion efficiency, often termed as the wall-plug efficiency, is less for these perovskite phosphors.

**Advances in Science and Technology to Meet Challenges**

The most important scientific advancement required is in the form of material design. A suitable lead-free halide perovskite composition, that combines i) high PLQY, ii) reasonable charge transport, and iii) stability required for LED operation, is still not available. In this scenario, achieving EL LEDs with good performance is highly challenging. Developing, pc-LEDs of lead-free perovskites appears to be a lesser challenge, since charge transport is not a requirement for pc-LED. Therefore, many lead-free perovskite compositions with light emission in the visible to short-wave infrared regions can be explored. Furthermore, since the phosphors emitting short-wave infrared radiation are relatively less compared visible-light phosphors, developing lead-free halide perovskites for short-wave infrared pc-LED might become more relevant for applications in near future. Note that the short-wave infrared radiations find applications in optical fiber communication, food processing industry, and security surveillance.

The first challenge is to improve the external quantum yield of lanthanide or $Cr^{3+}$ doped double perovskites, that emit short-wave infrared radiation. In different reported methods of synthesis, the doping has been found to be rather difficult.[164, 171] The dopant concentration in the precursor reaction mixture is about 100 times more than the dopant concentration in the final product. Developing better reactions for efficient doping, may be by introducing more reactive dopant precursors and/or reducing the nucleation/growth rate of host lattice, is required. An optimized reaction will incorporate sufficient number of dopants in the desired lattice sites enhancing the external PLQY and also will reduce the wastage of dopant precursors.

A short-wave infrared phosphors like $Er^{3+}$-doped $Cs_2AgInCl_6$ needs to be excited with high energy (3.8 eV, 330 nm), and emits at very low energy (0.8 eV, 1540 nm).[164] Therefore, wall-plug efficiency of pc-LED made by such phosphors will be less. To improve the wall-plug efficiency, either double perovskites with narrower bandgaps (~1.5 eV) needs to be doped with lanthanide ions, or a codopant that absorb light of ~1.5 eV (and then non-radiatively excite the lanthanide emitters) is needed. Attempts should also be made to make the lattice as rigid as possible, improving the thermal- and photo-stability of the phosphor.

Developing an easy strategy to coat the phosphors of desired thickness on commercial EL LED chip is also required. Preparing colloidal NCs of the phosphor materials, and formulating their ink composition, might become helpful for fabrication of pc-LEDs.

**Concluding Remarks**

There are different metal ions doped and undoped lead-free halide perovskite compositions that show intense PL in the wide range of blue to short-wave infrared (1540 nm) wavelengths. The spectral width can also be tailored from 1.6 eV for white-light emission to 0.01 eV for 1540 nm emission. So the lead-free halide perovskites are versatile PL materials. Unfortunately, the obtain charge transport through these materials are not enough. Also, stability under heat, light and water/moisture needs to be improved. The immediate future direction might be to address the stability issues, along with enhancing the wall-plug efficiency, such that the pc-LEDs of lead-free perovskites become viable candidates for applications. A long-term target is to achieve stable EL

LEDs of lead-free halide perovskites with EQE ~20%, but for that, out of the box material design strategies are required.

**Acknowledgements**

Habibul Arfin and Sajid Saikia acknowledge University Grant Commission (UGC) and Prime Minister's Research Fellowship (PMRF) India, respectively, for PhD research fellowship. Angshuman Nag acknowledges Science & Engineering Research Board (Swarnajayanti Fellowship, SB/SJF/2020-21/02) India.

## 10. High-brightness Perovskite Light-Emitting Diodes

*Chen Zou*[1] and *Lih Y. Lin*[2]

[1]Zhejiang University, China

[2]University of Washington, USA

**Status**

Since the first report of PeLEDs operated at room temperature that achieved 0.76% EQE in 2014,[9] the field of PeLED has been making rapid progress. In the direction of improved EQE, demonstrations of PeLEDs with >28% EQE have been reported.[19] For the applications of optical displays and lighting, an important performance metric is luminance or radiance of the LED. Especially in lighting applications, a minimum luminance of 10,000 cd m$^{-2}$ is usually required. Although PeLEDs have achieved high EQEs, their brightness, determined by the product of EQE and the corresponding injection current density, is still not satisfactory. Like LEDs made with other solution-processed materials such as OLEDs and QLEDs, the EQE of PeLEDs decreases rapidly under high injection current density and electric field, a phenomenon termed 'efficiency droop (or roll-off)'. For many PeLEDs, the current density corresponding to 50% EQE drop is in the range ≈ 10−100 mA cm$^{-2}$. Such a large efficiency droop limits their achievable brightness up to 100,000 cd m$^{-2}$. In addition, low brightness of PeLEDs further restricts their operational lifetime up to several hundred hours at 100 cd m$^{-2}$, impeding commercialization of PeLEDs in next-generation display and lighting. With a higher carrier mobility and easier fabrication process, perovskite has stronger potential in overcoming the challenge of efficiency-droop compared to OLEDs and QLEDs. Recently, many research efforts have been devoted in this direction through perovskite material optimization, charge injection balance, device structure and operation condition design. In 2018, Zou et al. achieved low efficiency-droop PeLEDs by increasing the quantum well width in quasi-2D perovskites to reduce Auger recombination.[97] This work attracted much attention and more studies were devoted in this direction. Later, Jiang et al. employed p-fluorophenethylammonium as organic spacer to generate quasi-2D perovskites, reducing the impact of Auger recombination.[56] As a consequence, a high luminance of 82,480 cd m$^{-2}$ was achieved due to suppressed efficiency roll-off. By replacing quasi-2D with bulk 3D perovskites and balancing the charge injection through modulating energy levels of ETLs, Sim et al. made a breakthrough with an extremely high luminance of 500,000 cd m$^{-2}$,[173] suggesting great potential of perovskites for high-brightness LEDs. In this section, we discuss the research issues and challenges in suppressing efficiency droop and achieving high-brightness PeLEDs, and summarize the advances that have been made.

**Current and Future Challenges**

Overcoming the challenge of efficiency roll-off and device instability under high injection current density is essential to advancing the luminance performance of PeLEDs. The origin of efficiency roll-off in PeLEDs is still being investigated, but in general it is believed that the following factors contribute to the phenomenon (Figure 17): (1) Imbalanced charge injection and transport. (2) Luminescence quenching due to non-radiative Auger recombination. (3) Material and device degradation caused by joule heating. (4) Ion migration in perovskites. They are briefly discussed below.

*Imbalanced charge injection and transport:* The mobility of holes is lower than that of electrons in perovskites. Furthermore, for most non-inverted PeLEDs, there is no energy barrier between common ETLs and perovskite layers while hole injection is not efficient due to the energy barrier between common HTLs and perovskite layers. Both can lead to excess carriers passing through the device without forming electron-hole pairs, as well as non-uniform carrier distribution in the emissive layer.

*Non-radiative Auger recombination:* The injected carriers recombine through three separate processes – trap-assisted monomolecular, bimolecular and Auger recombination. At high injection current density, Auger recombination which involves three particles dominates. In this process, the energy released through recombination of an electron-hole pair excites a third carrier rather than generating a photon, resulting in luminescence quenching and reduction of EQE.

*Joule heating:* Although the carrier mobility in PeLED is higher than that in OLEDs or QLEDs, it is still significantly lower than that in single-crystalline semiconductors. In perovskites with reduced dimensions such as QDs, 2D or quasi-2D perovskites, the conductance through the material is further impeded due to the embedded insulating ligands. As a result, joule heating is much higher in these solution-processed materials under the same injection current density. Combining this with the already poorer stability under elevated temperatures for solution-processed materials and thermal quenching of PL makes it imperative to find novel ways to manage this challenge.

*Ion migration:* The halogen ions in perovskites have low activation energy and can easily migrate under light illumination, heating or electric fields. This results in generation of defects and distortion of the perovskite crystal structure. Compared to covalent compounds, ion migration may be the most fundamental reason for perovskite's poor stability. Under constant DC electric field, the halide ions can also migrate across the interfaces into the charge transport layers and the electrodes, causing conductivity reduction and cathodic corrosion.

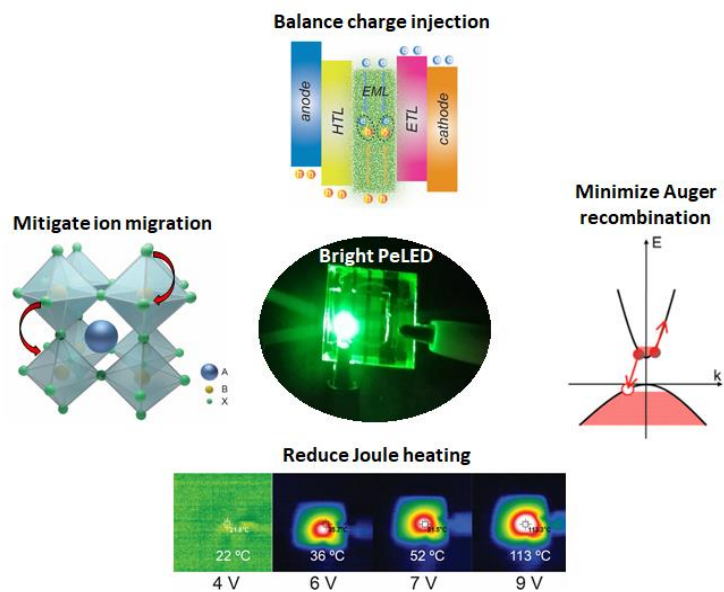

**Figure 17.** Challenges for achieving PeLEDs with high luminance: imbalanced charge injection and transport, non-radiative Auger recombination, Joule heating and ion migration.

**Advances in Science and Technology to Meet Challenges**

Various approaches have been explored to address the challenges listed above; some examples are illustrated in Figure 18. In Ref. [174], the authors investigated the behavior of MAPbI$_3$ PeLEDs under intense electrical excitation, and operated the device under short-pulse drive at current densities up to 620 A cm$^{-2}$. They found that in the high current density regime ($J > 10$ A cm$^{-2}$), the device showed less hysteresis and concluded that effect of ion migration is less significant than that of Joule heating and charge imbalance in causing EQE roll-off. Time-resolved PL measurements showed no evidence of Auger recombination loss. On the other hand, in another work efficiency roll-off in quasi-2D PeLEDs was probed through time-resolved PL measurements under various electric fields and optical excitation intensities.[97] The authors concluded that luminescence quenching was likely caused by Auger recombination, and reported suppressing the effect by increasing quantum well widths. Through this approach, the EQE remained high (~10%) under a current density of 500 mA cm$^{-2}$. The difference in the perovskites used, bulk 3D versus quasi-2D, might result in these two different conclusions. This was partially supported by the work reported in Ref. [75], where the authors observed bulk 3D perovskites exhibits more than 20 times lower Auger recombination rate compared to that of quasi-2D perovskites, leading to higher EL and EQE at high $J$ (>1 A cm$^{-2}$) from bulk 3D PeLEDs. In this work, different combinations of HTLs were also experimented to achieve an energy ladder for holes. Another approach to balancing charge injection is optimizing the thickness of charge transport layers and doping the ETL [175]. In both works, the devices with balanced charge injection showed significantly less EQE roll-off. In the direction of mitigating Joule heating, a strategy of applying a small current injection aperture was shown to effectively diffuse the heat to the surrounding region. The devices were also driven by pulsed current (2 µs pulse width and 0.2% duty cycle) to further reduce Joule heating and ion migration.[75] With 100 nm-diameter apertures, the PeLED could be operated up to ~1 kA cm$^{-2}$ without failure, and a luminance of 7.6 Mcd m$^{-2}$ was achieved at $J$ ~ 56 A cm$^{-2}$. By attaching heat spreader and heat sink as well as defining a narrow line-shape device geometry, a maximum operating current density of 2.5 kA cm$^{-2}$ was obtained under pulsed driving.[76] Through optimizing the device structure for high-speed operation, a radiance ~480 kW sr$^{-1}$ m$^{-2}$ was achieved at 8.3 kA cm$^{-2}$ using electrical pulses with 1.2 ns rise time.[176]

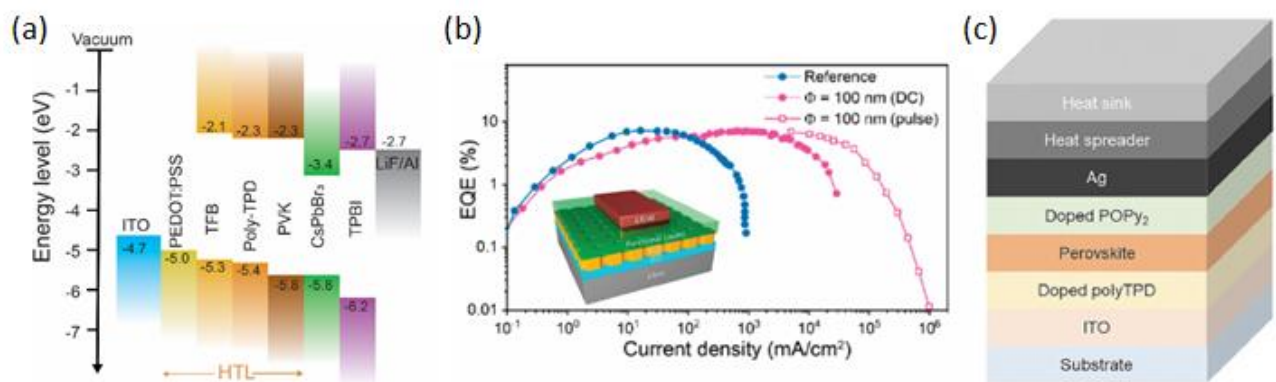

**Figure 18.** Example strategies for improving PeLED luminance. (a) Design HTLs to facilitate hole injection. (b) Incorporate current injection apertures and pulsed driving. (c) Employ heat spreader and heat sink in the device structure. (a,b) Reprinted with permission from Ref. [75], Copyright 2020 American Chemical Society. (c) Reprinted with permission from Ref. [76], Copyright 2018 John Wiley and Sons.

**Concluding Remarks**

In less than a decade, PeLEDs have made impressive progress in their efficiencies and luminance. This has been enabled by many research efforts aiming at improving the perovskite materials and device structures, as well as exploring various operating conditions. The demonstrated results show the promising prospect of perovskite light emitters for applications in lighting, optical displays and communications. In this section, we focus on the issue of achieving high luminance and briefly discuss the key challenges. We then summarize example scientific and technical advancements in this direction. There are many other outstanding works paving the progression path of PeLEDs that cannot be included in this article due to limited space.

To realize broad adoption of perovskite light-emitting technologies, long-term stability of materials and devices needs to be demonstrated. Techniques used in achieving high brightness will be beneficial to operational lifetime improvement of PeLEDs. The research community has been making steady strides in this direction, and the progress is not included in this section.

**Acknowledgements**

We gratefully acknowledge the support from the National Science Foundation (grant ECCS-1807397).

## 11. Perovskite White-light Sources

*Hengyang Xiang* and *Haibo Zeng*

Nanjing University of Science and Technology, China

**Status**

White-light sources, as a common basis for lighting, display, and other fields, are crucial to the development of human beings. Nowadays, the most mainstream light source is white LEDs (WLEDs), which are at a key stage of innovation to meet new demands, such as low cost, transparency, and flexibility/wearability.[177,178] Therefore, some light-emitting materials and devices with these above potentials have received extensive attention.[179] Perovskite is considered as a standout candidate owing to its low-cost solution-processable property and tunable spectral in whole visible light. The technical route of WPeLEDs mainly comes from previous white OLEDs and QLEDs. Their light-emitting mechanism mainly includes the down-conversion type (Figure 19a) and full EL type (Figure 19b), and the device structures mainly include red/green/blue (RGB) mixed single layer and RGB stacked multilayers. A timeline of the WPeLEDs' development in recent years is listed in Figure 19c.

In terms of down-conversion type, perovskites can be used as low-cost phosphors because of their tunable emission characteristic and high PLQY, and even blue PeLEDs have been developed as the excitation source to achieve WPeLEDs.[128] In terms of full EL type, a common strategy is to stack the R/G/B perovskites,[180,181] or mix them in a single-layer film.[182]

Besides the above R/G/B perovskite emitters, some perovskites with broad-spectrum luminescence properties, become a candidate for realizing white-light sources, such as self-trapped excitons (STEs) induced perovskites and their derivatives.[183] The luminescence based on STEs has the characteristics of large Stokes shift and broad spectral emission, therefore showing the potential as white-light sources.[184] Perovskites have soft lattice characteristics, and the electron-hole pairs generated after excitation can easily cause lattice distortion and be trapped by the lattice to form STEs. In recent years, some WPeLEDs have been successfully implemented in double perovskites (e.g., $Cs_2AgInCl_6$),[167] specially structured $ABX_3$ perovskites (e.g., $\delta$-$CsPbI_3$)[185] and perovskite derivatives (e.g., $CsCu_2I_3$, $Cs_3Cu_2I_5$).[186]

Element doping is another pathway for perovskite's broad-spectrum luminescence. Certain elements are often able to form new energy levels in the perovskite structure, such as Mn, Sm. In these ion-doped perovskites, the doping element can provide extra light-emitting centers, which can combine the intrinsic luminescence of the perovskite host to realize WPeLED.[187]

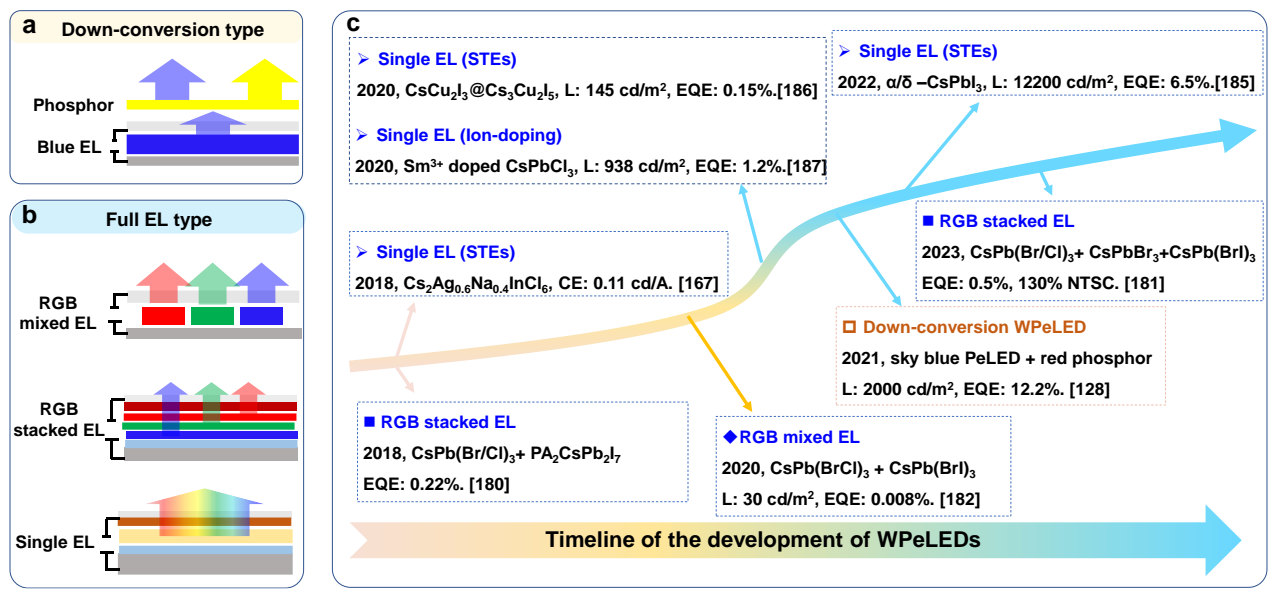

**Figure 19.** Research strategies and progress of WPeLEDs. Device structures for WPeLEDs based on the light-emitting mechanism of down-conversion type (a), full EL type (b), and (c) the timeline of the WPeLEDs' development in recent years.

**Current and future challenges**

As we list in the timeline (Figure 19c), although a variety of WPeLEDs have been developed, they are still facing challenges in device efficiency and operation stability, which are far away from the goals of the commercialization demands of lighting and display. The key point is the optoelectronic properties of perovskite materials, especially typical $ABX_3$ perovskites. Although they perform well in monochromatic PeLEDs,[32] their ionic crystal characteristics and mixed halogen characteristics hinder their realization of WPeLEDs.

On the one hand, solvent sensitivity issues and ion migration problems are obstacles when realizing the RGB mixed or stacked WPeLEDs. In the single-layer multi-color mixing structure, the direct mixing of multi-color perovskites is prone to phase separation, resulting in the failure to achieve the desired emission color. For vertically stacked devices, the stacking of multilayer films faces major technical difficulties on the solvent compatibility, including the erosion of the solvent on the lower layers, and even the decomposition of perovskites.

On the other hand, stability is an issue, including spectral stability and operational lifetime. In the ionic crystal, the ionic bond energy is weak, and the ion vibration and even migration will inevitably occur under the electric field. These ions' behaviors, especially in mixed halide perovskites, will lead to serious spectral shift. Moreover, these essential flaws limit the operational lifetime of PeLEDs, which is only a few hundred hours in current state.[159] Therefore, improving the ionic bond energy of perovskite crystals is an urgent task.

For STE typed perovskites, their poor electrical properties, including mismatched energy levels and carrier injection capabilities, are another issue for device performance.

**Advances in science and technology to meet challenges**

In order to solve the above issues, some strategies are being developed in perovskite materials and devices, especially for enhancing ionic bond energy. For example, using hydrogen bonds to anchor halogens and enhancing metal-halogen bonds to form more stable octahedral structures. In 2021, a hydrogen-bonded strategy was demonstrated in blue PeLEDs by amine-group (e.g., GA and FA) doping. Color-stable sky-blue PeLEDs were realized without shift under different biases.[188] In 2022, Chen et al. proposed a transition metal (e.g., Ni and Mn) doping strategy for $ABX_3$ perovskite to enhance the energy of metal-halogen bonds. Even with small doping levels (<4%), the energy barrier for ion migration can be increased fourfold by Mn and Ni substitution.[189] These strategies/technologies suppress the spectral drift and improve the operational lifetime, and will play an important role in WPeLEDs. Besides, surface-interface engineering is another feasible strategy for preventing halogen migration and perovskite decomposition. In 2021, Kuang et al. introduced a stable amide between the perovskite and the transport layer. It blocked the path of perovskite ion migration and chemical reaction with the transport layer, resulting in an efficient PeLED with a long operational lifetime of 682 h.[159]

To further prevent the contact between perovskites, some porous materials with water, heat and light stability are proposed to seal the perovskite into the pores, thereby improving the stability of the perovskites, such as perovskite metal-organic framework glasses,[190] $CsPbBr_3$:Sr/PbBr(OH) /molecular sieve composites,[191] $CsPbBr_3$ QDs in LTA zeolite,[192] and methacrylate-crosslinked perovskite NPs,[193] These ultrastable and highly efficient perovskites can be phosphors to meet the needs of down-conversion WPeLEDs and backlight displays.

For the STE-typed perovskites, the improvement of the electrical properties of STE materials is the key to improving the white-light EL stability. In addition to the phase transition synergy strategy proposed in α/δ-$CsPbI_3$, surface-interface engineering can improve carrier injection capability, enabling more carriers to reach the STE-typed EML, thereby improving the light-emitting performance. In 2021, Chen et al. introduced an additive (Tween) into the $CsCu_2I_3$/$Cs_3Cu_2I_5$, its ether bond has a strong interaction with $Cs^+$ in the system.[118] On the one hand, it can delay the crystallization process of copper-based halide and reduce defects; On the other hand, it can increase the surface potential of the film and improve carrier migration. It is beneficial to improving the injection and transport of carriers in the device. Based on this film, the prepared warm-white WPeLED achieved an EQE of 3.1% and a brightness of ~1600 cd m$^{-2}$ at 5.4 V.

**Concluding remarks**

Perovskites have outstanding optoelectronic properties and show great potential in realizing white-light sources. However, the ionic crystal characteristics of perovskites themselves lead to problems such as ion migration and phase separation, which limit the spectral stability and operation lifetime of the device. Therefore, the improvement of the performance of WPeLEDs is the main challenge at present. Some strategies, element doping, surface interface engineering, etc., had been proposed and achieved significant improvement. We can expect that the luminous efficiency and stability of WPeLEDs will be promoted quickly and gradually meet commercialization requirements.

**Acknowledgments**

This work was financially supported by the National Natural Science Foundation of China (52131304, 61725402, 62004101), the Fundamental Research Funds for the Central Universities (30920041117).

## 12. Perovskite-based hybrid light-emitting diodes

*Denghui Liu* and *Shi-Jian Su*

South China University of Technology, China.

**Status**

Recently, metal halide perovskites have attracted considerable research interest due to their fascinating optoelectronic properties in both lighting and display applications. Benefitting from the multi-function nature of perovskites and corresponding hybrid counterparts, attempts have been made to create a series of perovskite-based hybrid LEDs (Figure 20). For instance, small organic molecules or polymers can be introduced as functional additives to enhance film formation and radiative recombination of PeLEDs.[135,194] Also, perovskites can be hosted by the organics with high triplet energy and appropriate energy level to effectively localize charge carriers within emitting layer for enhanced light emission.[91] In addition to the perovskite emitters, other perovskite materials ($MAPbCl_3$) with high carrier mobility can potentially be served as HTL for enhanced hole injection and transportation of OLEDs.[195] Besides, WLED is very likely to be the development direction of perovskite technology since it is of importance for both lighting and display applications. Alternatively, perovskite-based hybrid LEDs have unique advantages in achieving tunable multi-color LEDs[196] and WLEDs due to the extensive selection of emitters and no need to consider the halogen exchange reaction. Thanks to the early exploration[197,198] and the accumulation of mature GaN LEDs, OLEDs and QLEDs technology, a few encouraging progresses have been made. Li et al.[196] successfully fabricated a Pe-GaN tandem LED with a tunable luminance and color via vertical integrating a bright and stable blue GaN LED and a green $MAPbBr_3$ PeLED. Wu et al.[199] used a delicate configuration to harvest the triplets of quasi-2D blue emission perovskite, a maximum forward-viewing EQE of 8.6% was achieved, exhibiting a considerable enhancement versus corresponding single-color sky-blue PeLED (4.6%). In addition, Lee et al. introduced a quantum-well-like charge-confinement structure for enhancing exciton formation in phosphorescent emitter via carrier trapping, delivering a recorded EQE of 10.81% via broadening exciton recombination zone.[200] Notably, the device efficiency and operational stability of these perovskite-based WLEDs were largely limited by the blue perovskite. Interestingly, Su et al. demonstrated a kind of perovskite-based hybrid WLEDs by layer-by-layer superimposing sky-blue organic p-i-n heterojunction unit on pure-red perovskite layer to manage the exciton generation region in the device.[201] As a result, a peak EQE of 7.35% was realized for the developed perovskite/organic hybrid WLEDs. More importantly, these devices exhibit stable EL spectra under various driving voltages and a greatly extended operating lifespan, showing a significant potential in high-performance perovskite-based WLEDs.

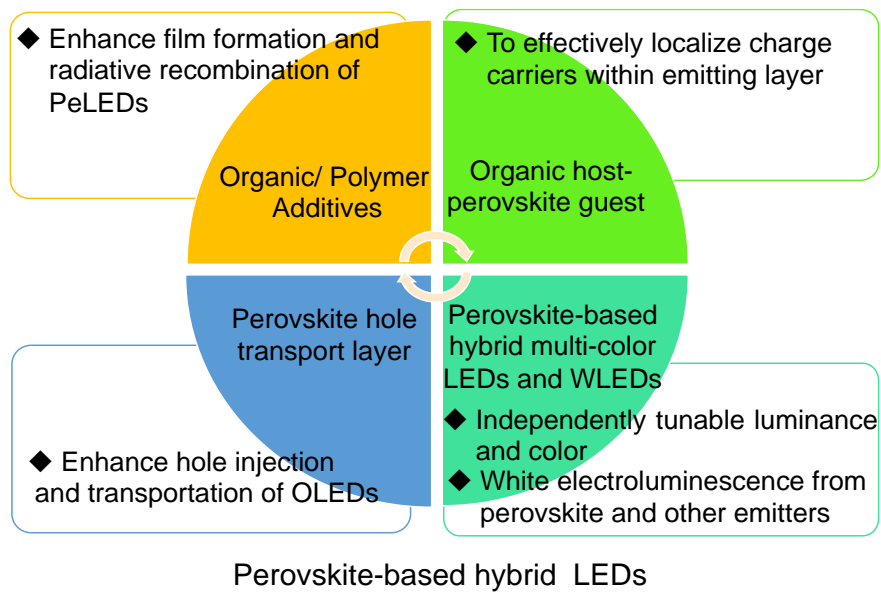

**Figure 20.** Typical perovskite-based hybrid LEDs.

**Current and Future Challenges**

Thus far, the research on perovskite-based hybrid LEDs has obtained preliminary results, while these explorations demonstrated a huge potential and feasibility of perovskite material system as low-cost, high-efficiency display and lighting source, it also reveals some issues and challenges on the road of commercialization. Despite high-efficiency PeLEDs can be achieved through incorporating organic or polymer additives, but the inner mechanisms and principles still remain to be further thoroughly understood, new technologies and intuitive evidences are urgently demanded to clarify ambiguous problems. In addition, organic host-perovskite guest systems show great potential in improving film quality and device performance. Nevertheless, as limited by their incompatible solubility, this processing route is commonly completed with vacuum evaporation technology. Moreover, perovskite materials with high carrier mobility and matched energy levels are used as HTLs of OLEDs, exhibiting comparable device performance with the ones using traditional organic HTLs. However, their operational stability still remains a great challenge for the next commercial application. As for the perovskite-based hybrid multi-color LEDs and WLEDs, the reported Pe-GaN tandem LEDs vertically integrated by GaN LED and PeLED usually rely on complex and expensive epitaxy growth techniques. The integration of perovskite and organic/inorganic QD emitters with complementary spectrum within an electroluminescent device has been confirmed to be a feasible approach, while the related processing compatibility, spectrum suitability, refractive index match and energy transfer regulation still need to be comprehensively considered. Generally speaking, single-EML perovskite-based hybrid WLEDs might be a more attractive strategy for its simplified device structure and processing route. However, the co-mixed system requires similar solubility, and currently, only traditional fluorescent materials with theoretical maximum internal quantum efficiency of 25% were adopted to achieve white emission. Also, more defects and increased unwanted energy transfer process might be introduced into the complex blended system, which made the high-performance single-EML perovskite-based hybrid WLEDs challenging. EQEs can be significantly improved by incorporating phosphorescent emitters with 100% exciton utilization efficiency in a multi-EML structure to fully harvest the excitons inside of the device. However, a sophisticated architecture and a complicated process are required for the device fabrication. Meanwhile, the total internal reflection induced by the large difference in the refractive index between the perovskite layer and other emitting layers, together with parasitic absorption of perovskite layer with high extinction, leading to the loss of generated photons, and thus decreased EQE, low luminance and poor device stability.

**Advances in Science and Technology to Meet Challenges**

To address the challenges to achieve low-cost, high-efficiency, stable perovskite-based hybrid LEDs is inseparable with the advances in corresponding science and technology. Here, possible solutions are given to overcome the aforementioned challenges involved in the perovskite-based hybrid LEDs, including processing compatibility, refractive index match, spectrum suitability and energy transfer regulation (Figure 21). Firstly, film surface modification and mature vacuum evaporation technologies would be helpful to deal with the processing compatibility issues of m-EML architecture, enabling the loading of subsequent deposited layer in wet and dry processes. In addition, novel emitters with appropriate solubility enable harvesting triplet excitons can be designed and synthesized to maximize the exciton utilization efficiency within the single-EML configuration. Secondly, the advances in full-color oriented perovskite nanoplatelets and vacuum evaporated organic host-perovskite guest systems might contribute to higher photon recycle efficiency without using light extraction technology. For the devices with high horizontal orientation and ultra-thin perovskite films, the photons from the perovskite and other emitting layers are easier to overcome the optical barrier and further outcoupled to the air. For the devices based on organic host-perovskite guest systems, the total internal reflection induced by the large refraction index between hybrid emitting layers leading to the loss of photons can be inhibited by the similar refractive index with the organic functional layer and adjustable thickness. The Förster energy transfer process can be regulated by the spectral overlap between the absorption of emitter and PL of host. Besides, the light emission is tightly connected with the spectral suitability. The complementary PL spectra of the emitters may enable white emission, and sufficient and insufficient energy transfer processes lead to single-color and white light emission, respectively. In view of the intrinsic characteristics of the adopted materials system, high-performance perovskite-based hybrid LEDs can be expected via reasonable management of the energy transfer process and spectral suitability to maximize the exciton utilization.

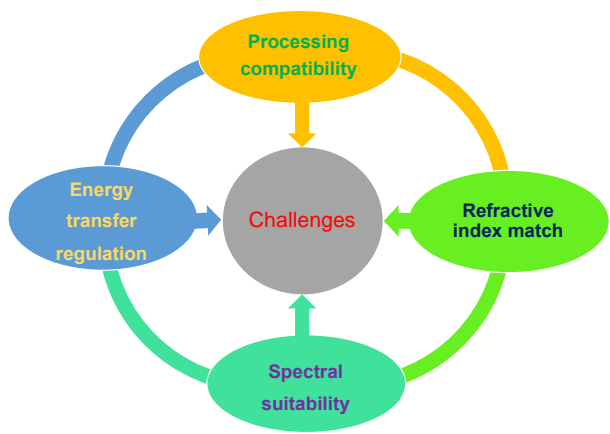

**Figure 21.** Advances in science and technology to meet challenges.

**Concluding Remarks**

In a brief summary, due to the advantageous photoelectronic properties of metal halide perovskite, the perovskite-based hybrid LEDs have received an encouraging progress in just few years, a bright prospect in the next-generation lighting and display equipment can be expected. In spite of this, many challenges and issues still need to be resolved, and tremendous efforts are needed to meet challenges in the future.

**Acknowledgements**

The authors greatly appreciate financial support from the National Natural Science Foundation of China (51625301, 91733302, and 51861145301), the Basic and Applied Basic Research Foundation of Guangdong Province (2019B1515120023), the Guangdong Provincial Department of Science and Technology (2016B090906003 and 2016TX03C175), and the Dongguan Innovative Research Team Program (2018607201002).

## 13. Using Perovskites as Fluorescent Powder and Films

*Chenhui Wang* and *Haizheng Zhong*

Beijing Institute of Technology, China.

**Status**

Perovskites QDs show great potential as alternative display materials to conventional QDs due to their intrinsic ionic characteristics with ultra-low formation energy and high defect tolerance. As a result, Perovskite QDs can be fabricated at room temperature through either LARP synthesis[202] or in-situ synthesis in polymeric[203] or glasses matrix,[204] which enables the fabrication of solution-processed high-fluorescent Perovskite QDs embedded composites for low-cost display applications.[205]

*Perovskite QD enhancement film (PeQDEF)*. QDEF has been widely used in liquid crystal displays (LCDs) for the enhancement of color performance and perceived brightness. In 2016, Zhong's group demonstrated the in-situ fabrication strategy of perovskite QDs by controlling the crystallization process of precursor during the polymer film formation.[203] This innovative strategy solved the problem of agglomeration when using traditional CdSe or InP QDs. The high transparency and PLQY of perovskite/polymer nanocomposites suggest the formation of uniformly distributed nanoscale Perovskite QDs without phase separation, which avoid several issues such as severe scattering loss in QDEF. The hybrid composite films with green $MAPbBr_3$ QDs and red rare earth phosphors obtained a wide color gamut with 83.1% of Rec. 2020. With the success of red emissive perovskite QDs, the integration of red and green dual-emissive film with blue Mini-LEDs achieves a high color quality LCD with a color space of 130% National Television System Committee (NTSC) 1931 and matching rate of 100%.[206]

Recently, in-situ fabrication strategy has been successfully extended to other fabrication technology including tape casting (Figure 22a), melt extrusion (Figure 22b), and spray drying (Figure 22c), which are more suitable for commercialized scale up fabrication. The as-productid PeQDEFs show enhanced color gamut, and meet the requirement of applications with a stable brightness and color coordinates during the 3,000 hours acceleration test. In 2020 and 2021, two batches of mass production with 500 TVs and 1,000 TVs have been accomplished by TCL and Zhijing NanoTech. Up to now (January 2023), the perovskite QDs integrated TV products approach up to 50,000.

*Perovskite QD color convertor (PeQDCC)*. The miniaturization of conventional LEDs is also a growing trend for emerging virtual reality, augmented reality, and ultrahigh-resolution display applications. PeQDCC can be arranged into microscale pixels using various patterning techniques and show great potential for the development of advanced micro-OLEDs or Micro-LEDs. Compared with CdSe or InP QDs, in-situ patterned perovskite QDs is one of the most practical ways to meet the requirement of high optical density and high resolution

Over the past few years, various patterning techniques have been developed.[207] Benefiting from the unique ionic characteristics and low formation energy of perovskites, perovskite precursor solutions can be in-situ printed onto a polymeric layer, and perovskite QDs patterns can be obtained with high PLQY up to 80% and they show broad applicability to a variety of perovskites and polymers.[208] Direct laser writing is also an efficient and simple method for patterning perovskite QDs during the formation process.[209] Photolithography is another patterning technique with high resolution, wide availability, and high throughput. Recently, a direct photolithography method to pattern in situ fabricated perovskite QDs based on polymerization catalyzed by lead bromide complexes was reported. Perovskite QD patterns show high resolution up to 2450 pixels per inch (PPI) as well as excellent fluorescence uniformity and good stability.[210]

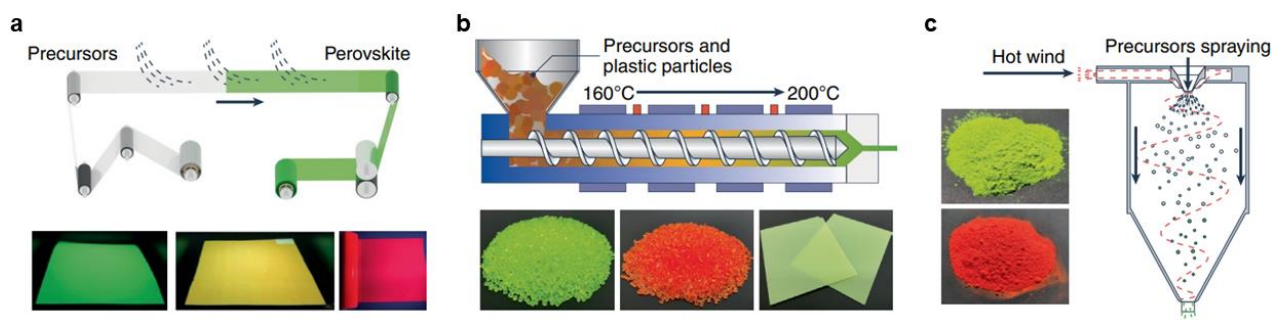

**Figure 22.** In-situ fabrication technology. (a) Tape casting process. (b) Melt extrusion process. (c) Spray drying process.

**Current and Future Challenges**

*Stability of PeQDEF*. In PeQDEF-based LCD, by appropriate encapsulation techniques, the reliability of perovskite QDs-based materials has already met the industrial requirements. Even so, the stabilities of perovskite QDs are still key issues to make further development for perovskite QDs applications. although the degradation mechanism of perovskite under $H_2O$, $O_2$, or illumination have been investigated, it is still a great challenge to improve the photostability against the high flux blue light intensity for micro-LED application.

*Patterning techniques for EL.* Although various patterning techniques have been applied for pattered perovskite QDs fabrication,[207] there are still remain great challenges. Perovskite precursors are compatible with inkjet printing, there are quality issues related to the coffee ring effect that still need to be solved. In addition, the introduction of sticky polymers for controlling the viscosity of droplets may be harmful to the radiative recombination of carrier injection in LEDs. Reduction of the general size of inkjet-printed units (<5 μm) is also essential to satisfy the resolution requirement of micro-miniaturization. Laser writing technique provide outstanding patterns; however, the quality of the units is limited due to the laser-induced damage. Photolithography patterning of perovskite QDs usually experiences several steps including exposure and development, which last long times and induced damage to perovskite QDs. The nanoimprinting fabrication of perovskite QDs require a hard or soft template with preformed structures. These issues block the development of PeQDCC patterning techniques.

*Environmental friendliness*. It is noted that the final products of PeQDEF contains a very low lead content of 30−200 ppm (400 ppm in natural agricultural soil for comparison). The industrial fabrication of perovskite QDs based products are more environmentally friendly. However, the development of lead-free perovskite emitters is still of great significance. Although a few promising lead-free narrow-band perovskite emitters have been reported (see Chapter 9), the efficiencies and stabilities are still lag far behind Pb analogues.

**Advances in Science and Technology to Meet Challenges**

Active-matrix PeLEDs are attractive advanced technology for display applications. Patterning perovskites into microscale pixels is the most important step for the fabrication of micro-PeLED devices. Despite the rapid development of perovskite patterning techniques, micro-PeLED devices are rarely reported, with a record EQE of only 6.8%.[210] Although high-efficiency in-situ fabricated PeLEDs have been widely demonstrated, it remains a great challenge to develop a fabrication process for achieving high efficiency, high resolution, and improved stability. In addition, there are some other key technical issues such as the inhomogeneous unit morphologies, poor charge transport, and undesirable recombination. For industrial application, there are also technical difficulties to integrate the PeLEDs with driven circuits.

**Concluding Remarks**

In conclusion, perovskite QDs with unique optical and electrical properties show great potential for display applications. The in-situ fabrication strategy is a milestone progress in the development of perovskite QDs toward display applications, which extends the commercialized scale up fabrication techniques and promotes the development of simple patterning technology. In-situ fabricated perovskite QDs have been successfully applied in LCD backlight, and also achieved significant progress as a color convertor. In the future, we need to pay more attentions to solve the problems of stability issues and system integrations. The breakthroughs on these issues will not only promote the commercialization of perovskite QDs in display technology, but also generate new application directions in other photonic technology.

**Acknowledgements**

This work was by supported by the National Key Research and Development Program for Young Scientists (2021YFB3601700) and Beijing Natural Science Foundation (Z210018).

## 14. Display based on Perovskite Light-Emitting Diodes

*Tong-Tong Xuan* and *Rong-Jun Xie*

Fujian Key Laboratory of Surface and Interface Engineering for High Performance Materials, College of Materials, Xiamen University, Xiamen 361005, China

**Status**

The rapid development of fifth-generation mobile communication, artificial intelligence, and the internet of things has opened up a new era of information technology. In this process of evolution, the display content has changed from simple text and pictures to ultrahigh resolution (UHD) photographs and videos. Therefore, displays should have outstanding performances, such as super high resolution, ultrawide color gamut, fast response, high contrast and brightness, and longevity. Unfortunately, commercial LCDs have poor color saturation, while OLED displays are limited by their short lifetime and high cost. Thus, to develop low-cost UHD displays has become a research focus.[211]

Metal halide perovskites exhibit excellent optoelectronic properties, such as near-unity PLQYs, narrow-band emission, tunable bandgap, high defect tolerance and high charge carrier mobility, which are superior to conventional semiconductor QDs and organic emitters.[40] Therefore, perovskites have been employed as an EML sandwiched between the HTL and the ETL to fabricate PeLEDs.[55] Although EL was seen at the liquid nitrogen temperature as early as 1994, Tan et al. firstly demonstrated infrared and visible emitting PeLEDs at room temperature in 2014. With extensive investigations on perovskites and solution-processed LEDs, the EQE of state-of-the-art PeLEDs has reached over 23% for pure green (525−535 nm) and pure red (620−640 nm) emission, respectively (Figure 23a). In addition, their narrow full width at half maximum (FWHM < 38 nm) and high color purity lead to an ultrawide color gamut of 140% of the NTSC. Therefore, PeLEDs (Figure 23b) can be used to achieve UHD displays with the broadcasting services (television) (Rec. 2020) standard announced by the International Telecommunication Union Radiocommunication sector intervention on complementary metal oxide semiconductor or thin film transistor backplanes.

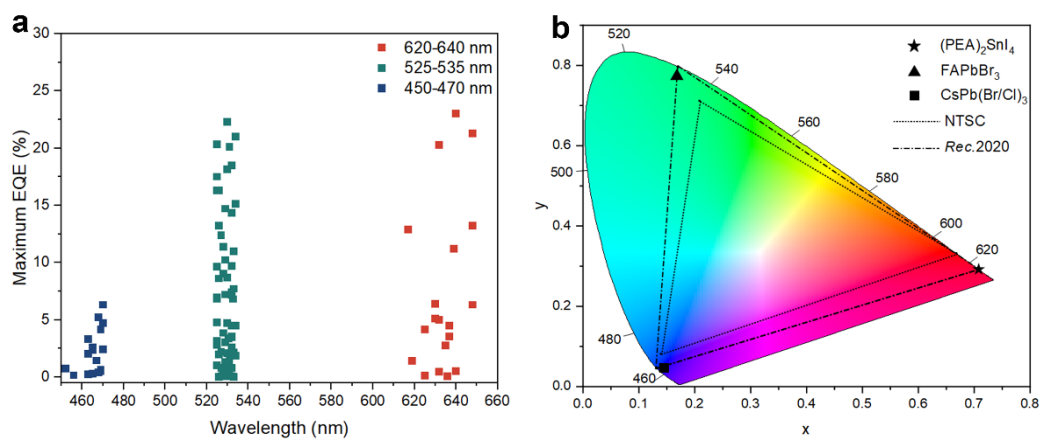

**Figure 23.** (a) Summary of the maximum EQEs of pure RGB PeLEDs. (b) Chromaticity coordinates of (PEA)$_2$SnI$_4$, FAPbBr$_3$, and CsPb(Br/Cl)$_3$ perovskite based pure RGB PeLEDs in the CIE 1931 color space, together with the NTSC, and Rec. 2020 standards.

**Current and Future Challenges**

Although PeLEDs have achieved considerable progresses in quantum efficiency for pure green and red PeLEDs, they still suffer from many fundamental limitations, which are summarized as follows:

(1) Color instability

Pure red and blue emission are typically realized by mixing halide ions or reducing the dimensionality of perovskites, which shows poor emissive spectral stability due to the electrical-field-induced phase segregation and sensitive surfaces leading to morphological changes.

(2) Poor EQE of pure blue PeLEDs

The EQEs (Figure 23a) of pure red and pure green PeLEDs have increased to more than 23%. However, the pure blue PeLEDs with an emission maximum of 450−470 nm still have a low EQE of less than 10%, which is far below the requirements of commercial displays (EQE > 20%). This is mainly attributed to low PLQYs of wide band gap perovskites, difficulty in charge injection, resonant energy transfer, and optical down conversion effects.[212]

(3) Efficiency roll-off

Similar to OLEDs and QLEDs, the EQE tends to decline associated with increase in brightness or current density of pure RGB PeLEDs, which is termed efficiency roll-off. The efficiency roll-off is originated from the perovskite emission quenching due to Joule heating, increased Auger recombination, or unbalanced charge injection.

(4) Short operation lifetime

A long operation lifetime is one of the prerequisites for commercial display applications of PeLEDs. Currently, state-of-the-art RGB PeLEDs have a $T_{50}$ of 112, 11, and 0.4 h at an initial luminance of 1000 cd m$^{-2}$, which is far lower than that of OLEDs and QLEDs (approximately $10^4$–$10^5$ h).[28]

(5) Toxicity of lead ions

Lead ions are not environmentally friendly and toxic. According to the 2022 European Union Restriction of Hazardous Substances rules, the maximum level of lead permitted in an electronic device is 1000 ppm. Therefore, there is an urgent need to develop lead-free perovskites for PeLED displays.

In addition, the spin-coating method has limitations in the fabrication of large-scale PeLED arrays for large-area UHD displays. Thus, it is necessary to make new breakthroughs for PeLED device fabrications.

**Advances in Science and Technology to Meet Challenges**

Perovskite EML is crucial for the performance of RGB PeLEDs. However, perovskites show poor stability against water, oxygen, light, heat and electric fields. Therefore, surface engineering, doping engineering, and/or construction of core/shell NCs or QDs in halide matrices have been applied to enhance luminescence properties, uniformity, compactness and stability of the perovskite EML.[213] Improvements in the perovskite EML can effectively reduce nonradiative recombination, inhibit ion migration and reduce charge leakage, thus enhancing the efficiency, brightness and stability of RGB PeLEDs. However, the efficiency of blue PeLEDs and lead-free perovskite-based PeLEDs still remains significant challenges. The composition of perovskites directly affects their luminescence properties and stability. Therefore, it is suggested to explore highly efficient and stable perovskites through first-principles calculations and machine learning to accelerate the discovery of high-performance PeLEDs.

Furthermore, unbalanced carrier injection in PeLEDs leads to carrier accumulation at the interface between the perovskite EML and HTL/ETL, which leads to Auger recombination, Joule heating and efficiency roll-off.[8] To address this issue, interface engineering between HTL, EML and ETL, and energy level engineering of the transport layer have been utilized to match the hole and electron carrier mobilities, enabling the balanced charge injection in PeLEDs.[214] Simultaneously, device operation for a prolonged period results in heat accumulation, which reduces the EL efficiency due to the separation of excitons. Therefore, increasing the rate of heat dissipation is a new avenue to enhance the EL performance and stability. Replacing common glass with high thermal conductivity substrates (such as silicon and sapphire) or adding high thermal conductivity nanomaterials (such as graphite sheets and polycrystalline diamond) into PeLEDs have been shown to improve heat dissipation. Thermal management not only improves the operation lifetimes of PeLEDs, but also reduces their efficiency roll-off, therefore resulting in higher brightness.

Moreover, to overcome the limitations of spin-plating processes, the vacuum thermal evaporation or inkjet printing has been proposed to fabricate large-area or mini/microscale PeLEDs for large-area UHD displays.[84] For example, Tang et al. fabricated a $CsPbBr_3$-based PeLED with a functional area of 40.2 cm$^2$ and a maximum EQE of 7.1%, and then realized high-resolution patterned perovskite films with 100 μm-size pixels.[90] Moreover, advanced micro/nano fabrication technologies, such as photoetching and nanoimprinting, can be used to prepare micro/nano scale PeLEDs to obtain super high PPI densities for display panels (>25000 PPI).

All of these strategies are summarized in Figure 24.

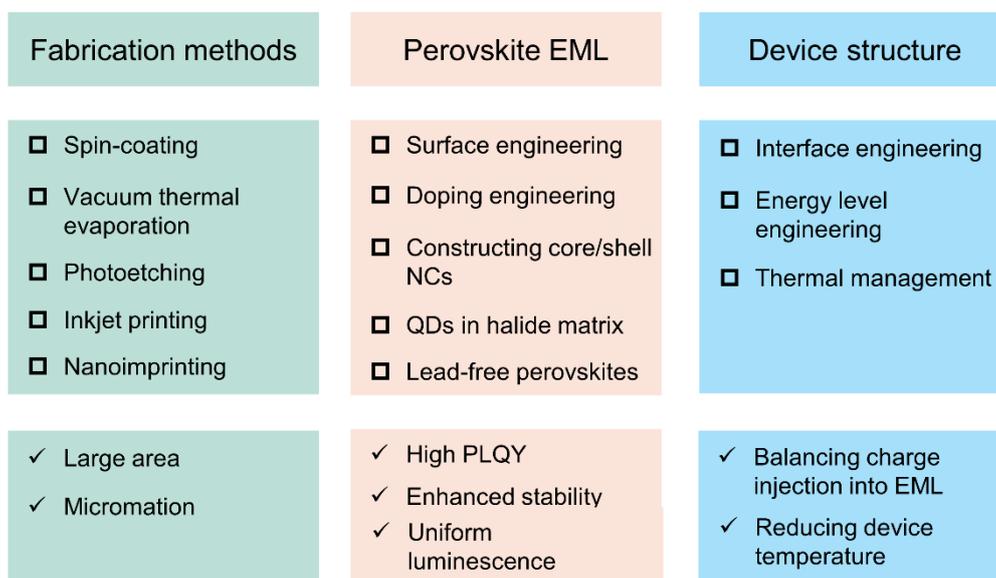

**Figure 24.** Summaries of strategies towards highly efficient and brightness RGB PeLEDs with ultrawide color gamut and long operation lifetime.

**Concluding Remarks**

RGB PeLED-based displays show higher color purity and wider color gamut than LCDs, and have long lifetime than OLEDs, which are recognized as one of the most advantageous competitors for future UHD displays. Within a few years, the EQE of RG PeLEDs has increased from 0.1% to above 23%. This remarkable improvement in efficiency benefits from an optimized fabrication process and in-depth understanding of the photoelectronic properties of perovskite materials and PeLED devices. However, many challenges must be overcome for commercial displays: (1) color instability; (2) low efficiency of pure blue PeLEDs; (3) efficiency roll-off; (4) short operational lifetime; and (5) toxicity of lead ions. Thus, much effort has been made in terms of device fabrication, design and preparation of perovskite EMLs, leading to continuous progresses in efficiency, brightness and reliability. In addition, with the introduction of advanced micro/nano processing technologies and chiral perovskite EMLs into PeLEDs, micro/nano-spin-PeLEDs with circularly polarized EL emissions have been developed, which are expected to achieve ultrahigh-resolution 3D displays. We believe that, although the pathway for the development of PeLED displays is bumpy, the future is bright.

**Acknowledgements**

This work was financially supported by the National Natural Science Foundation of China (Nos. U2005212, 51702373), the Natural Science Foundation of Fujian Province of China (No. 2020J01035), Shenzhen Science and Technology Innovation Committee (JCYJ20200109144614514), and the Fundamental Research Funds for the Central Universities (No. 2072020075).

## 15. Optical Communication Based on Perovskite Light-Emitting Diodes

*Chunxiong Bao*[1] and *Feng Gao*[2]

[1] Nanjing University, China.

[2] Linköping University, Sweden.

**Status**

Light has been widely used as a signal carrier for communication and plays an essential role in modern society. Light source for high-capacity optical communication is strongly relied on laser diodes with high modulation rates and emitting light power. LEDs, as another type of important solid light emitting devices that have been widely used in lighting, signal indicating and information display, have also been investigated as light sources for optical communication in recent decades, for their relatively low cost and little damage to human eyes. As emerging light emitters, perovskites have also been investigated as potential light source materials for optical communication, mainly as color converters and electroluminescent emitters.

Perovskite NCs usually have a short fluorescence decay time, which makes them potential color converters in white light source for optical communication.[215,216,217,218,219,220,221] O. M. Bakr et. al. introduced $CsPbBr_3$ NCs as a color converter in a visible light communication (VLC) system (Figure 25a).[215] Due to the short fluorescence decay time (~7.1 ns), the NCs exhibited high −3dB bandwidth of 491 megahertz (MHz), which is much larger than the conventional YAG:Ce phosphor (12.4 MHz), as shown in Figure 25b. Based on a laser diode and the $CsPbBr_3$ NCs color converter, they demonstrated a VLC system using on−off keying modulation scheme which exhibited a data transmission rate of 2 Gbps. Fu et. al. introduced perovskite NCs both as red and yellow color converters for micro-LED blue chip and formed a white light VLC system.[219] The red perovskite NCs showed a recorded −3dB modulation bandwidth of 822 MHz. By combining with the blue micro-LED chip and yellow perovskite NCs, they obtained a white light with CIE coordinates of (0.33, 0.35) and color rendering index of 75.7, and a final system data transmission rate of 1.7 Gbps. Meanwhile, perovskite NCs color converters also exhibit the potential applications in underwater optical communication[220] and mechanically flexible optical communication system.[221]

PeLEDs with perovskites as electroluminescent emitters have also been studied as light sources for communication. 2D Ruddlesden-Popper perovskite ($PEA_2PbBr_4$) nanoplate based deep-blue (center at 408 nm) LEDs were introduced as light sources for light communication.[222] A white light with CIE coordinate of (0.35, 0.38) was obtained by using perovskite green NCs and commercial red phosphor as color converters and a −3dB bandwidth of 20 kHz was obtained based on the white light source. Considering the short fluorescence decay time of the emitter (<1 ns), the slow speed of the light source can be attributed to the resistance-capacitance (RC) time constant due to the large area (9 mm$^2$) of the devices. Another interesting advantage of PeLEDs for communication is the potential of integrating functionalities of transmitter and receiver in one device (Figure 25a).[223,224] We reported a FAPbI$_3$ based PeLED exhibiting high performance as both light emitting and detection device. −3 dB bandwidth of 21.5 MHz and 65 MHz were obtained for 0.1 mm$^2$-area light emitting and detection devices, respectively.[223] Meanwhile, flexible PeLED fibers based on perovskite NCs was also reported simultaneously as transmitter and receiver for communication.[224]

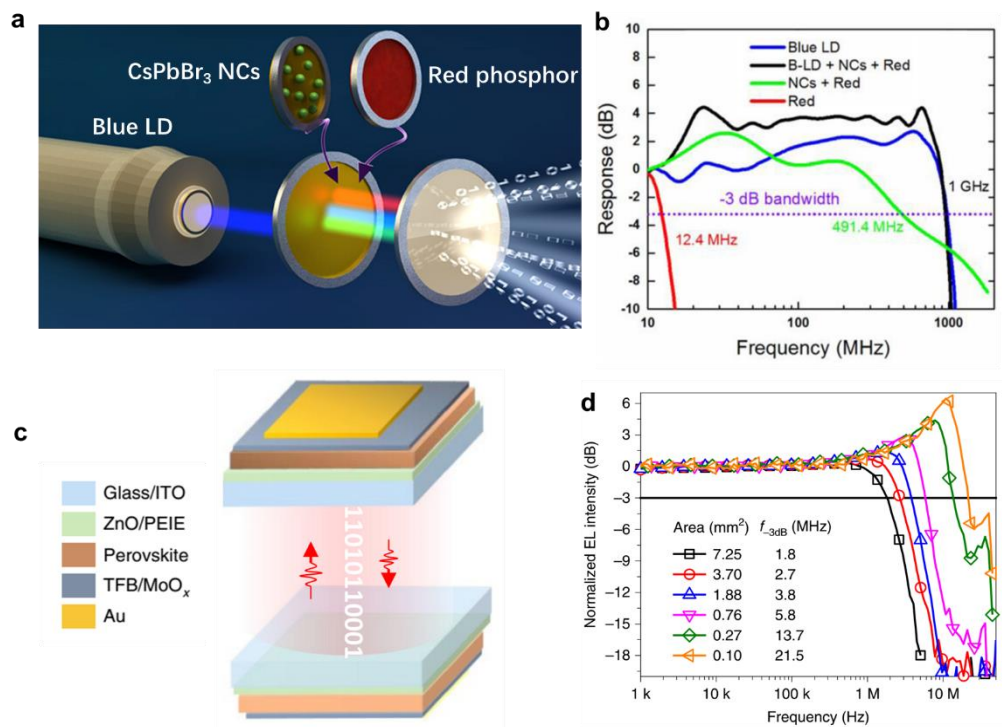

**Figure 25.** (a) Schematic of a VLC system using CsPbBr$_3$ NCs as color converters. (b) Frequency response curves of CsPbBr$_3$ NCs, red phosphor and blue laser diode chip. (c) Schematic of perovskite device used as simultaneously optical signal transmitter and receiver. (d) Frequency response of perovskite device with different device areas when working at light emitting mode. (a) and (b) are adapted form [215]; (c) and (d) are adapted from [223].

**Current and future challenges**

The response speed of LEDs is essential to the data transmission rate of the optical communication system. Perovskite NCs color converters have been reported showing −3dB bandwidth of hundreds of MHz which is much larger than commercial phosphors and comparable with or superior to emerging color converters (e.g., organic fluorophores and colloidal QDs). However, when compared to the excitation sources (laser diode or LED chips), the response speed of perovskite NC color converters is still a bottleneck of the data rate; further improvement of the response speed of perovskite NCs is urgently needed for the communication application. For perovskite electroluminescent devices, the challenge of the response speed is more severe. The systematical study of the response speed of the PeLEDs is rarely reported. The reported −3dB bandwidth of PeLEDs is still far behind that of the inorganic LED chips and OLEDs.

To reach decent signal/noise ratio in the receiver end, a decent emission power is required in the transmitter. As the high response speed can be usually obtained in small-area devices due to the RC time constant, a high light emitting power requires a high efficiency at a high drive current density. However, the obvious roll-off of the efficiency at high drive current density still hinders PeLEDs as high brightness light source for communication.

As typical application scenarios of LEDs in the communication field, VLC and visible light positioning (VLP) require a high-quality white light source, apart from the aforementioned high response speed and power. In the reported communication using perovskite color converters, the white lights were realized by the combination of different color perovskite NCs or phosphors and the excitation light source. The fast response perovskite NC color converters usually show sharp emission spectrum, which makes the composed white light a low color rendering index, limiting their application in VLC and VLP. Meanwhile, up to now, no communication-used white perovskite EL devices have been reported. Therefore, development of perovskite color converters and perovskite EL devices with fast response and broad emission spectrum are still challenging.

The long-term operation stability is also essential to a practical-used communication systems, which is another major challenge for perovskite color converters and EL devices. Up to date, both perovskite color converters and EL devices are still facing long-term stability issues, especially at high optical or electrical excitation levels, which will severely induce phase segregation and efficiency degradation in perovskite color converters and EL devices. Meanwhile, like other perovskite fields, lead toxicity issue is also a challenge for the practical application in communication.

**Advances in science and technology to meet challenges**

Increasing the excitation density can effectively decrease the fluorescence decay time of perovskite NCs. At low excitation regime, the increase of the excitation density can increase the probability for radiative recombination; and at high excitation density, the emergence of non-radiative pathways will obviously decrease the fluorescence decay time.[218] Moreover, based on the experience of inorganic and organic emitters, doping may be a promising strategy for decreasing the PL decay time of perovskite color converters.

For perovskite EL devices, charge transit time, RC time constant, and radiative decay time can be the main response speed limiting factor(s) depending on the properties of each function layer, device structures, and drive current. Usually, at present stage, reducing the RC time constant is an effective strategy to enhance the response speed of PeLEDs. For example, we increased the −3dB bandwidth of $FAPbI_3$ based PeLEDs from 1.8 MHz to 21.5 MHz via reducing the device area from 7.25 $mm^2$ to 0.1 $mm^2$ (Figure 25d).[223] The decrease of device area is a simple and direct way to reduce the capacitance of the devices but will also inevitably increase the internal series resistance. When the internal series resistance is comparable or greater than the external signal reading resistance (e.g., 50 Ω) the reducing device area will not help to reduce RC time constant. A measure to reduce the internal series resistance (e.g., increase the conductivity of the CTL and optimize the interface) should be taken into consideration to further decrease the RC time constant. When the RC time constant was ultimately reduced, the charge transit time and the radiative decay time would turn to the main speed limiting factors. Increasing the mobility of CTL and perovskite, and optimizing the energy level alignment are effective ways to reduce the transit time. The radiative decay time can be decreased using the aforementioned measures for color converters. As perovskite EL devices for communication have just emerged recently, the systematic study of the response speed of the PeLEDs have been rarely reported. The recent breakthrough of the response speed in OLEDs has provided an example: D. W. Samuel et. al. increased the −6dB bandwidth of OLEDs to 245 MHz through combining series of measures, such as reducing devices area, careful design of the CTL and contacts, and doping the emitter.[225]

Increasing the emitted light power of the PeLEDs requires a high efficiency at high drive current, which is usually limited by the efficiency roll-off phenomenon. Auger recombination and Joule heating induced by the high injection have been demonstrated as the main factors for the efficiency roll-off of PeLEDs.[97,226] Accordingly, there have been efforts to manage the Joule heating and realize high power emitting PeLEDs (with radiance of 2555 W $sr^{-1}m^{-2}$),[76] e.g., by increasing the dimension of the emitters to decrease the charge carrier density and thus relieving the Auger recombination and minimizing the roll-off, doping CTLs, optimizing device geometry, and attaching heat spreaders and sinks. The white light quality, long-term operation stability and lead issues are common challenge for PeLEDs, and strategies to meet these challenges have been intensely studied and discussed in other sections.

**Concluding remarks**

Perovskites have attracted attentions as light source materials for optical communication. As color converters, perovskite NCs have achieved a −3dB bandwidth of hundreds of MHz; as EL devices, PeLEDs have just emerged in recent years and showed a −3dB bandwidth of tens of MHz. We emphasize the potential of PeLEDs to realize the integration of light emission and detection in one device, offering a unique approach to inexpensive and integrated optical links. However, as an emerging technique, in order to reach practical applications, perovskite light sources are facing numbers of challenges, such as slow response speed, low emission power, low-quality white emission, instability, and lead toxicity issues. Some effective measures have been developed or borrowed from other emerging light source technologies to meet the challenges. Looking forward, we believe that continued breakthrough will make perovskite light sources promising alternatives for optical communication.

## 16. Electrically Pumped Lasing Based on Perovskite Light-Emitting Diodes

*Xiang Gao* and *Chuanjiang Qin*

Changchun Institute of Applied Chemistry, Chinese Academy of Sciences, China.

**Status**

A semiconductor laser is a kind of device that can emit coherent light with strong intensity and perfect directionality, consisting of gain medium, pump source, and optical cavity. Metal halide perovskites have been proven as excellent gain medium for optically pumped ultralow threshold lasers, following the rapid advances in PeLEDs. Since Xing et al. demonstrated the first amplified spontaneous emission (ASE) in MAPbX$_3$-based perovskite thin films at 150 K in 2014,[227] various types of perovskites have been reported to exhibit ASE and lasing under optical excitation. However, the ultimate goal of electrically pumped perovskite lasers have not been realized yet.

One of the critical steps toward electrically pumped perovskite lasers is to obtain continuous-wave (CW) optically pumped ASE and lasing. MAPbI$_3$-based CW optically pumped distributed-feedback (DFB) laser with threshold of 17 kW cm$^{-2}$ has been achieved from tetragonal-phase guest within larger-bandgap orthorhombic host matrix at T ≈ 100 K.[228] Besides, CW optically pumped ASE from single-phase have also been observed in 3D perovskite films at temperatures up to 120 K (Figure 26a), which potentially supports CW lasing at room temperatures.[229] The milestone occurs when the first room-temperature CW optically pumped lasing are realized in quasi-2D perovskites by combining a DFB cavity with high quality factor and managing the long-lived triplet excitons (Figure 26b).[230] All these perovskite gain mediums work well as emitters in the PeLED.

The integration of well-established PeLED and designed optical resonant cavity is an effective way to realize electrically pumped lasing. Highly luminescent thin films and efficient PeLEDs have been fabricated using quasi-2D perovskites, and state-of-the-art PeLED devices exhibit high EQEs of ~28% at low current densities (few mA cm$^{-2}$).[19] Compared to PeLEDs, considerably higher current densities (hundreds of A cm$^{-2}$) are required to attain population inversion in an electrically pumped perovskite laser diode. The optically pumped lasing from complete and functional LED structures has been achieved, as shown in Figures 26c and 26d.[231] The second-order DFB MAPbI$_3$ PeLEDs can lase under optical pumping with thresholds as low as 6 μJ cm$^{-2}$ at room temperature. These advances in optically pumped lasers and PeLEDs have paved the way toward the development of electrically pumped perovskite laser diodes.

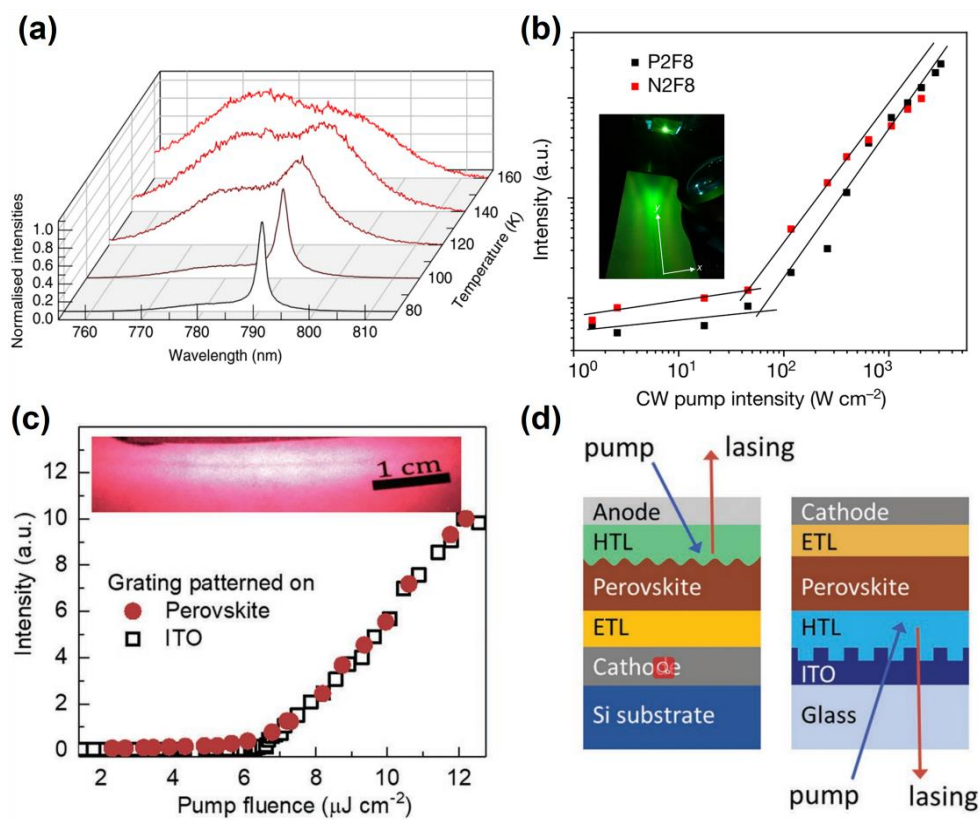

**Figure 26.** (a) ASE spectra under CW excitation at various temperatures and constant pump intensity. The amount of photons emitted into the ASE mode decreases with increasing temperature until ~120 K, after which spontaneous emission dominates. The excitation power was kept constant at 1512 W cm$^{-2}$. This figure is adapted from Ref. [229] (b) Lasing intensity as a function of CW pump intensity at room temperature. The threshold was about 45 W cm$^{-2}$. The inset graphically shows the image of the far-field pattern of CW lasing. The distance between the DFB substrate and the white paper was 10 cm. This figure is adapted from Ref. [230] (c) Integrated emission intensity as a function of excitation pulse fluence recorded for the top- and bottom-emitting devices shown in (d). The Si-based device is excited with 20 ps pulses at $\lambda_{pump}$ = 470 nm, while the glass/ITO-based device is excited with 40 ps pulses at $\lambda_{pump}$ = 532 nm. The inset photograph shows the far-field beam profile from the Si-based device projected on a white card ≈ 3 cm away. (c) and (d) are adapted from Ref. [231].

**Current and Future Challenges**

A well-known issue with organic lasers under electrical excitation is the selection of a high net gain organic material and an extremely low-loss resonator design under the external injection of carriers, and how to achieve laser devices with very large current injection and low excitation threshold. Although perovskites exhibit both high carrier mobility and high fluorescence quantum efficiency, the above problems still need to be overcame to achieve electrically pumped lasers. The demonstration of an electrically pumped laser from a PeLED device by using perovskite as an active layer still remains a colossal challenge for scientists in the field because of the intrinsically degradation of perovskites in oxygen and moisture, efficiency droop, and severe trap-assisted non-radiative recombination.

As mentioned in the first section, an electrically pumped perovskite laser device can be obtained by combining optical resonant cavity with a well-designed PeLED. However, only a few viable diode architectures have demonstrated lasing under optical pumping to date. Joule heating must be a serious problem that needs to be solved in the laser devices, because it can affect the performance of PeLEDs even at relatively low current density.[137] It will cause serious deterioration of the optical gain of the perovskites and increase the threshold current density. For the perovskite laser with a standard sandwich architecture, Joule heating might mainly origin from the poor conductivity of chemically processed charge transport layers and high contact resistance between adjacent layers in the device stack. Furthermore, the existence of injection barriers and the differences in carrier mobilities of different functional layers lead to the imbalanced injection and recombination of electrons and holes. As a result, severe current efficiency roll-off can usually be observed at high current densities.[232] Besides, Auger recombination (a three-particle nonradiative process) was proved as one source of EQE roll-off of PeLEDs. It will become the main competitor with population inversion at high charge carrier densities for electrically pumped perovskite lasing.[97] Therefore, the efficient injection and recombination of carriers are critical issues that need to be addressed for electrically pumped perovskite lasers. In addition, the reabsorption of lasing light and phonons of the nonradiative recombination also contributes to Joule heating.

**Advances in Science and Technology to Meet Challenges**

Perovskite gain mediums with low excitation threshold, photothermal and electric-field stability should be developed to meet the requirement of high current injection devices. Due to the advantages of facile solution fabrication engineering, natural quantum-well effect and high gain coefficient, quasi-2D perovskites are considered as auspicious emitters for PeLEDs with high EQE and optical gain medium for lasing operation.[233, 234]

To integrate an optical resonator into functional PeLEDs, external cavities, such as DFB and distributed Bragg reflector (DBR), are preferred, as the fabrication process of perovskite laser devices would be nonintrusive. In DBR structure, two types of cavity designs can be manufactured. In one case, the photons cycle in the perpendicular direction to the device layer (Figure 27a). The electrode needs to be added into the cavity, which would also result in low reflectance and absorption loss. In another case, photons cycle in the direction along the device layers (Figure 27b), referring to the design of GaN-based electrically pumped lasers.[235] This structure is not affected by the metal electrode absorption. For DFB structure, it is encouraging since an optically pumped laser has been realized in a complete operable PeLED structure using an embedded DFB resonator. The complete device structure is shown in Figures 27c and 27d.

Several thermal management strategies have been proposed to tackle the problem of Joule heating, including the use of highly thermal conductive substrates, heat spreaders and sinks, decreased device geometry, and pulsed excitation. Most likely, the first electrically pumped perovskite laser will require pulsed electrical operation at low temperature. MAPbI$_3$-based PeLEDs have recently been driven at current densities as high as 2.5 kA cm$^{-2}$ (Figure 27e), which is an important thermal management step toward electrically pumped perovskite lasers.[76] At such high current densities, the charge carrier injection imbalance may exacerbate EQE roll-off. Selection of suitable potential barrier layers between charge transporting and emitting layers is a common strategy to suppress the carrier injection imbalance and overflow in the PeLEDs. ETLs with high ionization energy and HTLs with low electron affinity can provide an energy barrier to prevent leakage of injected carriers from emitting layer to the transport layer, reducing EQE roll-off.[236] The Auger recombination might be overcame via new molecular design of perovskite gains and structure optimization of the electrically pumped laser diodes.

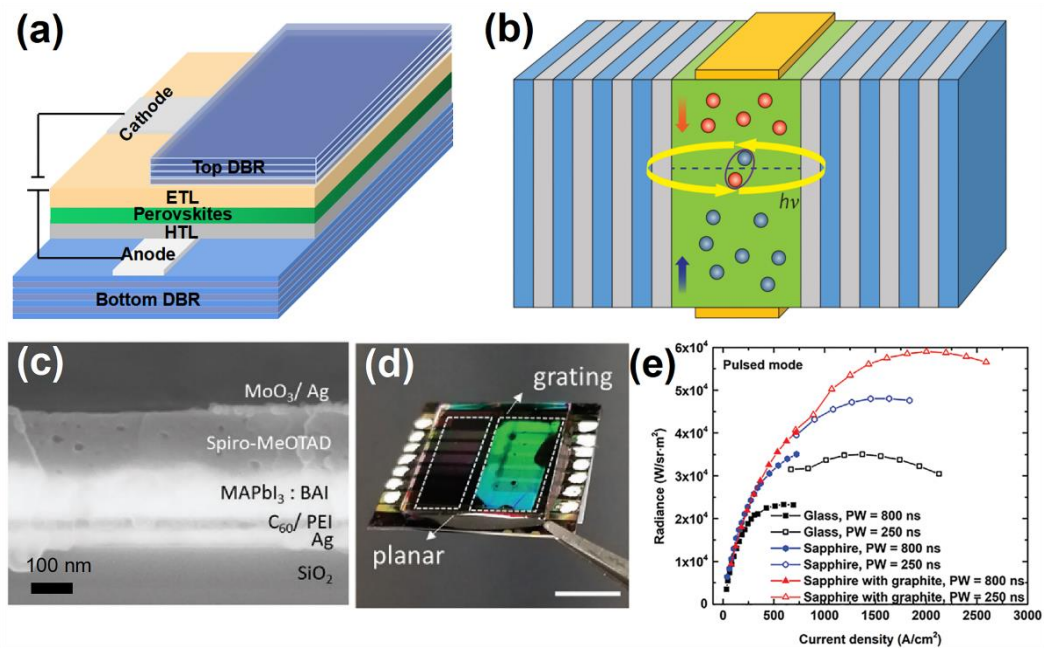

**Figure 27.** (a) The photons cycle in a direction perpendicular to the BDR layers. (b) Electrons (blue circles) and holes (red circles) are electrically injected from below and above, orthogonal to the cavity feedback, thereby minimizing optical losses. $hv$ is the cavity photon. This figure is adapted

from Ref. [235] (c) Cross-sectional scanning electron microscope image of the full device stack. This figure is adapted from Ref. [231] (d) Photograph of completed devices. Iridescence from the DFB grating is visible on the right half of the substrate, while the left half is not imprinted to serve as a planar control. The scale bar is about 1 cm. This figure is adapted from Ref. [231] (e) Radiance–current density curves of PeLEDs on glass or sapphire substrates, with or without a graphite heat spreader and copper heat sink, all driven in pulsed mode. This figure is adapted from Ref. [76]

**Concluding Remarks**

The field of electrically pumped perovskite lasers has aroused significant interests due to the fundamental advances in CW optically pumped lasers and PeLEDs. Researchers have attempted to achieve electrically pumped perovskite lasers, but to date, no true electrically pumped device has been proven based on PeLEDs. The roadmap discussed here demonstrates the multiple obstacles to the realization of electrically pumped perovskite lasers in a variety of ways, such as the integration of optical resonators in functional PeLEDs, Joule heating, charge carrier injection imbalance and Auger recombination. In order to obtain electrically pumped lasers, it is necessary to further investigate the behaviors under intense excitation and formulate strategies to reduce their negative influences. We firmly believe that electrically pumped lasing based on PeLEDs is potentially within reach by combining a well-designed resonator (DBR or DFB), a high-efficiency LED structure, a low-threshold perovskite gain, excellent Joule heating management and suitable charge transporting layers.

**Acknowledgements**

This work was supported by the National Natural Science Foundation of China under Grant 22075277 and the National Key Research and Development Program of China (No. 2019YFA0705900)

## 17. Circularly Polarized Emission of Perovskite Light-Emitting Diodes

*Young-Hoon Kim*[1,2] *and Matthew C. Beard*[1]

[1] National Renwable Energy Laboratory, USA.

[2] Hanyang University, Public of Korea.

**Status**

Since Tan et al. reported the first bright PeLEDs at room temperature EQE = 1% for green emission, EQE = 0.76% for near-infrared emission in 2014,[9] PeLEDs have grown dramatically and achieved EQE = 28.1% from perovskite polycrystalline bulk films[19] and 23.4% from colloidal perovskite NCs.[237] In addition to the rapid growth of normal PeLEDs, PeLEDs which can emit circularly polarized light (CPL) recently have gained intensive attention because polarized light can be applied to various technologies such as information storage, bioencoding and quantum computing.[238] Especially, CPL emitting from halide perovskites combines the advantages of narrow emission linewidth (FWHM ≤ 20 nm) and wide color gamut (≥ 95% in Rec. 2020 standard) with the high PL efficiency afforded by the synthesis of colloidal NCs,[239] which provide the great potential of CPL-emitting PeLEDs in demonstration of 3D displays, hologram display, virtual reality (VR), augmented reality (AR) and eXtended reality (XR).

In 2019, Wang and Vardeny et al. demonstrated the first circularly polarized EL (CP-EL) emission from methylammonium lead bromide films in PeLEDs by injecting spin-polarized charge carriers from metallic ferromagnetic $La_{0.63}Sr_{0.37}MnO_3$ electrodes at 10 K under magnetic field (200 mT) (Figure 28).[240] PeLEDs achieved ≈ 1% of polarization degree of EL, $P_{CP\text{-}EL}$ which is calculated using the equation

$$P_{CP-EL} = \frac{I_{left} - I_{right}}{I_{left} + I_{right}}$$

where $I_{left}$ and $I_{right}$ are the EL intensities of left- and right-CPL, respectively. This pioneering achievement demonstrated that PeLEDs can have the potential to emit CPL.

In 2021, Kim and Beard et al. reported the first CP-EL of any type that operates at room temperature and without external magnetic fields or ferromagnetic metal contacts (Figure 29).[241] Chiral 2D metal halide perovskites which incorporate chiral organic molecules into their crystal framework resulted in ≈ >80% spin-polarized hole current injected into a non-chiral emitter layer, through the chiral-induced spin selectivity (CISS) effect. The spin-orientation and resulting light polarization is determined by the handedness of the chiral organic molecules. The resulting spin-polarized holes were injected into a non-chiral metal-halide perovskite NC emitting layer and subsequently radiatively recombined with electrons with properly aligned spin direction, emitting CP-EL with $P_{CP-EL} \approx 2.6\%$. The CP-EL at room temperature without external magnetic fields or ferromagnetic metals demonstrated the potential of PeLEDs in future 3D/hologram/VR/AR/XR displays which can visualize metaverse and in other various technologies using CPL.

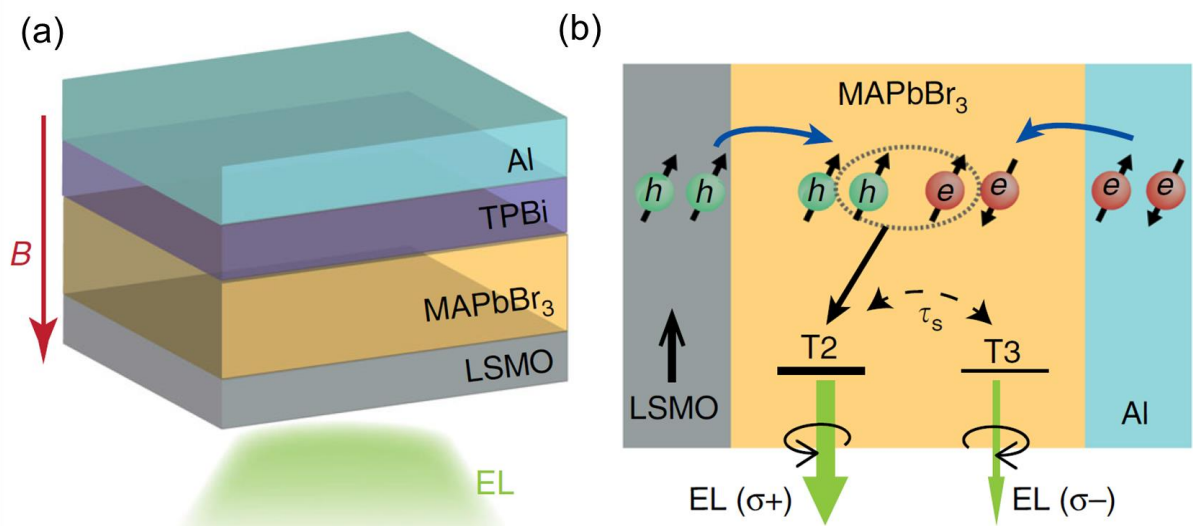

**Figure 28.** Schematic illustration of devices (a) and mechanism (b) of spin-polarized charge injection through ferromagnets and CP-EL emission in PeLEDs at 10K under magnetic fields. Reprinted with permission from [240], Copyright 2019 Wiley-VCH.

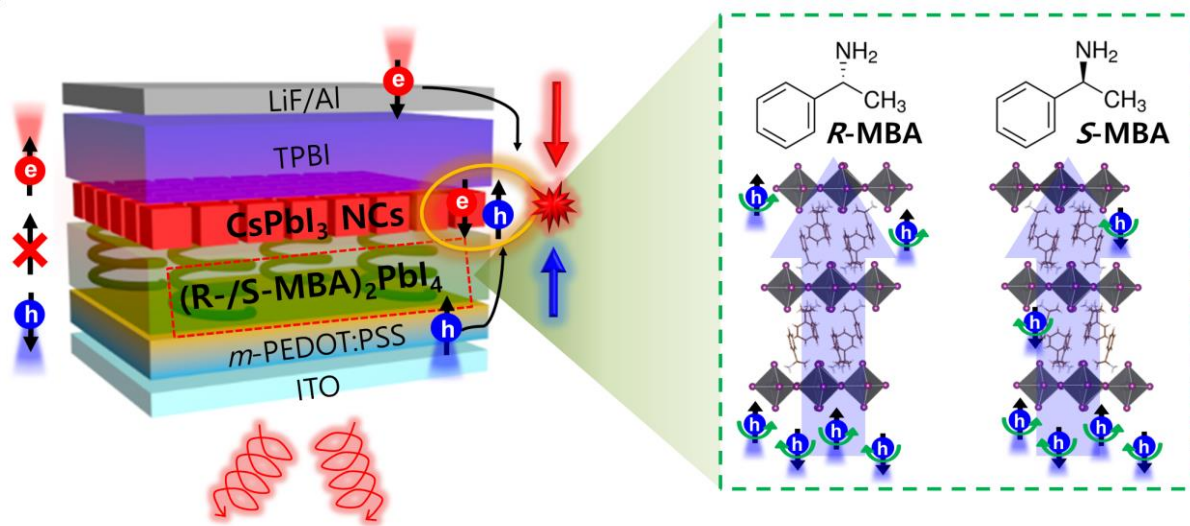

**Figure 29.** Schematic illustration of spin-polarized charge injection through chiral perovskite framework and CP-EL emission in PeLEDs at room temperature without magnetic fields or ferromagnets. Reprinted with permission from [241], Copyright 2021 AAAS.

**Current and Future Challenges**

The most important factor that determines performance of spin-PeLEDs is $P_{CP\text{-}EL}$. Although highly spin-polarized charge carriers (spin-polarization efficiency >80%) were injected to the emitting layer in PeLEDs through the CISS effect, the PeLEDs showed $P_{CP\text{-}EL} \leq 2.6\%$ which is far below that of inorganic spin LEDs based on GaAs multi-quantum-well emitting layer ($P_{CP\text{-}EL} \approx 95\%$) that contain undesirable magnetic elements.[242] $P_{CP\text{-}EL}$ in PeLEDs is directly related to the ratio of the spin lifetime and charge carrier lifetimes ($\tau_{spin}/\tau_{carrier}$),[241] which indicates that light emitting materials in PeLEDs should have long spin lifetime comparable to the luminescence lifetime in order to emit efficient CPL. Currently colloidal perovskite NCs are reported to have spin lifetime $\leq$ 10 ps which is much shorter than the carrier lifetime ($\approx 1-10$ ns),[243] limiting the $P_{CP\text{-}EL}$ in PeLEDs at room temperature.

A second challenge is to enhance the EQE of CP-EL PeLEDs while maintaining high $P_{CP\text{-}EL}$. Although in previous literature PeLEDs increased $P_{CP\text{-}EL}$ by employing mixed halide on the NC emitting states in order to increase the spin lifetime,[241] the resulting EQE was limited due to ion-migration between CISS chiral perovskite polycrystalline films and emitting perovskite NCs and far below that of state-of-the-art normal PeLEDs (EQE ~ 28.1%).[19]

Poor device stability of CP-EL PeLEDs should also be addressed to stimulate interest from researchers. Because halide perovskites are formed in brittle ionic bonding, they can degrade and generate defect states easily upon exposure to joule heating during device operation and under other environmental factors. Furthermore, ion migration accelerated upon electric field significantly reduces the device lifetime. However, we point out that operating device data of CP-EL PeLEDs has not been reported or studied. Recently, single layer graphene deposited between two perovskite NC layers with varying halide was able to block halide migration while allowing for charge to pass through the layer, stable LEDs emitting two colors were demonstrated.[244]

**Advances in Science and Technology to Meet Challenges**

Colloidal perovskite NCs offer many opportunities to increase luminescence efficiency and $P_{CP\text{-}EL}$ in PeLEDs, afforded by excellent chemical tunability of crystal structure and surface ligand attachments. The spin lifetime in the colloidal perovskite NCs is directly related to the $P_{CP\text{-}EL}$ in PeLEDs. The factors that control the spin-lifetime in perovskite NCs is not well understood. A long spin lifetime as well as high luminescence efficiency can be obtained by suppressing spin scattering with defects or grain boundaries. Therefore defect passivation of NCs through various methods such as post-ligand treatment, doping and formation of core/shell structure can increase the both EQE and $P_{CP-EL}$ in PeLEDs. Furthermore, replacing heavy elements (e.g., Pb for B-site cations, I for X-site cations) with light elements (e.g., Mn, Co, Cu for B-site cations, Cl and Br for X-site cations) can weaken spin-orbit coupling and thus increase spin lifetime in perovskite NCs.

Furthermore, inspired by the development of CP-EL OLEDs, we suggest that formation of helically assembled perovskite NCs in emitting layer can further increase the $P_{CP-EL}$ by selective scattering and birefringence in the emitting layers which align the luminescence when passing through the chiral emitting medium. These new systems can be realized by imbedding NC into chiral organic host matrix and can maximize the $P_{CP-EL}$ in PeLEDs. Furthermore, we expect that CP-EL can be enhanced by adhering chiral organic ligands onto the surface of perovskite NCs.

**Concluding Remarks**

In addition to the unprecedentedly fast growth of EQE in PeLEDs, PeLEDs have shown great promise as a light-emitting source of CPL due to their chemical tunability and controllable spin-orbit coupling. By analogy with inorganic spin-LEDs based on GaAs-based light emitters and OLEDs emitting CP-EL as a result of light scattering in the chiral emitting medium, we anticipate that CP-EL PeLEDs have two different strategies to further increase the $P_{CP-EL}$; i) increasing spin lifetime in the emitting materials: ii) having chiral or helical assembly of perovskites in the emitting layer. These strategies are expected to demonstrate the possibility of PeLEDs in future displays, information storage/transfer and quantum computing.

**Acknowledgements**

Support as part of the Center for Hybrid Organic Inorganic Semiconductors for Energy, an Energy Frontier Research Center, funded by the Office of Science within the US Department of Energy is acknowledged. This work was authored in part by the National Renewable Energy Laboratory, operated by Alliance for Sustainable Energy, LLC, for the U.S. Department of Energy (DOE) under Contract No. DE-AC36-08GO28308. The views expressed in this document do not necessarily represents the views of the DOE or the U.S. Government. This work was also supported by the National Research Foundation of Korea (NRF) grant funded by the Korea government (MSIT) (2022R1C1C1008282).

**Data availability statement**

No new data were created or analyzed in this study.